\documentclass[aip,jcp]{revtex4-1}
\pdfoutput=1

\usepackage{natbib}
\usepackage[version=3]{mhchem} % Formula subscripts using \ce{}
\usepackage{bm} % Bold fonts in math environment \bm{}
\usepackage{threeparttable} % Table with notes
\usepackage{dcolumn}
	\newcolumntype{d}[1]{D{.}{.}{#1}}
\usepackage{booktabs} % To thicken table lines
\usepackage{xspace}
\usepackage{braket}
\usepackage{comment}

\usepackage{color, colortbl}
	\newcommand{\chl}{\cellcolor{yellow}} 
\usepackage{soul}
\renewcommand{\hl}[1]{#1}
\renewcommand{\chl}[1]{#1}
\renewcommand{\rowcolor}[1]{}

\usepackage{titlesec}
\titlelabel{\thetitle.\quad}
\usepackage{extarrows}

\usepackage{todonotes}

\usepackage{algorithm}  
\usepackage[noend]{algpseudocode}
\algblockdefx[Name]{Case}{EndCase}%
[1]{\textbf{#1}}%
\usepackage{hyperref} % hyperref must be places before cleveref 
\usepackage{cleveref}	
	\crefname{figure}{Figure}{Figures}
	\crefname{table}{Table}{Tables}
	\crefname{equation}{Eq.}{Eqs.}
	\crefname{section}{Section}{Sections}
	\crefname{subsection}{Section}{Sections}
	\crefname{subsubsection}{Section}{Sections}
	\crefname{algorithm}{Algorithm}{Algorithms}
	\creflabelformat{equation}{#2#1#3}
	\crefrangelabelformat{equation}{#3#1#4 to~#5#2#6}

\newcommand*{\elatt}{\ensuremath{E_{\text{latt}}}\xspace}
\newcommand*{\angstrom}{\mbox{\normalfont\AA}\xspace}

\newcommand*{\mHartree}{mE$_\mathrm{h}$\xspace}
\newcommand*{\uHartree}{$\mu$E$_\mathrm{h}$\xspace}
\newcommand*{\kcal}{kcal mol$^{-1}$\xspace}
\newcommand*{\abinitio}{{\it ab initio}\xspace}
\newcommand*{\I}{\mathrm{i}}
\newcommand*{\bigO}[1]{\mathcal{O}({#1})}
\newcommand*{\etal}{\textit{et al.}\xspace}
\newcommand*{\sub}[1]{\ensuremath{_{\text{#1}}}}
\newcommand*{\super}[1]{\ensuremath{^{\text{#1}}}}
\newcommand*{\mpqc}{{\scshape MPQC}\xspace}
\newcommand*{\MPQC}{Massively Parallel Quantum Chemistry (\mpqc)\xspace}

\newcommand{\dao}[2]{{#1}\bm{#2}}
\newcommand{\local}[3]{{#1}_{(#2, #3)}}
\renewcommand{\tensor}[1]{\ensuremath{\bm{#1}}}
\renewcommand{\vec}[1]{\ensuremath{\bm{#1}}}
\newcommand{\frob}{\mathop{\mathrm{Frob}}}

\newcommand{\rangelimit}[1]{\ensuremath{\bm{R}_{#1}^{\text{max}}}}
\renewcommand{\onlinecite}[1]{\hspace{-1 ex} \nocite{#1}\citenum{#1}} 

\draft % marks overfull lines with a black rule on the right

\begin{document}

% Use the \preprint command to place your local institutional report number 
% on the title page in preprint mode.
% Multiple \preprint commands are allowed.
%\preprint{}

\title{Efficient Evaluation of Exact Exchange for Periodic Systems via Concentric Atomic Density Fitting} %Title of paper

\author{Xiao~Wang}
\affiliation{Department of Chemistry, Virginia Tech, Blacksburg, Virginia 24061, USA}
\affiliation{Center for Computational Quantum Physics, Flatiron Institute, New York, New York 10010, USA}
\author{Cannada~A.~Lewis}
\affiliation{Department of Chemistry, Virginia Tech, Blacksburg, Virginia 24061, USA}
\author{Edward~F.~Valeev}
\email{efv@vt.edu}
\affiliation{Department of Chemistry, Virginia Tech, Blacksburg, Virginia 24061, USA}

\date{\today}

\begin{abstract}

The evaluation of exact (Hartree--Fock, HF) exchange operator is a crucial ingredient for the accurate description of electronic structure in periodic systems through \abinitio and hybrid density functional approaches. An efficient formulation of periodic HF exchange in LCAO representation presented here is based on the concentric atomic density fitting (CADF) approximation, a domain-free local density fitting approach in which the product of two atomic orbitals (AOs) is approximated using a linear combination of fitting basis functions centered at the same nuclei as the AOs in that product. Significant reduction in the computational cost of exact exchange is demonstrated relative to the conventional approach due to avoiding the need to evaluate four-center two-electron integrals, with sub-millihartree/atom errors in absolute Hartree-Fock energies and good cancellation of fitting errors in relative energies. Novel aspects of the evaluation of the Coulomb contribution to the Fock operator, such as the use of real two-center multipole expansions and spheropole-compensated unit cell densities are also described.
\end{abstract}

% insert suggested PACS numbers in braces on next line

\maketitle %\maketitle must follow title, authors, abstract and \pacs

% Body of paper goes here. Use proper sectioning commands. 
\section{Introduction}
\label{sec:intro}

There has been dramatic recent progress in reduced-scaling many-body formalisms for {\em molecular} electronic structure problem.\cite{Ayala1999,Ziokowski2010,Fedorov2005,friedrich_fully_2009,Kobayashi2008,Li2009,riplinger_efficient_2013,werner_efficient_2011,schutz_low-order_1999,ma_explicitly_2018,riplinger_sparse_2016} Such approaches allow robust simulation of electronic structure of large systems that surpasses the accuracy of mainstream methods, i.e. the Kohn--Sham density functional theory (KS DFT), by allowing to systematically eliminate analytic and numerical approximations.
\hl{Although reduced-scaling formalisms, including many-body methods, have been well established in the solid state community,{\cite{del_ben_electron_2013,wilhelm_large-scale_2016,gruneis_natural_2011,kaltak_low_2014,marsili_large-scale_2017,umari_accelerating_2011,neuhauser_breaking_2014,ljungberg_cubic-scaling_2015,booth_towards_2013,ihrig_accurate_2015,hu_adaptively_2017}}
their development applicable to periodic solids is a relatively unexplored frontier in quantum chemistry when compared to the molecular setting, the few pioneering efforts notwithstanding.{\cite{ayala_atomic_2001,f.izmaylov_resolution_2008,pisani_local-mp2_2005,lorenz_local_2011,pisani_cryscor:_2012,schafer_quartic_2017,rebolini_divideexpandconsolidate_2018,usvyat_periodic_2018}}} A significant barrier to straightforward extension of the molecular methods to the periodic setting is the different numerical representations dominant in each setting. The linear combination of atomic orbitals (LCAO) approach is dominant in molecular computations with open-boundary conditions
but it is not commonly used in the periodic context where the use of plane waves is dominant;
\hl{the use of bases with (strictly-)local support has been demonstrated in periodic solids (linear muffin-tin orbitals (LMTO){\cite{skriver_lmto_1984}}, finite/spectral elements{\cite{Jia:2009vm}}, and numerical atomic orbitals{\cite{blum_ab_2009,levchenko_hybrid_2015}}).}
The primary advantage of the LCAO representation is its compactness for atomistic matter, in which the atomic electronic structure is largely preserved; the compactness of representation and the quasi-local support of most important AOs are the key to practical reduced-scaling formalisms for many-body methods.
\hl{Thus, the use of the LCAO representation provides an appealing alternative to the many-body formalisms in solids.}

This paper's focus is the mean-field problem in the LCAO representation for periodic solids, as a precursor to reduced-scaling many-body methods in such representation. Although the basic LCAO machinery
in the periodic setting has long been established,\cite{Ladik:1973il,HARRIS1975147,pisani_exact-exchange_1980,Saunders:1992iz} several recent efforts have focused on how to improve the efficiency of operator evaluation in the LCAO representation by {\em density fitting} (DF).\cite{varga_long-range_2008,f.izmaylov_resolution_2008,burow_resolution_2009,maschio_fast_2007,usvyat_fast_2007,lorenz_local_2011,levchenko_hybrid_2015,sun_gaussian_2017} 
DF is key to fast computation in the LCAO representation, whether it is at the mean-field or many-body level. Long before the AO basis set approaches numerical
completeness the AO product space becomes highly redundant; density fitting is a way to eliminate the redundancy and thus to greatly reduce the cost of operator evaluation (and for AOs that are represented
purely numerically\cite{blum_ab_2009,levchenko_hybrid_2015} density fitting is the only practical way to evaluate operators).\cite{ihrig_accurate_2015}
Another way to view density fitting is as a physically-motivated factorization of the 2- and higher-body operator representations; by separating the tensor modes of the operator
representation the density fitting reduces the cost of operator manipulations such as transformation to the basis of single-particle eigenstates, which is key to reduce the cost of integral
transformation in modern reduced-scaling many-body methods.\cite{Werner03,Neese2009,riplinger_efficient_2013,Pinski2015,riplinger_sparse_2016,werner_efficient_2011,ma_explicitly_2018}

In this paper we describe how to improve the efficiency of the exact exchange operator construction in the periodic LCAO setting
using concentric atomic density fitting (CADF) that has been successfully utilized in molecular electronic structure by us\cite{hollman_semi-exact_2014,hollman_fast_2017} and others.\cite{merlot_attractive_2013,manzer_efficient_2015,ihrig_accurate_2015,rebolini_comparison_2016}
In \cref{sec:theory}, we briefly review the basics of density fitting approximation in electronic structure,
the periodic LCAO HF method, and the implementation of CADF periodic exact exchange algorithm (CADF-K).
\hl{Computations on various 1-, 2-, and 3-dimensional periodic systems} are discussed in \cref{sec:computational_details,sec:results}. Conclusions and future work are given in \cref{sec:conclusions}.

\section{Formalism}
\label{sec:theory}

For the purpose of making this manuscript as self-contained as reasonable, we start out by quickly recapping the basics of density fitting and periodic LCAO mean-field methods.

\subsection{Density Fitting}
\label{sec:df}

Density fitting, also for historical reasons known in
quantum chemistry as the resolution of the identity (RI), approximates product quasi-densities by a linear combination of {\em density-fitting} (also referred to generically as {\em auxiliary}) basis functions.
The modern DF formalism was introduced and analyzed by Whitten for the purpose of approximating the 2-body Coulomb integrals\cite{whitten_coulombic_1973,jafri_electron_1974} (for earlier DF-like approximations see Ref. \onlinecite{Harris:1966ky} and references therein). At the same time Baerends \etal\cite{baerends_self-consistent_1973} used \cref{eq:df} in the context of evaluation of the Coulomb potential of a density represented as a linear combination of 2-center products in terms of 1-electron functions. After much subsequent work\cite{sambe_new_1975,beebe_simplifications_1977,dunlap_approximations_1979,dunlap_firstrow_1979}
DF entered the mainstream of quantum chemistry due to the seminal work by Feyereisen \etal\cite{feyereisen_use_1993,vahtras_integral_1993} and Ahlrichs \etal\cite{eichkorn_auxiliary_1995,eichkorn_auxiliary_1997,bauernschmitt_calculation_1997,weigend_ri-mp2:_1998}.
Careful engineering
of the density fitting basis sets enables us to make the DF error of {\em global} fitting sufficiently small and largely canceling in practically-relevant energy differences;
the computational speed up can be as high as one to two orders of magnitude for small-to-medium sized molecules\cite{eichkorn_auxiliary_1997,weigend_efficient_2002,weigend_approximated_2009,bostrom_ab_2009}.

To highlight the issues with straightforward application of DF to periodic systems, consider fitting 2-center AO products $|\mu\nu) \equiv \phi_\mu^* ({\bf r}) \phi_\nu ({\bf r})$ by its approximate representation $|\widetilde{\mu\nu})$ as a linear combination of 1-center density-fitting basis functions $\{| X ) \} $:
\begin{align}
\label{eq:df}
	|\mu\nu) \overset{\text{DF}}{\approx} |\widetilde{\mu\nu}) \equiv \sum_X C_{\mu,\nu}^X |X).
\end{align}
The set of $\{| X ) \} $ in \cref{eq:df} is a $\mu\nu$-independent system-spanning set in {\em global} fitting
and $\mu\nu$-dependent set in {\em local} fitting;
clearly, DF in an infinite (e.g. periodic) system must necessarily be local to keep the number of fitting coefficients finite.
The fitting coefficients are determined by minimizing the ``norm'' of the error density, $\delta_{\mu, \nu} \equiv |\mu\nu) - |\widetilde{\mu\nu})$, ``weighted'' by an arbitrary operator $\hat{W}$. If the fitting basis is complete, 
then minimizing $(\delta_{\mu,\nu}| \hat{W} |\delta_{\mu,\nu})$ is equivalent to solving
\begin{align}
\label{eq:df1}
\hat{W} |\delta_{\mu,\nu}) = 0.
\end{align}
For any positive $\hat{W}$, such as the Coulomb operator $\hat{V} |f) \equiv \int \frac{f({\bf r}')}{|{\bf r} - {\bf r}'|} \, \text{d}{\bf r}'$,
this is only possible by making $ |\delta_{\mu,\nu}) = 0$, i.e. $ |\mu\nu) =  |\widetilde{\mu\nu})$. 
For a finite fitting basis $|X)$ minimizing $(\delta_{\mu,\nu}| \hat{W} |\delta_{\mu,\nu})$ is equivalent to solving the linear system:
\begin{align}
\label{eq:df2}
\begin{split}
\forall X: (X| \hat{W} |\delta_{\mu,\nu}) = 0 \Rightarrow (X| \hat{W}|\mu\nu) = \sum_Y C_{\mu,\nu}^Y (X|\hat{W}|Y),
\end{split}
\end{align}
which only involves 2- and 3-center integrals, hence is efficient.
Comparison of \cref{eq:df1,eq:df2}  reveals the dual role played by the fitting basis in density fitting in a finite basis:
$\{ |X) \}$ defines not only the expansion space for the fitted density but also the projection space (in the sense of the collocation method)
for solving the exact fitting problem \eqref{eq:df1}. For example, the Coulomb fitting ($\hat{W} = \hat{V}$) via \cref{eq:df2}
can be viewed as ensuring that the Coulomb potential of the error density $\delta_{\mu,\nu}$
is zero ``at'' every collocation function $|X)$.
Also, unless $\hat{W}=1$ ``density fitting'' is a misnomer: in an incomplete basis
minimizing $(\delta_{\mu,\nu}| \hat{W} |\delta_{\mu,\nu})$ is equivalent to minimizing the 2-norm of the ``potential'' generated by $\hat{W}^{1/2}$
from the error density $|\delta_{\mu,\nu})$; since $\hat{V}^{1/2} |f) \equiv \pi^{-3/2} \int \frac{f({\bf r}')}{|{\bf r} - {\bf r}'|^2} \, \text{d}{\bf r}'$,
the Coulomb fitting is equivalent to minimizing the error in the {\em electric field}.

The error in a 4-center Coulomb integral can be easily seen as quadratic in the density fitting errors (i.e. its DF approximation is {\em robust}\cite{dunlap_robust_2000,Dunlap:2000iw}) if (a) $\hat{W} = \hat{V}$, and (b) the same fitting basis set is used
for the bra and the ket:
\begin{align}
\label{eq:df3}
(\rho\sigma | \hat{V} |\mu\nu) - (\widetilde{\rho\sigma} | \hat{V} | \widetilde{\mu\nu}) = (\delta_{\rho,\sigma} | \hat{V} | \delta_{\mu,\nu}) + (\delta_{\rho,\sigma}| \hat{V} | \widetilde{\mu\nu}) + (\widetilde{\rho\sigma} | \hat{V} | \delta_{\mu,\nu}) \xlongequal{\hat{W} = \hat{V}, \{X\}_{\rho,\sigma} = \{X\}_{\mu,\nu}} (\delta_{\rho,\sigma} | \hat{V} | \delta_{\mu,\nu}),
\end{align}
where each component of $(\delta_{\rho,\sigma}| \hat{V} | \widetilde{\mu\nu})$ and $(\widetilde{\rho\sigma} | \hat{V} | \delta_{\mu,\nu})$ vanishes according to \cref{eq:df2}. 
Chemists' notation is used for the two-electron integrals here and in the rest of this paper. Condition (b) is satisfied automatically in global Coulomb fitting, hence its ubiquitousness. For local fitting, or global fitting of an indefinite operator $\hat{O}$, a manifestly
robust approximation\cite{dunlap_robust_2000} can be used:
\begin{align}
\label{eq:df4}
(\rho\sigma | \hat{O} |\mu\nu) & \overset{\rm robust\,DF}{=} 
(\rho\sigma | \hat{O} | \widetilde{\mu\nu}) + (\widetilde{\rho\sigma} | \hat{O} | \mu\nu ) - (\widetilde{\rho\sigma} | \hat{O} | \widetilde{\mu\nu}).
\end{align}

One of the challenges for local DF is making it accurate without introducing artifacts. Fundamentally, the problem with local fitting is that
the potential generated by $|\widetilde{\mu\nu})$ is only accurate where the fitting functions are located. Thus the local DF will introduce large errors in
integrals $(\rho\sigma | \hat{V} |\mu\nu)$ with large bra-ket distances. This issue is avoided in {\em domain-based} local DF\cite{Werner03,schutz_linear_2003} 
by fitting $|\rho\sigma)$ and $|\mu\nu)$ using set of $|X)$ specialized
to every $(\rho\sigma, \mu\nu)$ combination; the resulting DF approximation is robust.
This makes both $|\widetilde{\rho\sigma})$ and $|\widetilde{\mu\nu})$ optimal for approximating $(\rho\sigma | \hat{V} |\mu\nu)$ accurately for any $\rho\sigma-\mu\nu$
distance.
The domain-based local fitting has been used for periodic systems in the context of local LCAO MP2 methods by Schutz, Usvyat, and co-workers.\cite{maschio_fast_2007,usvyat_fast_2007,pisani_periodic_2008,maschio_fitting_2008,maschio_local_2011}
The domain fitting greatly increases the number of fitting problems to be solved relative to the global and purely local approaches
and thus pushes the crossover w.r.t. the global fitting to relatively large systems.
Also note that the domain DF requires integrals $(X|\hat{V}|\mu \nu)$ where $\{|X)\}$ span fitting sets for all possible $|\rho\sigma)$ products, i.e. potentially
the entire system.
Thus purely local fitting (where $\{|X)\}$ are ``local'' to each $|\mu\nu)$) would be preferred if it can be made accurate.

The accuracy of purely local fitting can be systematically tuned by several approaches:
\begin{itemize}
\item by selecting sufficiently complete fitting sets, whether by brute force \hl{{\cite{levchenko_hybrid_2015,ihrig_accurate_2015}}} or by augmenting the fitting basis with plane waves as recently done by Sun and co-workers\cite{sun_gaussian_2017}. Note the artifactually slow decay of fitting errors that occurs in 1-dimensional systems\cite{Gill:2005jw}.
\item by improving the accuracy of long-range potential via (a) constraining the fitted density to match the leading multipole moments as the exact density
or, similarly, (b) including distant functions in the collocation fitting set in \cref{eq:df2} \cite{Tew:2018cf};
\item by using the robust DF approximant (\cref{eq:df4}) to reduce the error; however, this can lead to artifacts such as loss of positivity of the Coulomb operator.\cite{merlot_attractive_2013}
\end{itemize}
The accuracy of the long-range tail of the potential generated by the fitted density is especially important for the periodic systems
in the context of computing the Coulomb potential due to its slow decay. This means at least constraining the charge of the fitted density to ensure the exact electrical neutrality
of the unit cell and thus ensure that the Coulomb potential is finite.\cite{jaffe_gaussian_1996,varga_density_2005,varga_long-range_2008,burow_resolution_2009,lazarski_density_2015,franchini_accurate_2014}

The long-range behavior of the exact exchange operator is less problematic, at least in insulators, due to the rapid decay of the 1-particle density matrix. Therefore it is reasonable to assume that the use of
purely local density fitting would be a viable strategy even in periodic systems. Here we extend our previous efforts\cite{hollman_semi-exact_2014,hollman_fast_2017} on fast {\em strongly}-local density fitting for exchange operator construction in molecules to the periodic setting. By strong locality we mean that the fitting space for product $|\mu_1\mu_2\dots)$ is composed only of the fitting basis functions on the same atoms as $\mu_i$; we dub such
local DF approximation {\em concentric atomic} DF. The idea of CADF-like fitting goes back to the work of Baerends and co-workers,\cite{baerends_self-consistent_1973} who used it to fit atomic components
of the 1-particle density. Recently the concentric fitting approach, referred to as pair atomic resolution of the identity (PARI), was investigated by Merlot \etal\cite{merlot_attractive_2013} in the context of Coulomb fitting;
the authors were first to point out the violation of positivity by the {\em robust} PARI approximant that led to catastrophic failures. Hollman and co-workers\cite{hollman_semi-exact_2014} showed that the accuracy
of the CADF could be dramatically improved and the nonpositivity artifacts could be partially alleviated by combining
CADF approximation for most integrals with exact evaluation of 4-center integrals, dubbed semiexact CADF (seCADF),\cite{hollman_semi-exact_2014}
as well as developed a rigorously screened {\em linear scaling} version of CADF exact exchange builder.\cite{hollman_fast_2017}
Manzer \etal\cite{manzer_efficient_2015} demonstrated that the nonpositivity violations do no occur when robust PARI is only used for the exchange operator.

The above applications of CADF/PARI in molecular context were recently mirrored in the molecular\cite{ihrig_accurate_2015} {\em and} periodic\cite{levchenko_hybrid_2015} setting by Scheffler and co-workers, using
non-Gaussian LCAO representation based on numeric atomic orbitals that have strictly local support.
Unlike the earlier Gaussian-based {\em robust} CADF efforts, the use of {\em nonrobust} local RI approach for fitting numerical AO products
causes the need for \hl{the construction of high-quality fitting basis sets with improved accuracy.}\cite{ihrig_accurate_2015}
Application of robust CADF to the exact exchange construction in periodic systems has not yet been accomplished to the best of our knowledge, in Gaussian or non-Gaussian LCAO representations, hence this is the primary objective of this work.

\subsection{Periodic Hartree-Fock}
\label{sec:overview}
\hl{The periodic Hartree-Fock procedure implemented in the current work is rather standard and can be found in the literature since the 1980s.{\cite{HARRIS1975147,pisani_exact-exchange_1980,pisani_hartree-fock_1988,dovesi_periodic_2000}}} 
The Hartree-Fock crystalline orbitals (CO) in the LCAO framework are expanded,
\begin{align}
\psi_{i, \vec{k}} (\vec{r}) &= \sum_{\mu} C^i_{\mu, \vec{k}}\phi_{\mu, \vec{k}} (\vec{r}),
\end{align}
in terms of Bloch AOs, which in turn are translation-symmetry-adapted linear combinations of the AOs:
\begin{align}
\phi_{\mu, \vec{k}} (\vec{r}) &\equiv \sum_{\vec{R}} e^{	\I \vec{k}\cdot\vec{R}} \chi_{\mu} (\vec{r} - \vec{R}).
\end{align}
The CO coefficients $C^i_{\mu, \vec{k}}$ in this work are determined by solving the {\em canonical} Hartree-Fock equations projected on the AOs in the reference cell:
\begin{align}
\tensor{F_k} \tensor{C_k} &= \tensor{S_k} \tensor{C_k} \tensor{\epsilon_k}, \label{eq:hf}
\end{align}
where \tensor{F_k}, \tensor{S_k}, and \tensor{\epsilon_k} are Fock matrix, overlap matrix, and CO (band) energies, respectively.
In this work these equations were solved for a uniform grid of \vec{k} points in the first Brillouin zone, without exploiting any additional symmetries.
The Fock matrix in \cref{eq:hf} is a linear combination of conventional AO (direct space) Fock matrices:
\begin{align}
F_{\mu, \nu, \vec{k}} &= \sum_{\vec{R}} e^{\I \vec{k}\cdot\vec{R}} F_{\mu, \dao{\nu}{R}}, \label{eq:F_k} \\
F_{\mu, \dao{\nu}{R}} &\equiv \langle \mu | \hat{F} | \dao{\nu}{R} \rangle, 
\end{align}
where a compound index $\dao{\mu}{R}$ is used to represent the $\mu$-th AO located in cell \vec{R}, i.e. $\chi_{\mu} (\vec{r} - \vec{R})$. For AOs in the reference cell, $\vec{R} = \vec{0}$ and thus \vec{R} will be omitted for simplicity.
Note that the Fock matrix is Hermitian, i.e. $F_{\mu, \nu, \vec{k}} = F_{\nu, \mu, \vec{k}}^*$, which requires the AO Fock matrix to obey
\begin{align}
\label{eq:F_Hermiticity}
F_{\mu, \dao{\nu}{R}} = F_{\nu, \dao{\mu}{(-R)}}^*.
\end{align}
This condition is closely related to the requirement of translational invariance: $F_{\mu, \dao{\nu}{R}} = F_{\dao{\mu}{(-R)}, \nu}$.

The closed-shell Fock matrix in AO basis is expressed in terms of the usual kinetic energy, nuclear and electronic Coulomb potential, and exchange contributions:\cite{dovesi_periodic_2000}
\begin{align}
F_{\mu, \dao{\nu}{R}} &= T_{\mu, \dao{\nu}{R}} + V_{\mu, \dao{\nu}{R}} + 2 J_{\mu, \dao{\nu}{R}} - K_{\mu, \dao{\nu}{R}}, \label{eq:F_comp}\\
T_{\mu, \dao{\nu}{R}} &= - \frac{1}{2} \langle \mu | \nabla_{\vec{r}}^2 | \dao{\nu}{R} \rangle, \label{eq:T}\\
V_{\mu, \dao{\nu}{R}} &= \sum_{\vec{G}} \sum_{A=1}^{N\sub{atom}} \langle \mu | \frac{Z_A}{|\vec{r} - \vec{A} - \vec{G}|} | \dao{\nu}{R} \rangle, \label{eq:V} \\
J_{\mu, \dao{\nu}{R}} &= \sum_{\vec{G}} \sum_{\vec{L}} \sum_{\rho, \sigma} \left( \mu \dao{\nu}{R}| \frac{1}{r_{12}} | \dao{\sigma}{(G+L)} \dao{\rho}{G}  \right) D_{\dao{\rho}, \dao{\sigma}{L}} , \label{eq:J}\\
K_{\mu, \dao{\nu}{R}} &= \sum_{\vec{G}} \sum_{\vec{L}} \sum_{\rho, \sigma} \left( \mu \dao{\rho}{G} | \frac{1}{r_{12}} |   \dao{\sigma}{(G+L)} \dao{\nu}{R} \right) D_{\dao{\rho}, \dao{\sigma}{L}}, \label{eq:K}
\end{align}
where \vec{A} denotes the nuclear position of atom $A$ in the reference cell and $D_{\rho,\dao{\sigma}{L}}$ density matrix. 
Indices \vec{R}, \vec{G}, and \vec{L}, span the infinite set of lattice vectors and are associated with pseudo-overlap distribution, Coulomb operator, and density representation, respectively. Note that the translational invariance
of \tensor{D} ,
\begin{align}
D_{\dao{\rho}{G},\dao{\sigma}{(G+L)}} &= D_{\rho, \dao{\sigma}{L}},\label{eq:D_translt}
\end{align}
is exploited in \cref{eq:J,eq:K}. The AO (direct-space) density matrix $D_{\rho, \dao{\sigma}{L}}$ is constructed from the CO coefficients as
\begin{align}
D_{\rho, \dao{\sigma}{L}} = \frac{1}{\Omega} \int \mathrm{d}\vec{k}  \, e^{\I \vec{k}\cdot\vec{L}} \sum_i^{\mathrm{occ}} {C^{i}_{\rho, \vec{k}}} \left(C^{i}_{\sigma, \vec{k}}\right)^*, \label{eq:D}
\end{align}
where $\Omega$ is the volume of the first Brillouin zone of the reciprocal lattice.
Columns of $F_{\mu, \dao{\nu}{R}}$ (and $S_{\mu, \dao{\nu}{R}}$) are transformed to Bloch AO basis for each \vec{k}
before solving the Hartree-Fock equations (\cref{eq:hf}).
The Hartree--Fock energy is obtained as
\begin{align}
	E^\text{HF} &= \sum_{\vec{R}}\sum_{\mu, \nu} D_{\mu, \dao{\nu}{R}} \left(F_{\mu, \dao{\nu}{R}} + T_{\mu, \dao{\nu}{R}} + V_{\mu, \dao{\nu}{R}} \right). \label{eq:energy}
\end{align}

All contributions to the AO Fock matrix for a gapped system can be evaluated to arbitrary finite precision $\epsilon$ with finite effort despite the infinite ranges of indices \vec{R}, \vec{G},  and \vec{L}.
First, finite support of Gaussians limits the range of \vec{R} in \cref{eq:T,eq:V,eq:J} and \vec{L} in \cref{eq:J} to polylog of precision (i.e. $\propto -\log^k\epsilon$). Summation over \vec{G} in the Coulombic contributions
(\cref{eq:V,eq:J}) is slow and must be carried out to sufficiently long range to make \tensor{V} and \tensor{J} Hermitian according to \cref{eq:F_Hermiticity}. However, fast techniques for the Coulomb potential evaluation, such as the fast multipole method (FMM),
are well established and will not be discussed here; in this work we used multipole approximation to evaluate the Coulomb potential efficiently
(see Appendix \ref{sec:appendix} for technical details of real-valued multipole expansion employed here).
Lastly, \vec{G} and \vec{(G+L-R)} in \cref{eq:K} have polylog ranges due to finite support of Gaussians, whereas the range of the \vec{L} sum is also
polylog in {\em gapped} systems due to the exponential decay of  the density, $D_{\rho, \dao{\sigma}{L}}$\cite{kohn_analytic_1959,cloizeaux_analytical_1964,he_exponential_2001,taraskin_spatial_2002}. However, in metallic systems convergence of the \vec{L} sum in \cref{eq:K} will be slow (polynomial in precision)\cite{goedecker_decay_1998,ismail-beigi_locality_1999,taraskin_spatial_2002-1}. The remaining question is that of cost.
The computational effort of periodic LCAO HF is primarily dominated by the evaluation of the four-center two-electron integrals, particularly in \cref{eq:K} (explicit evaluation of the integrals in \tensor{J}, denoted as 4c-J,
is only needed for the near-field contribution, i.e. small values of $|\bf{G}|$, and can be reduced further by density fitting).
The next section will describe how the computational cost of \cref{eq:K}  can be greatly reduced by local density fitting.

\subsection{Concentric Atomic Density Fitting for the Exchange Operator}
\label{sec:cadf_k}

The concentric atomic density fitting (CADF) is a local approximation to \cref{eq:df}
\begin{align}
|\mu\nu) = \sum_{X\in (\mu, \nu)} C_{\mu,\nu}^{\local{X}{\mu}{\nu}} |\local{X}{\mu}{\nu}), \label{eq:C_local}
\end{align}
where $\local{X}{\mu}{\nu}$ are the density fitting functions that are concentric with functions $\mu$ {\em or} $\nu$. Nonrobust CADF approximation,
\begin{align}
(\mu\nu|\rho\sigma) \overset{\text{nonrobust~CADF}}{=} \sum_{\substack{X\in (\mu, \nu) \\ Y\in (\rho, \sigma)} } C_{\mu,\nu}^{\local{X}{\mu}{\nu}} (\local{X}{\mu}{\nu} |\local{Y}{\rho}{\sigma}) C_{\rho,\sigma}^{\local{Y}{\rho}{\sigma}} , \label{eq:nonrobust_cadf}
\end{align}
is positive definite, \hl{but inaccurate with fitting basis sets of practical size}, 
hence robust CADF is used in this work:
\begin{align}
(\mu\nu|\rho\sigma) \overset{\text{CADF}}{=} \sum_{X\in (\mu, \nu)} C_{\mu,\nu}^{\local{X}{\mu}{\nu}} (\local{X}{\mu}{\nu} | \rho\sigma) + \sum_{Y\in (\rho, \sigma)}  (\mu\nu | \local{Y}{\rho}{\sigma} ) C_{\rho,\sigma}^{\local{Y}{\rho}{\sigma}} - \sum_{\substack{X\in (\mu, \nu) \\ Y\in (\rho, \sigma)} } C_{\mu,\nu}^{\local{X}{\mu}{\nu}} (\local{X}{\mu}{\nu} |\local{Y}{\rho}{\sigma}) C_{\rho,\sigma}^{\local{Y}{\rho}{\sigma}} . \label{eq:robust_df}
\end{align}
Although robust CADF can violate positivity,\cite{merlot_attractive_2013}  it is safe to use for \tensor{K} (\cref{eq:K}).\cite{manzer_efficient_2015} Inserting \cref{eq:robust_df} to \cref{eq:K} leads to the following expression (Einstein's convention is assumed for \cref{eq:K_df_1,eq:K_df_2,eq:K_df}):
\begin{align}
\begin{split} \label{eq:K_df_1}
K_{\mu, \dao{\nu}{R}} \overset{\text{CADF}}{=} \, & \Big[ C_{\mu,\dao{\rho}{G}}^{\local{X}{\mu}{\dao{\rho}{G}}} (\local{X}{\mu}{\dao{\rho}{G}} |  \dao{\sigma}{(G+L)} \dao{\nu}{R}) + (\mu \dao{\rho}{G} | \local{Y}{\dao{\sigma}{(G+L)}}{\dao{\nu}{R}}) C_{\dao{\sigma}{(G + L)}, \dao{\nu}{R}}^{\local{Y}{\dao{\sigma}{(G+L)}}{\dao{\nu}{R}}} \\
& - C_{\mu, \dao{\rho}{G}}^{\local{X}{\mu}{\dao{\rho}{G}}} (\local{X}{\mu}{\dao{\rho}{G}}  | \local{Y}{\dao{\sigma}{(G+L)}}{\dao{\nu}{R}}) C_{\dao{\sigma}{(G+L)}, \dao{\nu}{R}}^{\local{Y}{\dao{\sigma}{(G+L)}}{\dao{\nu}{R}}} \Big] D_{\rho, \dao{\sigma}{L}}
\end{split} \\
\begin{split} \label{eq:K_df_2}
= & 2 \, C_{\mu,\dao{\rho}{G}}^{\local{X}{\mu}{\dao{\rho}{G}}} (\local{X}{\mu}{\dao{\rho}{G}} | \dao{\sigma}{(G+L)} \dao{\nu}{R}) D_{\rho, \dao{\sigma}{L}} \\
& - C_{\mu,\dao{\rho}{G}}^{\local{X}{\mu}{\dao{\rho}{G}}} (\local{X}{\mu}{\dao{\rho}{G}}|\local{Y}{\dao{\sigma}{(G+L)}}{\dao{\nu}{R}}) C_{\dao{\sigma}{(G+L)}, \dao{\nu}{R}}^{\local{Y}{\dao{\sigma}{(G+L)}}{\dao{\nu}{R}}} D_{\rho, \dao{\sigma}{L}}
\end{split} \\
= & Q_{\mu, \dao{\sigma}{L'}}^{\dao{X}{G'}} F_{\dao{\sigma}{L'},\dao{\nu}{R}}^{\dao{X}{G'}}, \label{eq:K_df}
\end{align}
where
\begin{align}
Q_{\mu, \dao{\sigma}{L'}}^{\dao{X}{G'}} &\equiv \sum_{\vec{G}}\sum_{\rho} C_{\mu,\dao{\rho}{G}}^{\local{X}{\mu}{\dao{\rho}{G}}} D_{\rho, \dao{\sigma}{L}}, \label{eq:Q} \\
F_{\dao{\sigma}{L'},\dao{\nu}{R}}^{\dao{X}{G'}} &\equiv 2 \, (\local{X}{\mu}{\dao{\rho}{G}} | \dao{\sigma}{(G+L)} \dao{\nu}{R}) - \sum_{Y\in(\dao{\sigma}{(G+L)}, \dao{\nu}{R})} (\local{X}{\mu}{\dao{\rho}{G}}|\local{Y}{\dao{\sigma}{(G+L)}}{\dao{\nu}{R}}) C_{\dao{\sigma}{(G+L)}, \dao{\nu}{R}}^{\local{Y}{\dao{\sigma}{(G+L)}}{\dao{\nu}{R}}}. \label{eq:cadf_F}
\end{align}
Note the appearance of lattice vector indices \vec{G'} and \vec{L'} that distinguish cell indices of $X$ and $\sigma$ on the left-hand side of \cref{eq:Q,eq:cadf_F} from cell indices
\vec{G} and \vec{L} on the right-hand side. The appearance of \vec{G'} and \vec{L'} in \cref{eq:Q} can be understood if we express \tensor{Q} using translationally-redundant density representation:
\begin{align}
Q_{\mu, \dao{\sigma}{L'}}^{\dao{X}{G'}} &\equiv \sum_{\vec{G}}\sum_{\rho} C_{\mu,\dao{\rho}{G}}^{\local{X}{\mu}{\dao{\rho}{G}}} D_{\dao{\rho}{G}, \dao{\sigma}{(G+L)}}. \label{eq:Q_untranslated}
\end{align}
Due to the dependence of $\local{X}{\mu}{\dao{\rho}{G}}$ and $\dao{\sigma}{(G + L)}$ on \vec{G}, summation over \vec{G} produces nonzero elements of $Q$ with functions $X$ and $\sigma$ spread
in a range of cells, with the respective ranges labeled by \vec{G'} and \vec{L'}.
In contrast, in \cref{eq:cadf_F} indices \vec{G'} and \vec{L'} are introduced simply as notational convenience.
Note that only translationally unique components of $F_{\dao{\sigma}{L'},\dao{\nu}{R}}^{\dao{X}{G'}}$ need to be evaluated:
\begin{align}
F_{\dao{\sigma}{L'},\dao{\nu}{R}}^{\dao{X}{G'}} =  F_{\dao{\sigma}{(L'-R)},\nu}^{\dao{X}{(G'-R)}}. \label{eq:unique_F}
\end{align}

Note that \cref{eq:K_df_2} assumes the equality of first two terms on the right-hand side of \cref{eq:K_df_1}. This assumption is satisfied given a sufficiently large lattice range for the \tensor{G} and \tensor{L} sums.
Indeed, consider product $|\mu\dao{\nu}{R_1})$ and density $D_{\rho,\dao{\sigma}{R_2}}$. As discussed above, the maximum range of \vec{R_1} and \vec{R_2} are polylog in precision due to
finite support of Gaussians and rapid decay of the density in {\em gapped} systems. The two range extents are denoted as \rangelimit{o} and \rangelimit{d}, respectively, i.e. $|\vec{R_1}| \leq |\rangelimit{o}|$ and $|\vec{R_2}| \leq |\rangelimit{d}|$. 
In this work \rangelimit{o} is determined before SCF
starts while \rangelimit{d} is updated in every SCF iteration based on the specified precision thresholds (see \cref{sec:computational_details}). Therefore, ranges of lattice vectors in \cref{eq:K_df_1,eq:K_df_2,eq:K_df,eq:Q,eq:cadf_F} are: 
\begin{itemize}
	\item $|\vec{G}| \leq |\rangelimit{o}|$ and $|\vec{L}| \leq |\rangelimit{d}|$, because \vec{G} and \vec{L} are involved in pseudo-overlap and density representations, respectively.
	\item $|\vec{R}| \leq 2\,|\rangelimit{o}| + |\rangelimit{d}|$, because cell \vec{R} is linked to cell \vec{0} via
	\begin{align}
		\mu \overset{|\rangelimit{o}|}{\longleftrightarrow}  \dao{\rho}{G} \overset{|\rangelimit{d}|}{\longleftrightarrow}  \dao{\sigma}{(G+L)}  \overset{|\rangelimit{o}|}{\longleftrightarrow}  \dao{\nu}{R}. \label{eq:connect}
	\end{align} 
	\item $|\vec{G'}| \leq |\rangelimit{o}|$. Since $\{\dao{X}{G'}\}$ spans the union of density fitting functions used to represent all non-negligible $(\mu, \dao{\rho}{G})$ pairs, sums over \vec{G'} and \vec{G} have the same extent.  
	\item $|\vec{L'}| \leq |\rangelimit{o}| + |\rangelimit{d}|$. Since $\{\dao{\sigma}{L'}\}$ spans the union of all possible $\dao{\sigma}{(G+L)}$, which must be connected to $\dao{\mu}{0}$ via \cref{eq:connect}, sums over \vec{L'} and $(\vec{G} + \vec{L})$ have the same extent.
\end{itemize}

The CADF coefficients are precomputed before the SCF iterations by solving the usual fitting linear system,
\begin{align}
\left( \local{Y}{\mu}{\dao{\rho}{G}} | \mu\dao{\rho}{G} \right) = & \sum_{X\in(\mu,\dao{\rho}{G})}  M^{\local{Y}{\mu}{\dao{\rho}{G}}}_{\local{X}{\mu}{\dao{\rho}{G}}} C_{\mu,\dao{\rho}{G}}^{\local{X}{\mu}{\dao{\rho}{G}}} , \label{eq:C_local_solid}
\end{align}
with the Coulomb fitting ``metric'' $ M^{\local{Y}{\mu}{\dao{\rho}{G}}}_{\local{X}{\mu}{\dao{\rho}{G}}}  \equiv \left(\local{Y}{\mu}{\dao{\rho}{G}}|\local{X}{\mu}{\dao{\rho}{G}}\right)$.
Fitting coefficients $C_{\dao{\nu}{R},\dao{\sigma}{(G+L)}}^{\local{Y}{\dao{\sigma}{(G+L)}}{\dao{\nu}{R}}}$ in \cref{eq:cadf_F} are then obtained by translation.
Note that the number of non-negligible $(\mu, \dao{\rho}{G})$ pairs for each AO $\mu$ is polylog in precision, hence the fitting coefficient tensor is sparse.
Tensor \tensor{Q} is also sparse in insulators due to the exponential decay of the density.
However, note that tensor \tensor{F} is not sparse due to its slow (polynomial) decay. Nevertheless, evaluation of \cref{eq:K_df} is fast
because only a sparse subset of \tensor{F} contributes to the contraction with sparse \tensor{Q} in \cref{eq:K_df}. Indeed, in \cref{eq:K_df}, we can see that for a specific pair of ($\dao{X}{G'}$, $\dao{\sigma}{L'}$), no contribution to $K_{\mu,\dao{\nu}{R}}$ will be made if $Q_{\mu, \dao{\sigma}{L'}}^{\dao{X}{G'}}$ is negligible for all $\mu$. Thus, based solely on the sparsity of \tensor{Q}, a prescreening list of ($\dao{X}{G'}$, $\dao{\sigma}{L'}$) pairs, named $L_{X \sigma}$, can be formed prior to the construction of \tensor{F} to reduce the computational cost. This list defines which blocks of \tensor{F} are computed (see \cref{sec:computational_details} for the details of tensor blocking).

\hl{ Of course, the CADF approximation is not sufficient to arrive at an efficient method since it is necessary to also screen out small
contributions in every step of the CADF-K algorithm. Although it is possible to evaluate a block-sparse tensor expression like the CADF-K expression (}\cref{eq:K_df_1} \hl{) such that all contributions above certain threshold $\epsilon$ are included, the necessary analysis is complicated (see Ref. } \onlinecite{hollman_fast_2017} \hl{for how this can be done for the nonperiodic CADF-K). Here we sacrifice the ability to control the precision of the exchange matrix by a single threshold and instead
introduce separate thresholds for exploiting sparsity in individual steps. Since for the efficiency reasons
the AO tensors in our implementation are {\em tiled} (blocked) not by atoms but by sets of one or more atoms (up to an entire unit cell), the screening is adapted to work approximately the same no matter how the tiling is chosen. The following thresholds are utilized:}
\begin{itemize}
	\item \hl{$\epsilon_{\mathrm{schwarz}}$: Magnitudes of three- and four-center ERIs are estimated using the standard (Cauchy-Bunyakovsky-)Schwarz bound{\cite{haser_improvements_1989}}:}
		\begin{align}
			| (X\vec{G'}| \mu \rho\vec{G}) | \leq& (X|X)^{1/2} ( \mu \rho\vec{G}| \mu \rho\vec{G})^{1/2}, \\
			| (\mu \nu\vec{R}| \sigma\vec{(G+L)} \rho\vec{G}) | \leq& (\mu \nu\vec{R}|\mu \nu\vec{R})^{1/2}  (\sigma\vec{L} \rho|\sigma\vec{L} \rho)^{1/2},
		\end{align}
	\hl{respectively. The bounds for the individual integrals are then straightforwardly combined to produce bounds for the Frobenius norms of tiles of the ERI tensors. A tile of integrals is neglected if its estimated Frobenius norm is below $\epsilon_{\mathrm{schwarz}}$ times the number of integrals of the tile. This is equivalent to skipping tiles with \textit{per-element} (scaled) Frobenius norms less than $\epsilon_{\mathrm{schwarz}}$; in other words, the skipped tiles contain integrals with average magnitude below $\epsilon_{\mathrm{schwarz}}$.}
	\item $\epsilon_{S}$: For any {\em non-concentric} orbital basis set (OBS) shell pair ($\mu$, $\dao{\nu}{R}$), if the \hl{per-element} Frobenius norm of its overlap integral is below $\epsilon_{S}$, i.e.
		\begin{align}
		\frob_{\mu'\in \mu, \nu'\in \nu}\{\langle \mu' |\dao{\nu'}{R}\rangle\} < \epsilon_{S}, \label{eq:epsilon_s_def}
		\end{align} 
	we will neglect integrals associated with the corresponding pseudo-overlap distributions, such as $|\mu \dao{\nu}{R})$ in \cref{eq:T,eq:V,eq:J} and $|\mu \dao{\rho}{G})$ in \cref{eq:C_local_solid}. \vec{R}'s range limit, \rangelimit{o}, is determined when all ($\mu$, $\dao{\nu}{R}$) pairs for a specific \vec{R} satisfy \cref{eq:epsilon_s_def}.
	\item $\epsilon_{D}$: \hl{If the per-element Frobenius norm of all tiles in density $D_{\rho, \dao{\sigma}{L}}$ for a given unit cell} \vec{L} is below $\epsilon_{D}$, \hl{that unit cell's} density is considered to be negligible; \hl{this screening defines} the \rangelimit{d} \hl{range}.
	\item $\epsilon_{F}$: \hl{This parameter controls which tiles of} \tensor{F} (\cref{eq:cadf_F}) \hl{are computed by defining}  the prescreening list $L_{X\sigma}$ (see \cref{alg:cadf}).
	\item $\epsilon_{\text{sparse}}$: \hl{A tile of tensor} \tensor{Q} (\cref{eq:Q}) or \tensor{F} (\cref{eq:cadf_F}) \hl{is omitted if the per-element Frobenius norm is below} $\epsilon_{\text{sparse}}$.
\end{itemize}
The values of these parameters used in this work are detailed in \cref{sec:computational_details}.

\hl{Note that $\epsilon_{\text{sparse}}$ is used to screen out the insignificant tiles, so that they are not used in subsequent computations. The difficult problem is how to predict which tiles are significant before computing them. For the AO integrals this can be done by using Schwarz inequalities or other more sophisticated estimators},\cite{Hollman:2015ca} \hl{but for results of tensor contractions this
is less straightforward. The standard TiledArray approach for screening block-sparse tensor contractions}\cite{Calvin:2015ka} 
\hl{estimates the norms of the result by using the submultiplicative property of the Frobenius norm; this, in effect, assumes that all contraction contributions to the result tile add up constructively (i.e. have same sign). The \textit{insignificant} result tiles whose estimated per-element Frobenius norm is less than $\epsilon_{\text{sparse}}$ are then not computed, but small contributions to \textit{significant} result tiles are still computed. Here we use a hybrid approach: for the significant result tiles all contributions whose per-element Frobenius norm are estimated to be less than $\epsilon_{\text{sparse}}/\sqrt{n_\mathrm{contrib}}$ are not evaluated, where $n_\mathrm{contrib}$ is the total number of contributions from the tensor contraction to the particular given tile. Replacing $\sqrt{n_\mathrm{contrib}}$ by $n_\mathrm{contrib}$ would \textit{guarantee} that the omitted contributions affect the result's norm by less than $\epsilon_{\text{sparse}}$; the use of $\sqrt{n_\mathrm{contrib}}$ does not provide strict guarantee, but does account for the cancellation of small omitted contributions assuming that they are normally distributed around zero. This allows to reduce the number of evaluated contributions without any observable effect on the accuracy.}

Our periodic CADF-K implementation is summarized in \cref{alg:cadf}. The orbital basis set and density fitting basis set will be denoted as OBS and DFBS, respectively, in the following.

\begin{algorithm}[H]
\begin{footnotesize}
	\caption{Density-Based Periodic CADF-K$^a$}\label{alg:cadf}
	\begin{algorithmic}[1]
		\Procedure{CADF-K}{}	
			\Case{before the beginning of SCF loops}
				\State Read in orbital and density fitting basis sets
				\State Evaluate fitting coefficients \tensor{C}
					\Comment{\cref{eq:C_local_solid}}
				\State Evaluate two-center ERI, $(\dao{X}{G''}|Y)$ where $|\vec{G''}| \leq 2\,|\rangelimit{o}| + |\rangelimit{d}|$
				\State Evaluate integral-direct three-center ERI, $(\dao{X}{G''}|\nu\dao{\sigma}{G'})$				
			\EndCase
			\Case{in each SCF loop}
				\State \tensor{D} $\gets$ density matrix from previous iteration
				\State Evaluate \tensor{Q}
					\Comment{\cref{eq:Q}}
				\State $L_{X \sigma} = $ \Call{ForceShape}{\tensor{Q}}
					\Comment{See function definition below}
				\State Evaluate translationally unique \tensor{F} via \Call{BuildF}{$L_{X \sigma}$}
					\Comment{See function definition below}
				\State Evaluate \tensor{K}
					\Comment{\cref{eq:K_df}}
			\EndCase
		\EndProcedure
		\Function{ForceShape}{\tensor{Q}}
			\For{$X$ in DFBS translated by \vec{G'}}
				\For{$\sigma$ in OBS translated by \vec{L'}} 
					\For{$\mu$ in OBS}
						\If{$\overline{||Q_{\mu,\dao{\sigma}{L'}}^{\dao{X}{G'}}||}_F > \epsilon_F$}
							\State Add $(\dao{X}{G'}, \dao{\sigma}{L'})$ to $L_{X \sigma}$
							\State \textbf{break}
%						\Else
%							\State \textbf{continue}
						\EndIf
					\EndFor
				\EndFor
			\EndFor
			\State \Return $L_{Y \rho}$
		\EndFunction
		\Function{BuildF}{$L_{X \sigma}$}
			\For{$(\dao{X}{G'}, \dao{\sigma}{L'})$ in $L_{X \sigma}$}
				\For{\vec{R} within range ($2\,|\rangelimit{o}| + |\rangelimit{d}|$)}
					\If{$|\vec{L'}-\vec{R}| \leq |\rangelimit{o}|$ }
						\For{$\nu$ in OBS}
							\State Construct $F_{\nu, \dao{\sigma}{(L'-R)}}^{\dao{X}{(G'-R)}}$
								\Comment{\cref{eq:cadf_F}}
						\EndFor
					\EndIf
				\EndFor
			\EndFor 
			\State \Return \tensor{F}
		\EndFunction
	\end{algorithmic}
$^a$ All AO indices refer to the tiles of the corresponding AO dimensions. $\overline{||\cdot||}_F$ is the per-element Frobenius norm of the tensor tile.
\end{footnotesize}
\end{algorithm}

\section{Computational Details}
\label{sec:computational_details}
Periodic Hartree--Fock with CADF-K has been implemented in  the \MPQC package (version 4.0.0).\cite{mpqc4paper} Implementation of sparse tensor algebra is provided by our
TiledArray tensor framework\cite{calvin_tiledarray:_2016} supported by the MADNESS task-based runtime\cite{harrison_madness:_2016}. 
These components were compiled with \hl{GCC 8.3.0} compiler.  
All computations were performed on a commodity cluster at Virginia Tech, each node of which has two 12-core Intel Xeon E5-2680 v3 processors (theoretical peak performance of one node is 960 GFlops/s) and 128 GB of RAM, \hl{unless otherwise stated}.

\hl{We chose polyacetylene, polyethylene, single-walled carbon nanotube (4, 0), hexagonal boron nitride monolayer (h-BN), urea monolayer, urea crystal, lithium hydride (LiH), and diamond as test cases representative of relevant dimensionalities and bond types}. The structures were obtained from references \onlinecite{avitabile_low_1975,forner_numerical_1997,dresselhaus_physics_1995,kim_geometric_2017,swaminathan_crystal_1984,osti_1281384,osti_1199672}. \hl{In the present work, Def2-SVP{\cite{weigend_balanced_2005}} orbital basis set paired with Def2-SVP-J{\cite{weigend_accurate_2006}} density fitting basis set was utilized for all systems except for LiH and diamond, for which more computationally feasible basis sets were adopted; a double-$\zeta$ quality basis set derived from molecular cc-pVDZ{\cite{dunning_gaussian_1989}} by Lorenz and co-workers {\cite{lorenz_local_2012}} (denoted as CR-cc-pVDZ) was chosen for LiH, and Pople's 3-21G{\cite{binkley_self-consistent_1980}} for diamond. 
To test the convergence of CADF-K with respect to density fitting basis sets, we also investigated regular cc-pVDZ{\cite{dunning_gaussian_1989}} OBS paired with cc-pV\textit{X}Z-RI (\textit{X} = D, T, Q, 5, 6){\cite{weigend_efficient_2002,hattig_optimization_2005}} DFBS.}

\cref{tab:thresh} lists the {\it default} and {\it tight} values for various thresholds.  
The former set is used in all our calculations unless specified otherwise, while
the latter is recommended when one expects high-precision results.
\begin{table}[!ht]
	\caption{Default and tight values for screening thresholds used in this work.}\label{tab:thresh} 
	\begin{ruledtabular}
		\begin{tabular*}{0.9\textwidth}{l @{\extracolsep{\fill}} c c l}
			\toprule
			Threshold & Default & Tight & Description \\
			\hline
			\rowcolor{yellow}
			$\epsilon_{\mathrm{schwarz}}$ & $10^{-12}$ & $10^{-16}$ & Schwarz threshold \\
			$\epsilon_S$ & $10^{-10}$ & $10^{-14}$  & overlap threshold \\
			$\epsilon_D$ & $10^{-8}$ & $10^{-12}$ & density truncation threshold  \\
			$\epsilon_F$ & \chl $10^{-10}$ & \chl $10^{-12}$ & force shape threshold \\
			$\epsilon_{\text{sparse}}$ & $10^{-10}$ & \chl $10^{-12}$ & tensor sparse threshold \\
			\bottomrule
		\end{tabular*}
	\end{ruledtabular}
\end{table}

\section{Results and Discussion}
\label{sec:results}

\subsection{Computational Complexity with respect to Lattice Summation Ranges}
\label{sec:scaling}

The computational complexity of molecular electronic structure methods is traditionally expressed as the operation count (or, alternatively, time-to-solution)
as a function of the system size $N$ in the asymptotic regime of $N\to\infty$. It is somewhat less common to discuss complexity of molecular computations
with respect to precision $\epsilon$, defined as the difference between the exact solution for a given property (e.g. Hartree-Fock energy)
and its approximate value obtained in practice. Precision is thus limited by the discretization (``basis set'') error as well as other numerical approximations such as
screening. The effects of both kinds of numerical approximations cannot be easily separated for most representations.

The cost of computations on infinite systems can only be meaningfully analyzed in terms of scaling with the unit cell size and with respect to
the precision.\footnote{Since we only focus on the most time-consuming step, namely the construction of direct-space exchange matrix (\cref{eq:K}), for a chosen unit cell,
we do not analyze the cost with respect to the number of sampled \vec{k} points in the first Brillouin zone ({\it N}\sub{k}, see, for example, \cref{eq:D}); in any case,
the number of \vec{k} points can always be reduced to 1 by increasing unit cell size appropriately.}
Here we focus only on the latter, and specifically focus on the effect on finite lattice summation ranges on precision; the effects of the
CADF approximation and various screening thresholds on the precision is considered in subsequent sections.

Without any screening the cost of the exchange operator evaluation is $\bigO{N_{cell}^3}$,
where {\it N}\sub{cell} is the number of unit cells included in the lattice summations in \cref{eq:K} over \vec{R}, \vec{G}, and \vec{L}.
Due to the localized nature of Gaussian basis functions, the \vec{G} and \vec{R} series involved in the pseudo-overlap distributions $\{\mu,\dao{\rho}{G}\}$ and $\{\dao{\rho}{R},\dao{\sigma}{(G+L)}\}$ decay exponentially to zero with the increase of \vec{G} and \vec{R}. Therefore, the \vec{L} summation, which depends on the decay of density matrix, usually predominates the computation of exchange. 
Nevertheless with decaying density matrix the best possible scaling exponents with respect to {\it N}\sub{cell} is 0 (constant).

\begin{figure}[t!]
	\begin{center}
		\includegraphics[width=0.7\textwidth]{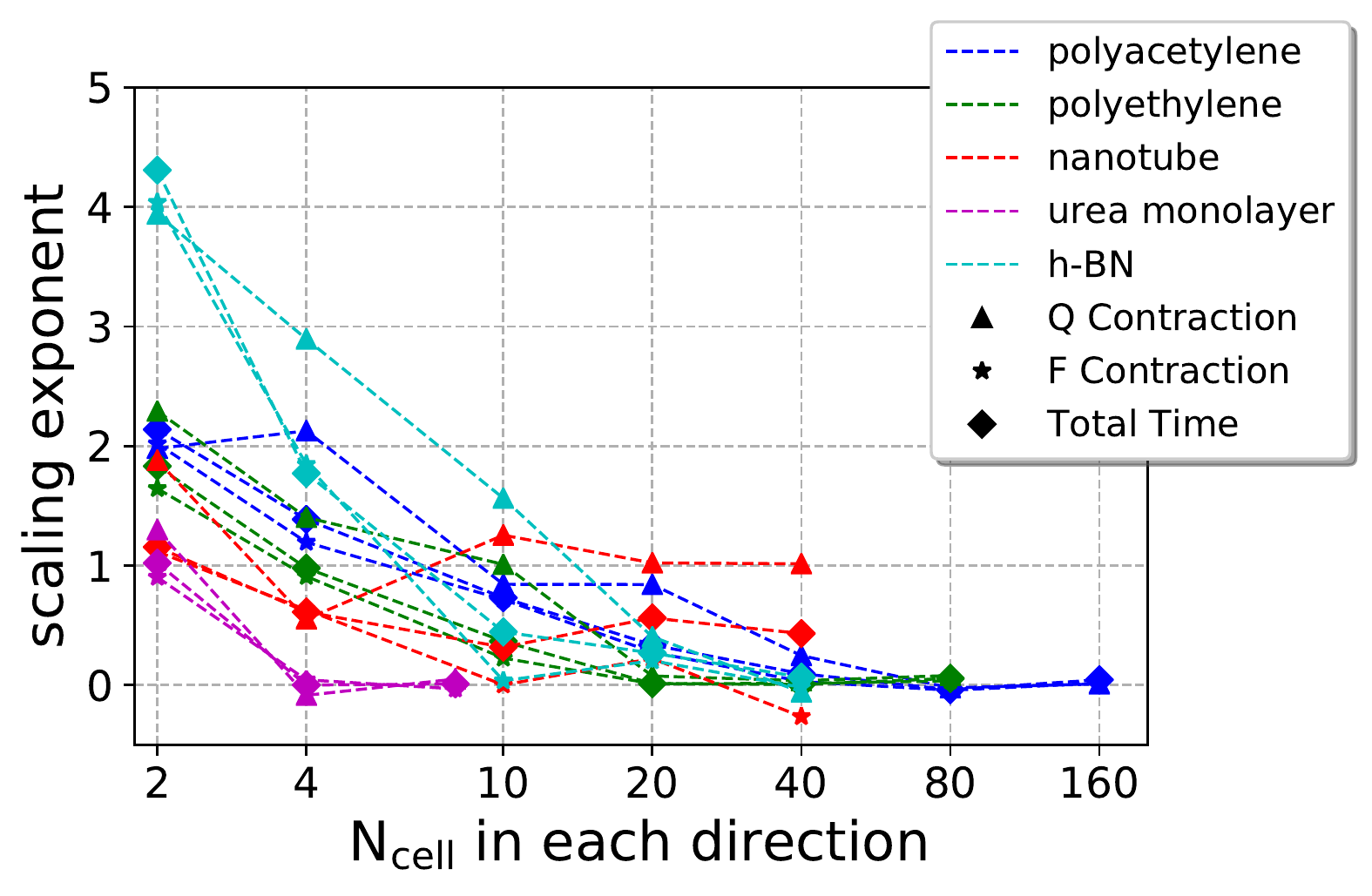}
		\caption{Effective scaling exponent of the CADF-K cost, including the most expensive \tensor{Q} and \tensor{F} contractions (see \cref{eq:Q,eq:cadf_F} for details), as well as the total time, in periodic Hartree-Fock for the five selected systems. Def2-SVP/Def2-SVP-J basis pair is used.}
		\label{fig:exp_vs_ncells}
	\end{center}
\end{figure}

\cref{fig:exp_vs_ncells} shows the {\em effective} scaling exponent of {\it N}\sub{cell} for the periodic CADF-K builder using the default cutoff thresholds.
The effective scaling exponent $k$ is defined by a ``two-point fit'' of the time-to-solution $C(N)$ measured with size parameters $N_i$ and $N_{i+1}$:
\begin{align}
	k_{N_{i-1},N} \equiv \text{log}_{\frac{N_i}{N_{i-1}}} \frac{C(N_i)}{C(N_{i-1})}.
\end{align}
Polyethylene (C$_2$H$_4$)$_n$ and polyacetylene (C$_2$H$_2$)$_n$ represent prototypical one-dimensional covalent periodic systems with large gaps.
Computational costs of their exchange term quickly decay to zero after including a finite number of unit cells (20 for polyethylene and 80 for polyacetylene) due to the rapid (exponential) decay of
contributions to all three lattice sums with respect to the lattice index magnitude.
The slower decay in polyacetylene is due to its smaller band gap and thus less localized density matrix, resulting in a slower convergence of the $\vec{L}$ summation.

Single-walled nanotube (n, 0) has been reported to have finite but small band gaps both in experiment and theory.\cite{ouyang_energy_2001,matsuda_definitive_2010} The slow decay of its density matrix (reflected in the \vec{L} summation) will dominate the calculation in long distance. As expected, the asymptotic scaling of nanotube with {\it N}\sub{cell} approaches linear after 40 unit cells in each direction, equivalent to 171.6 \angstrom from the reference cell. Note that scaling behaviors of \tensor{Q} and \tensor{F} are different, i.e. the former stays linear while the latter drops to zero after 10 cells. This is because \tensor{Q} and \tensor{F} have direct and indirect dependence on density \tensor{D}, respectively. For threshold $\epsilon_{D} = 10^{-8}$, density decreases to a small value after 10 cells but it is still above $\epsilon_{D}$. As a result, the operation count of \tensor{Q} contraction (\cref{eq:Q}) grows with \tensor{D}.
On the other hand, the incremental elements of \tensor{Q} are so small that function {\scshape ForceShape} in \cref{alg:cadf} yields an unchanged shape for \tensor{F}, leading to a scaling exponent of zero for computation of \tensor{F}. 
Numerical analysis indicates that we might have overcomputed elements of \tensor{Q} due to greatly overestimating norms of its tiles. 
An improved norm-estimating approach for \tensor{Q}, especially in small-gap systems, is left for the future work.

The two-dimensional test cases, urea \hl{monolayer} and h-BN, are prototypical molecular and covalently-bonded crystals, respectively. The loosely packed structure of molecular crystals usually leads to large band gaps and fast decay of density matrices. \cref{fig:exp_vs_ncells} shows that for urea \hl{monolayer}, the scaling exponent quickly decays to zero after $N\sub{cell} = 4$. The covalently-bonded h-BN has similar structure as graphene but with a wide band gap of about 5 eV.\cite{park_band-structure_1987,si_divacancies_2009,huang_defect_2012,kim_geometric_2017} The total CADF-K time of h-BN stops growing after $N\sub{cell} = 40$. Similar to the nanotube case, \tensor{Q} decays slower than \tensor{F} for h-BN because we overestimated its tile norms, but finally drops to zero.

\subsection{Precision Analysis}
\label{sec:error}

In this section we discuss the effect of the CADF approximation and screening thresholds on precision.
{\it Default} values of thresholds (see \cref{tab:thresh}) are used in \cref{sec:df_error}, ensuring that the incompleteness of DFBS is the main source of error, i.e. further reduction of thresholds by a factor of 10 does not change the Hartree--Fock energy by more than 1 \uHartree. 
In \cref{sec:cutoff_error}, to study the effect of cutoff thresholds individually, other parameters are chosen from the {\it tight} set.

\subsubsection{Precision of CADF}
\label{sec:df_error}
\begin{figure}[t!]
	\begin{center}
		\includegraphics[width=0.7\textwidth]{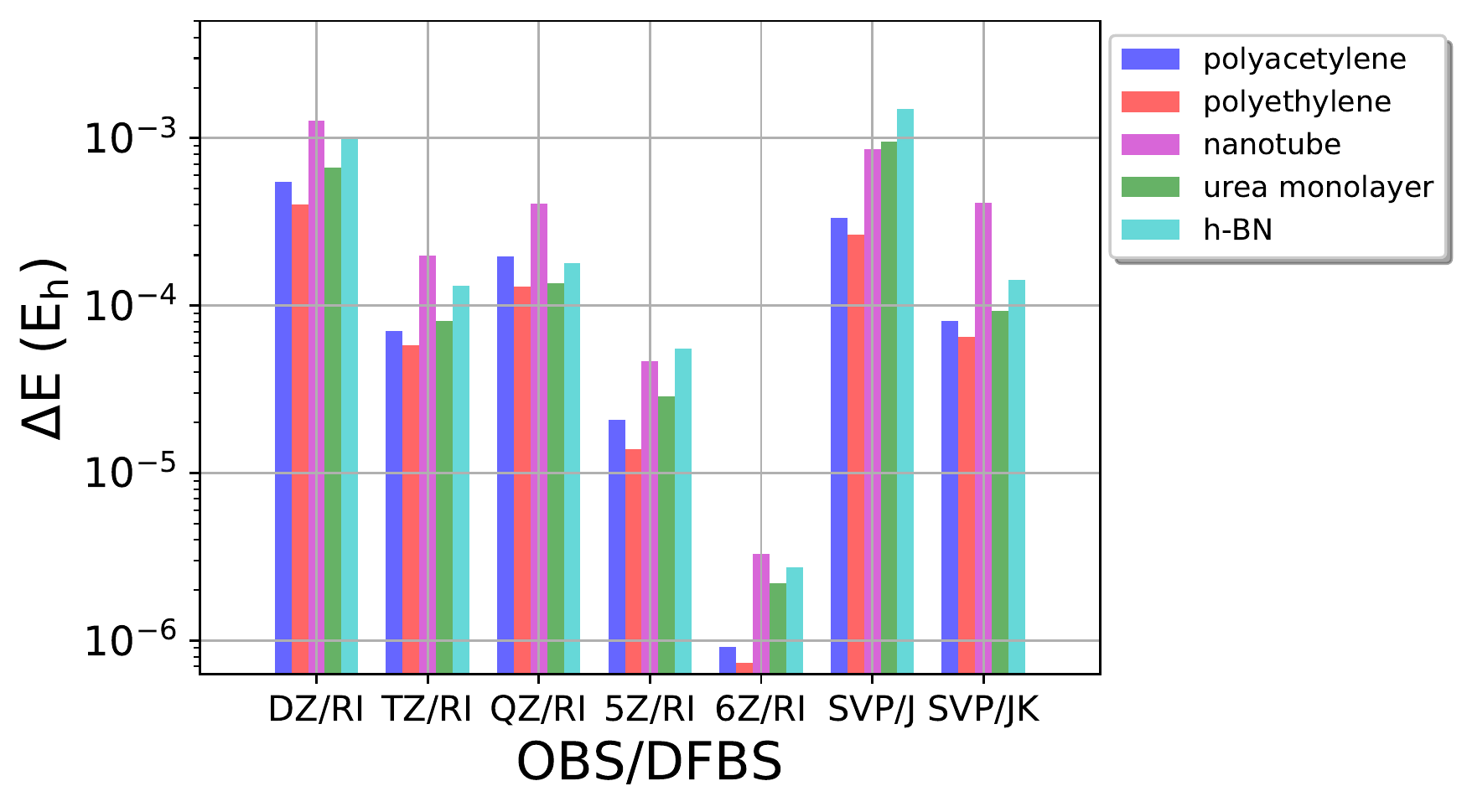}
		\caption{\hl{Errors per atom in total Hartree--Fock energies (in E$_{h}$) for CADF-K with respect to 4c-K for the selected one- and two-dimensional systems. The $X$Z/RI ($X$ = D, T, Q, 5, 6) labels on the $x$ axis denote cc-pVDZ/cc-pV$X$Z-RI OBS/DFBS pairs, and SVP/J(K) is Def2-SVP/Def2-SVP-J(K) pairs. Multipole-accelerated 4c-J is used in all calculations.}}
		\label{fig:error_aux}
	\end{center}
\end{figure}

\cref{fig:error_aux} shows the \hl{absolute} HF energy error per atom due to the use of CADF-K relative to the exact evaluation using four-center ERIs (4c-K) for various OBS/DFBS pairs. For the tested systems, errors per atom are comparable to typical DF errors (all close to or below 1 \mHartree) with the tendency to reduce as the DFBS increases. Note that the CADF error does not reduce monotonically, \hl{i.e. cc-pVQZ-RI yields larger errors than cc-pVTZ-RI}. This is likely due to the fact that the cc-pV$X$Z-RI ($X$ = D, T, Q, 5, 6) basis sets are designed to minimize the fitting errors for the matching
cc-pV$X$Z basis, and not to form a systematic series. \hl{Particularly, with the cc-pVDZ/cc-pV6Z-RI basis pair, CADF errors can be controlled well below 10 {\uHartree} per atom for tested systems.}

Similar accuracy has been achieved for cc-pVDZ/cc-pVDZ-RI and Def2-SVP/Def2-SVP-J pairs. The latter will be used in the rest of the article for practical reasons regarding CPU time and memory usage.

\subsubsection{Precision vs. screening thresholds}
\label{sec:cutoff_error}
Among the cutoff thresholds listed in \cref{tab:thresh}, $\epsilon_D$ for truncating density matrix and $\epsilon_{F}$ for predetermining the sparsity shape of tensor \tensor{F} have the most significant impact on the efficiency and accuracy of CADF-K, and thus will be studied in more detail. 

\begin{figure}[t!]
	\begin{center}
		\includegraphics[width=0.9\textwidth]{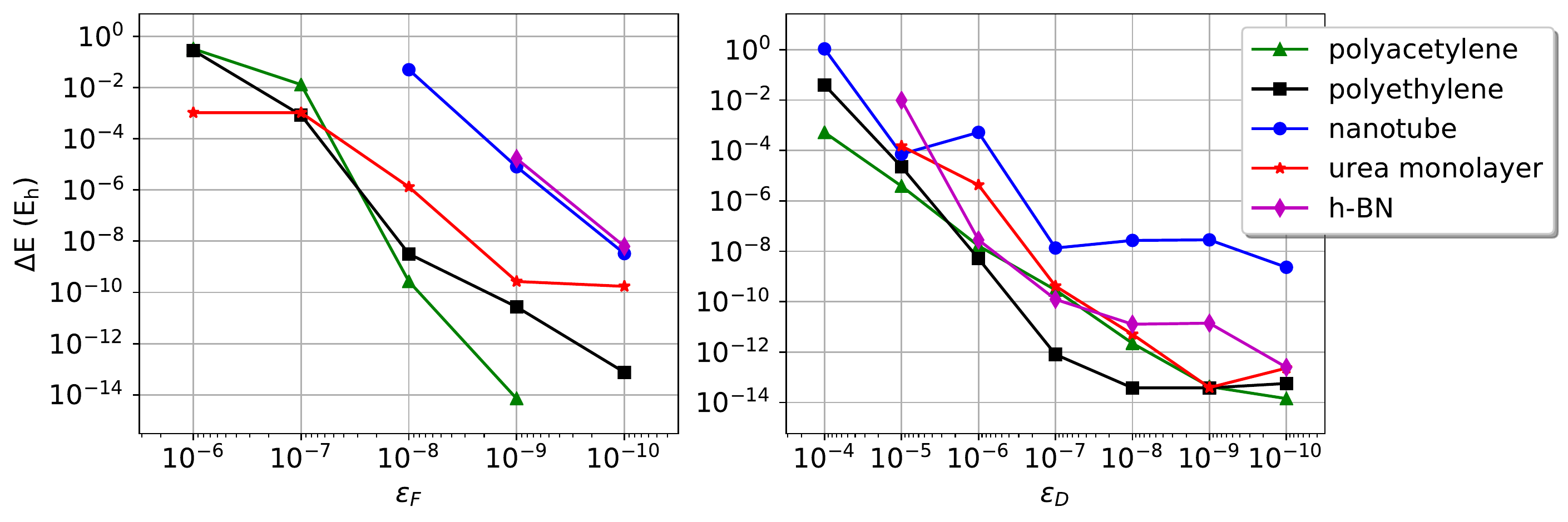}
		\caption{\hl{Errors per atom (in E$_h$) in total Hartree-Fock energy with CADF-K for various $\epsilon_{D}$ (left) and $\epsilon_{F}$ (right) values with respect to the energies obtained with CADF-K and the {\it tight} cutoff thresholds}}
		\label{fig:performance_tcut}
	\end{center}
\end{figure}

\cref{fig:performance_tcut} shows variation of the CADF-K error with the two screening thresholds. We can see that the CADF-K error decreases quickly with both cutoffs, \hl{yielding results with the precision close to or well below 0.01 {\uHartree} for $\epsilon_{D} \leq 10^{-8}$ and $\epsilon_{F} \leq 10^{-10}$}. To balance the accuracy and computational cost, \hl{we use $\epsilon_{D} = 10^{-8}$ and $\epsilon_{F} = 10^{-10}$ as our default cutoffs for the following results, unless otherwise stated}. 

\subsection{Performance of CADF-K vs. 4c-K}
\label{sec:speed}

\cref{tab:walltime_final} presents average wall times per iteration for the CADF-K and 4c-K builders \hl{in the eight selected systems, including three-dimensional ones}. CADF-K demonstrates speedups of \hl{up to 13}$\times$ relative to 4c-K.
\hl{Among the eight model systems, the greatest speedup is observed for urea crystal. This is because urea crystal adopts a looser force shape threshold $\epsilon_{F} = 10^{-8}$ (see Table {\ref{tab:thresh}} for explanation) in order to satisfy the memory requirement of the cluster (it is safe to loose $\epsilon_{F}$ in this case since the introduced error is well below the density fitting error).} 

\hl{To demonstrate the accuracy of our method in terms of relative energies, here we define the (electronic) lattice dissociation energy, {\elatt}, as the energy difference between the cell structure in a periodic form and that in an isolated form, i.e.}
\begin{align}
	\elatt = (E_{\text{cell}}^{\text{periodic}} - E_{\text{cell}}^{\text{isolated}})/N_{\text{atom}}, \label{eq:elatt}
\end{align}
\hl{where $N_{\text{atom}}$ is the number of atoms in a cell. Note that {\elatt} is not lattice energy or cohesive energy. From the last two columns of Table {\ref{tab:walltime_final}}, we can see that the error in lattice dissociation energy ($\Delta \elatt$) due to the CADF approximation is at most 0.5} \kcal per atom, or \hl{at most} $1\%$ of 4c-K \elatt. We point out that the CADF errors can be better controlled if we use a more converged DFBS (see Fig. {\ref{fig:error_aux}} of the present work and Fig. 6 of the work by Ihrig and co-workers {\cite{ihrig_accurate_2015}}).

\begin{table}[!ht]
	\caption{\hl{Comparison of performance and precision of CADF-K vs. 4c-K. {\super{a}}}}\label{tab:walltime_final}
	\begin{threeparttable}
		\begin{ruledtabular}
			\begin{tabular*}{0.9\textwidth}{l @{\extracolsep{\fill}} d{-1} d{-1} d{-1} d{-1} d{-1}}
				\toprule
				System  & \multicolumn{2}{c}{Wall time (in seconds)} & \multicolumn{1}{c}{Speedup} & \multicolumn{1}{c}{$\Delta \elatt$\tnote{b}} & \multicolumn{1}{c}{\% error\tnote{c}} \\
				\cline{2-3}
				& \multicolumn{1}{c}{4c-K\tnote{d}}   & \multicolumn{1}{c}{CADF-K} &  &  & \\
				\hline
				polyacetylene	& 1.02	 	& 0.22		& 4.71  &  0.059 	& 0.32 \\  	
				polyethylene	& 1.01	 	& 0.16		& 6.26  &  0.047 	& 0.34 \\  	
				nanotube		& 187.58 	& 22.11		& 8.48  &  0.028 	& 0.07 \\  	
				h-BN			& 182.74 	& 29.66		& 6.16  &  0.192 	& 0.22 \\  	
				urea monolayer	& 22.97	& 3.85	& 5.96  &  0.009 	& 0.68 \\  	
				urea crystal	& 542.49 	& 41.64\tnote{e}		& 13.03 &  0.026 	& 1.06 \\   	
				LiH				& 827.54  &	346.63 &	2.39 & 0.071 &	0.09 \\  
				diamond			& 904.72 	& 364.30\tnote{e}	& 2.48  &  0.495 	& 0.35 \\  				
				\bottomrule
			\end{tabular*}
		\end{ruledtabular}  
		\begin{tablenotes}
			\item[a] Wall time is measured as the time per iteration. Multipole accelerated 4c-J is used for both 4c-K and CADF-K. Def2-SVP is used as the orbital basis set for the first six systems, CR-cc-pVDZ for LiH, and 3-21G for diamond. Def2-SVP-J is used as the density-fitting basis set in all cases in CADF-K. Calculations are performed on Intel Xeon E5-2680 v4 processors @ 2.4 GHz with 28 cores.
			\item[b] $\Delta \elatt$ is the difference between \elatt (in \kcal atom$^{-1}$) computed by CADF-K and that computed by 4c-K.
			\item[c] Percent error of \elatt computed by CADF-K with respect to 4c-K \elatt. 
			\item[d] Permutational and translational symmetries of four-center ERIs are exploited in 4c-K.
			\item[e] $\epsilon_{F} = 10^{-8}$ is used in the CADF-K calculation of urea crystal and diamond ({\it default} values for other thresholds) in order to satisfy the memory requirement of the cluster.
		\end{tablenotes}
	\end{threeparttable}
\end{table}     

Periodic Hartree-Fock based on Gaussian AOs has also been implemented in other solid-state quantum chemistry packages, including, but not limited to, Crystal\cite{dovesi_quantum-mechanical_2018}, Gaussian\cite{frisch_gaussian_manual,izmaylov_efficient_2006}, CP2K\cite{hutter_cp2k:_2014}, and PySCF\cite{sun_pyscf:_2018}. A strict apple-to-apple comparison between various packages is impossible because of their diverse implementation techniques, different optimization levels, and licensing restrictions. Here in order to show that the demonstrated speedups of our CADF-K algorithm are indeed meaningful, i.e. our 4c-K implementation in MPQC is efficient, we perform a comparison of timing between the 4c-K in \mpqc and that in Crystal, as well as an accuracy comparison with Crystal and Gaussian to prove the correctness of our code (see the footnotes of \cref{tab:walltime_compare_with_others} for the
relevant computational details). \hl{Polyacetylene, polyethylene, h-BN, LiH, and diamond are selected as examples. From Table {\ref{tab:walltime_compare_with_others}} we can see that for all tested systems, the 4c-K implementation in MPQC is as efficient as Crystal. Consistent estimate of lattice dissociation energies has been achieved among the three packages at sub-kcal mol$^{-1}$ atom$^{-1}$ accuracy ($10^{-4}$ kcal mol$^{-1}$ atom$^{-1}$ if one compares MPQC to Gaussian). Note that the timings obtained from Crystal are based on tightened integral thresholds in order to reach a similar level of accuracy as Gaussian and MPQC.}

\begin{table}[!ht]
	\caption{\hl{Comparison of performance and precision of the 4c-K implementations in {\mpqc} and reference codes{\super{a}}}}\label{tab:walltime_compare_with_others}
	\begin{threeparttable}
		\begin{ruledtabular} 
			\begin{tabular*}{0.9\textwidth}{l @{\extracolsep{\fill}} d{-1} d{-1} d{-1} d{-1} d{-1}}
				\toprule
				System  & \multicolumn{1}{c}{Gaussian\tnote{b}} & \multicolumn{2}{c}{Crystal\tnote{c}} & \multicolumn{2}{c}{MPQC\tnote{d}} \\
				\cline{2-2} \cline{3-4} \cline{5-6}
				& \multicolumn{1}{c}{\elatt} & \multicolumn{1}{c}{Time}   & \multicolumn{1}{c}{\elatt} & \multicolumn{1}{c}{Time}   & \multicolumn{1}{c}{\elatt} \\
				\hline
				polyacetylene	& -26.65105  & 15.3    & -26.62745  & 4.9    & -26.65105 \\
				polyethylene	& -13.37271  & 22.9    & -13.37273  & 5.8    & -13.37271 \\
				h-BN			& -87.80406  & 2838.7  & -87.79822  & 1886.6 & -87.80407 \\     
				LiH             & -83.29362  & 25769.9 & -83.29275 & 6358.8 & -83.29341 \\
				diamond         & -118.15346 & 23377.3 & -118.14671 & 7223.9 & -118.15378 \\       
				\bottomrule
			\end{tabular*}
		\end{ruledtabular} 
		\begin{tablenotes}
			\item[a] Def2-SVP is used as the orbital basis set for the first three systems, CR-cc-pVDZ for LiH, and 3-21G for diamond. Calculations are performed on Intel Xeon E5-2680 v4 processors @ 2.4 GHz with 1 core. The wall times are given in seconds and \elatt in kcal mol$^{-1}$ atom$^{-1}$.
			\item[b] Gaussian09 is used with the keyword \textbf{SCF=Tight}.
			\item[c] Space symmetry and the bipolar approximation are turned off in Crystal. ERI cutoffs and shrink factors are optimized to achieve maximal efficiency without a significant loss of accuracy. Note that for h-BN tightened integral cutoffs, \textbf{TOLINTEG} 20 20 20 20 40, were used in order to converge the SCF iterations. The reported time of Crystal includes the computation of Coulomb term, which is assumed insignificant relative to the cost of exchange.    
			\item[d] Default cutoffs are used in \mpqc.
		\end{tablenotes}     
	\end{threeparttable}     
\end{table}     

\section{Conclusions}
\label{sec:conclusions}

We have presented an efficient density-based implementation of exact exchange for periodic systems via the concentric atomic density fitting (CADF) approximation based on atom-centered Gaussian-type orbitals. Both molecular and periodic systems can be treated on an equal footing.  We have shown the performance of periodic CADF-K on \hl{several systems with different dimensionality} and obtained a significant improvement relative to the 4c-K algorithm. Errors in lattice dissociation energy introduced by the CADF approximation are below the chemical accuracy. Further efficiency improvements could be conceivably achieved by using maximally localized Wannier functions\cite{zicovich-wilson_general_2001,casassa_symmetry-adapted_2006,marzari_maximally_1997,zeiner_bloch_1998} in the context of an orbital-based CADF-K algorithm.\cite{manzer_efficient_2015}

% If in two-column mode, this environment will change to single-column format so that long equations can be displayed. 
% Use only when necessary.
%\begin{widetext}
%$$\mbox{put long equation here}$$
%\end{widetext}

% If you have acknowledgments, this puts in the proper section head.

\begin{acknowledgments}
This research was supported by the U.S. National Science Foundation (awards 1550456 and 1800348). We thank the Advanced Research Computing (ARC) at Virginia Tech for providing computing resources. X.W. thanks Dr. Chong Peng for insightful discussions. The Flatiron Institute is a division of the Simons Foundation.
\end{acknowledgments}

~\\
~\\
{\bf Data Availability Statement}
~\\

The data that support the findings of this study are available from the corresponding author upon request.

\appendix

\section*{Appendix: Coulomb potential evaluation using real multipole expansion}
\label{sec:appendix}

Fast evaluation of Coulomb potential is typically performed via the Ewald method\cite{ewald_evaluation_1921,Saunders:1992iz,toukmaji_ewald_1996,kittel_introduction_1996} or the fast multipole method (FMM);\cite{greengard_fast_1987,white_derivation_1994,strain_achieving_1996,challacombe_periodic_1997,kudin_linear-scaling_2000,kudin_revisiting_2004} however, other efficient approaches exist.\cite{Genovese:2006eb,Beylkin:2008im,Jia:2009vm} In the current work, we implemented a straightforward multiscale \hl{multipole-accelerated} approach as described in this section; this approach is similar to
the method of Ref. \onlinecite{Sierka:2003bs}, except we use \hl{(1)} {\em real}\cite{perezjorda_concise_1996} {\em bipolar}\cite{Carlson:1950ca,Buehler:1951ca} (or, two-center\cite{Solovyov:2007bk}) multipole expansions\hl{, and (2) leading multipole moments of the unit cell's total charge density are compensated to ensure absolutely convergent values for total energy and potential as well as the equivalence with the Ewald method}. This appendix documents the relevant formulas.

Our goal is to evaluate matrix representation of the total Coulomb potential, i.e. the sum of \cref{eq:V,eq:J}.
The ``Coulomb'' summations (over \vec{G}) are first divided into the crystal near field (CNF) and far field (CFF) according to the partitioning criteria defined in Ref. \onlinecite{lazarski_density_2015}. The Coulomb interaction $(\rho_{\vec{0}}|\rho_{\vec{G}})$ between charge distributions of the reference unit cell \vec{0} and a distant unit cell \vec{G} is computed via 4c-J (\cref{eq:J}) if \vec{G} is in CNF, otherwise it is evaluated using multipole expansion. The use of complex multipole moments is common in the literature,\cite{greengard_fast_1987,white_derivation_1994,strain_achieving_1996,challacombe_periodic_1997,kudin_fast_1998,kudin_fast_1998-1,kudin_revisiting_2004} however, if the basis is real
there is no reason to use the complex-valued formalism.
Real spherical multipole moments and their interaction kernels are defined as\cite{perezjorda_concise_1996} 
\begin{align}
O_{l,m}(\vec{r}) &= 
\begin{cases}
\frac{|\vec{r}|^l}{(l + m)!} P_l^{m}(\text{cos}\theta) \text{cos}(m\phi)\text{, if } m \geq 0 \\
\frac{|\vec{r}|^l}{(l + m)!} P_l^{m}(\text{cos}\theta) \text{sin}(m\phi)\text{, if } m < 0
\end{cases} , \\
M_{l,m}(\vec{r}) &= 
\begin{cases}
\frac{(l - m)!}{|\vec{r}|^{(l+1)}} P_l^{m}(\text{cos}\theta) \text{cos}(m\phi)\text{, if } m \geq 0 \\
\frac{(l - m)!}{|\vec{r}|^{(l+1)}} P_l^{m}(\text{cos}\theta) \text{sin}(m\phi)\text{, if } m < 0 
\end{cases} , \label{eq:multipole_M}
\end{align}
where (r, $\theta$, $\phi$) are spherical polar coordinates of \vec{r} and $P_l^{m}(\text{cos}\theta)$ denotes associated Legendre polynomials about cos$\theta$. Now we introduce functions $N_{l,m}^{\pm} (\vec{r})$ ($N$ = $O$ or $M$) as
\begin{align}
N_{l,m}^{+}(\vec{r}) &=
\begin{cases}
N_{l,m}(\vec{r}) \text{, if } m \geq 0 \\
(-1)^m N_{l,-m}(\vec{r}) \text{, if } m < 0    
\end{cases} , \\
N_{l,m}^{-}(\vec{r}) &=
\begin{cases}
(-1)^{(m+1)} N_{l,-m}(\vec{r}) \text{, if } m > 0 \\
0 \text{, if } m = 0 \\
N_{l, m} (\vec{r}) \text{, if } m < 0
\end{cases} .
\end{align}
Note that $O_{l,m}^{\pm}$ and $M_{l,m}^{\pm}$ have the same expressions as Eqs. 3-6 in Ref. \onlinecite{perezjorda_concise_1996}.
Coulomb potential created at \vec{P} by a distant unit charge at \vec{r} whose potential is multipole expanded around \vec{Q} can be expressed as
\begin{align}
L_{l, m}(\vec{P} - \vec{r}) &=
\begin{cases}
\sum_{l'=0}^{\infty} \sum_{m'=-l'}^{l'} \Big[ M_{l+l',m+m'}^{+}(\vec{P} - \vec{Q}) O_{l',m'}^{+}(\vec{r} - \vec{Q}) \\
\hspace{7em} + M_{l+l',m+m'}^{-}(\vec{P} - \vec{Q}) O_{l',m'}^{-}(\vec{r} - \vec{Q}) \Big] \text{, if } m \geq 0 \\
\sum_{l'=0}^{\infty} \sum_{m'=-l'}^{l'} \Big[ M_{l+l',m+m'}^{-}(\vec{P} - \vec{Q}) O_{l',m'}^{+}(\vec{r} - \vec{Q}) \\
\hspace{7em} + M_{l+l',m+m'}^{+}(\vec{P} - \vec{Q}) O_{l',m'}^{-}(\vec{r} - \vec{Q}) \Big] \text{, if } m < 0    
\end{cases} . \label{eq:multipole_L}
\end{align}
The Coulomb interaction energy between sets of point charges \{$q_i$\} and \{$q_j$\} (e.g. two unit cells) becomes
\begin{align}
E^{coul} &= \sum_{i,j} \frac{q_i q_j}{|\vec{r_i} - \vec{r_j}|} \\
&= \sum_{l=0}^{\infty} (-1)^{l} \sum_{m=-l}^{l} (2 - \delta_{m,0}) \bar{O}_{l,m}(\vec{P}) \bar{L}_{l,m}(\vec{P}), \label{eq:multipole_energy}
\end{align}
where
\begin{align}
\bar{O}_{l,m}(\vec{P}) &\equiv \sum_{i} q_i O_{l,m}(\vec{r_i} - \vec{P}) \label{eq:multipole_N}
\end{align}
are the total (real) multipole moments of  \{$q_i$\} centered at \vec{P},
and 
\begin{align}
\bar{L}_{l,m}(\vec{P}) &\equiv \sum_{j} q_j L_{l,m}(\vec{P} - \vec{r_j}) \label{eq:potential_L}
\end{align}
are the charge-including local potentials at \vec{P} due to distant \{$q_j$\} (centered at \vec{Q})
and evaluated via \cref{eq:multipole_L}.
For continuous charge distribution the summation over point charges in \cref{eq:multipole_N,eq:potential_L} becomes a trace with 1-RDM, e.g.:
\begin{align}
\bar{O}_{l,m}(\vec{P}) \equiv \sum_{\vec{R}}\sum_{\mu,\nu}\left(\mu|O_{l,m}(\vec{r}-\vec{P})|\dao{\nu}{R}\right) D_{\mu,\dao{\nu}{R}} \label{eq:multipole_O}.
\end{align} 
Evaluation of the real multipole moment integrals over Gaussian basis was implemented in the open-source Libint\cite{Libint2} library following
the recurrence formulas in Ref. \onlinecite{perezjorda_concise_1996}.

\hl{
To obtain meaningful (absolutely convergent) total and orbital energies it is mandatory to compensate the unit cell's electric dipoles (for 2-dimensional crystals) and quadrupoles (for 3-dimensional crystals). This is achieved by adding charges and point dipoles to the surface of the unit cell in such a way that the net charge density in the bulk of the crystal is not changed, while the electric dipole and quadrupole moments of the unit cell vanish. The compensation can be viewed as modifying the boundary conditions such that the net charge density in the bulk of the crystal remains unchanged; the compensating charges/dipoles would end up on the surface of a finite crystal composed of the multipole-compensated unit cells and thus help cancel the shape-dependent contributions to the energy and potential.{\cite{HARRIS1975147,deLeeuw:1980kw,Saunders:1992iz}}
Charge compensation of the dipoles in the context of FMM was introduced by Kudin and Scuseria as soon as they started to utilize
(C)FMM for LCAO calculations on 3-D solids.{\cite{kudin_fast_1998}}
Here we compensate not only the dipoles but also the Cartesian quadrupoles. The latter automatically
ensures not only the absolute convergence of the potential (e.g. orbital energies) and
the equivalence to the Ewald limit of the potential due to the vanishing
trace of the second moment (spheropole) of the charge density.{\cite{Euwema:1975ee,Saunders:1992iz}}
Our approach is formulated for a monoclinic (Bravais) lattice.
}

\hl{
Dipole moments are compensated by placing charges of $\pm q_i$ at the center of opposing facets; namely, for each principal axis $i$
($i=0,1,2$ for a 3-d lattice) charges of $q_i$ and $-q_i$ are placed at $\vec{a}_j/2 + \vec{a}_k/2$ and $\vec{a}_i + \vec{a}_j/2 + \vec{a}_k/2$, respectively, where $j=\mod(i+1,3)$, $k=\mod(i+2,2)$, and $\vec{a}_i$ are the unit cell vectors. The magnitudes of charges
needed to compensate net dipole $\vec{\mu}$ are given by}
\begin{align}
q_i = & \vec{\mu} \cdot \tilde{\vec{b}}_i
\end{align}
\hl{with $\tilde{\vec{b}}_i  \equiv \vec{b}_i/(2 \pi) $ being the reduced reciprocal lattice vectors:}
\begin{align}
\vec{b}_i & \equiv 2 \pi \frac{\vec{a}_j \times \vec{a}_k}{V},
\end{align}
\hl{where $V \equiv \vec{a}_0 \cdot (\vec{a}_1 \times \vec{a}_2)$ is the unit cell volume.}

\hl{
Quadrupole moments
are compensated by placing point dipoles $\pm \vec{\mu}_i$ at the center of opposing facets.
Specifically, dipoles  $\vec{\mu}_i$ and $-\vec{\mu}_i$ are placed at
$\vec{a}_j/2 + \vec{a}_k/2$ and $\vec{a}_i + \vec{a}_j/2 + \vec{a}_k/2$, respectively.
The compensating dipoles needed to compensate unit cell's net Cartesian quadrupole tensor
$\vec{q}$ (evaluated with the dipole-compensating charges) are given by
}
\begin{align}
\vec{\mu}_i \equiv & \frac{1}{2} \left( \tilde{\vec{b}}_i^\dagger \vec{q} \tilde{\vec{b}}_i \right) \vec{a}_i + \left( \tilde{\vec{b}}_i^\dagger \vec{q} \tilde{\vec{b}}_j \right) \vec{a}_j .
\end{align}
\hl{
If needed, this formula can be straightforwardly generalized to higher-order multipoles.
For simplicity, point dipoles were approximated by 2 opposite charges separated by 0.1 a.u.
}

% Create the reference section using BibTeX:
\bibliography{references}

%merlin.mbs aipnum4-1.bst 2010-07-25 4.21a (PWD, AO, DPC) hacked
%Control: key (0)
%Control: author (8) initials jnrlst
%Control: editor formatted (1) identically to author
%Control: production of article title (-1) disabled
%Control: page (0) single
%Control: year (1) truncated
%Control: production of eprint (0) enabled
\begin{thebibliography}{144}%
\makeatletter
\providecommand \@ifxundefined [1]{%
 \@ifx{#1\undefined}
}%
\providecommand \@ifnum [1]{%
 \ifnum #1\expandafter \@firstoftwo
 \else \expandafter \@secondoftwo
 \fi
}%
\providecommand \@ifx [1]{%
 \ifx #1\expandafter \@firstoftwo
 \else \expandafter \@secondoftwo
 \fi
}%
\providecommand \natexlab [1]{#1}%
\providecommand \enquote  [1]{``#1''}%
\providecommand \bibnamefont  [1]{#1}%
\providecommand \bibfnamefont [1]{#1}%
\providecommand \citenamefont [1]{#1}%
\providecommand \href@noop [0]{\@secondoftwo}%
\providecommand \href [0]{\begingroup \@sanitize@url \@href}%
\providecommand \@href[1]{\@@startlink{#1}\@@href}%
\providecommand \@@href[1]{\endgroup#1\@@endlink}%
\providecommand \@sanitize@url [0]{\catcode `\\12\catcode `\$12\catcode
  `\&12\catcode `\#12\catcode `\^12\catcode `\_12\catcode `\%12\relax}%
\providecommand \@@startlink[1]{}%
\providecommand \@@endlink[0]{}%
\providecommand \url  [0]{\begingroup\@sanitize@url \@url }%
\providecommand \@url [1]{\endgroup\@href {#1}{\urlprefix }}%
\providecommand \urlprefix  [0]{URL }%
\providecommand \Eprint [0]{\href }%
\providecommand \doibase [0]{http://dx.doi.org/}%
\providecommand \selectlanguage [0]{\@gobble}%
\providecommand \bibinfo  [0]{\@secondoftwo}%
\providecommand \bibfield  [0]{\@secondoftwo}%
\providecommand \translation [1]{[#1]}%
\providecommand \BibitemOpen [0]{}%
\providecommand \bibitemStop [0]{}%
\providecommand \bibitemNoStop [0]{.\EOS\space}%
\providecommand \EOS [0]{\spacefactor3000\relax}%
\providecommand \BibitemShut  [1]{\csname bibitem#1\endcsname}%
\let\auto@bib@innerbib\@empty
%</preamble>
\bibitem [{\citenamefont {Ayala}\ and\ \citenamefont
  {Scuseria}(1999)}]{Ayala1999}%
  \BibitemOpen
  \bibfield  {author} {\bibinfo {author} {\bibfnamefont {P.~Y.}\ \bibnamefont
  {Ayala}}\ and\ \bibinfo {author} {\bibfnamefont {G.~E.}\ \bibnamefont
  {Scuseria}},\ }\href {\doibase 10.1063/1.478256} {\bibfield  {journal}
  {\bibinfo  {journal} {J. Chem. Phys.}\ }\textbf {\bibinfo {volume} {110}},\
  \bibinfo {pages} {3660} (\bibinfo {year} {1999})}\BibitemShut {NoStop}%
\bibitem [{\citenamefont {Zi{\'{o}}{\l}kowski}\ \emph
  {et~al.}(2010)\citenamefont {Zi{\'{o}}{\l}kowski}, \citenamefont
  {Jans{\'{i}}k}, \citenamefont {Kj{\ae}rgaard},\ and\ \citenamefont
  {J{\o}rgensen}}]{Ziokowski2010}%
  \BibitemOpen
  \bibfield  {author} {\bibinfo {author} {\bibfnamefont {M.}~\bibnamefont
  {Zi{\'{o}}{\l}kowski}}, \bibinfo {author} {\bibfnamefont {B.}~\bibnamefont
  {Jans{\'{i}}k}}, \bibinfo {author} {\bibfnamefont {T.}~\bibnamefont
  {Kj{\ae}rgaard}}, \ and\ \bibinfo {author} {\bibfnamefont {P.}~\bibnamefont
  {J{\o}rgensen}},\ }\href {\doibase 10.1063/1.3456535} {\bibfield  {journal}
  {\bibinfo  {journal} {J. Chem. Phys.}\ }\textbf {\bibinfo {volume} {133}},\
  \bibinfo {pages} {014107} (\bibinfo {year} {2010})}\BibitemShut {NoStop}%
\bibitem [{\citenamefont {Fedorov}\ and\ \citenamefont
  {Kitaura}(2005)}]{Fedorov2005}%
  \BibitemOpen
  \bibfield  {author} {\bibinfo {author} {\bibfnamefont {D.~G.}\ \bibnamefont
  {Fedorov}}\ and\ \bibinfo {author} {\bibfnamefont {K.}~\bibnamefont
  {Kitaura}},\ }\href {\doibase 10.1063/1.2007588} {\bibfield  {journal}
  {\bibinfo  {journal} {J. Chem. Phys.}\ }\textbf {\bibinfo {volume} {123}},\
  \bibinfo {pages} {134103} (\bibinfo {year} {2005})}\BibitemShut {NoStop}%
\bibitem [{\citenamefont {Friedrich}\ and\ \citenamefont
  {Dolg}(2009)}]{friedrich_fully_2009}%
  \BibitemOpen
  \bibfield  {author} {\bibinfo {author} {\bibfnamefont {J.}~\bibnamefont
  {Friedrich}}\ and\ \bibinfo {author} {\bibfnamefont {M.}~\bibnamefont
  {Dolg}},\ }\href {\doibase 10.1021/ct800355e} {\bibfield  {journal} {\bibinfo
   {journal} {J. Chem. Theory Comput.}\ }\textbf {\bibinfo {volume} {5}},\
  \bibinfo {pages} {287} (\bibinfo {year} {2009})}\BibitemShut {NoStop}%
\bibitem [{\citenamefont {Kobayashi}\ and\ \citenamefont
  {Nakai}(2008)}]{Kobayashi2008}%
  \BibitemOpen
  \bibfield  {author} {\bibinfo {author} {\bibfnamefont {M.}~\bibnamefont
  {Kobayashi}}\ and\ \bibinfo {author} {\bibfnamefont {H.}~\bibnamefont
  {Nakai}},\ }\href {\doibase 10.1063/1.2956490} {\bibfield  {journal}
  {\bibinfo  {journal} {J. Chem. Phys.}\ }\textbf {\bibinfo {volume} {129}},\
  \bibinfo {pages} {044103} (\bibinfo {year} {2008})}\BibitemShut {NoStop}%
\bibitem [{\citenamefont {Li}, \citenamefont {Piecuch},\ and\ \citenamefont
  {Gour}(2009)}]{Li2009}%
  \BibitemOpen
  \bibfield  {author} {\bibinfo {author} {\bibfnamefont {W.}~\bibnamefont
  {Li}}, \bibinfo {author} {\bibfnamefont {P.}~\bibnamefont {Piecuch}}, \ and\
  \bibinfo {author} {\bibfnamefont {J.~R.}\ \bibnamefont {Gour}},\ }\href
  {\doibase 10.1063/1.3108393} {\bibfield  {journal} {\bibinfo  {journal} {AIP
  Conf. Proc.}\ }\textbf {\bibinfo {volume} {1102}},\ \bibinfo {pages} {68}
  (\bibinfo {year} {2009})}\BibitemShut {NoStop}%
\bibitem [{\citenamefont {Riplinger}\ and\ \citenamefont
  {Neese}(2013)}]{riplinger_efficient_2013}%
  \BibitemOpen
  \bibfield  {author} {\bibinfo {author} {\bibfnamefont {C.}~\bibnamefont
  {Riplinger}}\ and\ \bibinfo {author} {\bibfnamefont {F.}~\bibnamefont
  {Neese}},\ }\href {\doibase 10.1063/1.4773581} {\bibfield  {journal}
  {\bibinfo  {journal} {J. Chem. Phys.}\ }\textbf {\bibinfo {volume} {138}},\
  \bibinfo {pages} {034106} (\bibinfo {year} {2013})}\BibitemShut {NoStop}%
\bibitem [{\citenamefont {Werner}\ and\ \citenamefont
  {Sch{\"u}tz}(2011)}]{werner_efficient_2011}%
  \BibitemOpen
  \bibfield  {author} {\bibinfo {author} {\bibfnamefont {H.-J.}\ \bibnamefont
  {Werner}}\ and\ \bibinfo {author} {\bibfnamefont {M.}~\bibnamefont
  {Sch{\"u}tz}},\ }\href {\doibase 10.1063/1.3641642} {\bibfield  {journal}
  {\bibinfo  {journal} {J. Chem. Phys.}\ }\textbf {\bibinfo {volume} {135}},\
  \bibinfo {pages} {144116} (\bibinfo {year} {2011})}\BibitemShut {NoStop}%
\bibitem [{\citenamefont {Sch{\"u}tz}, \citenamefont {Hetzer},\ and\
  \citenamefont {Werner}(1999)}]{schutz_low-order_1999}%
  \BibitemOpen
  \bibfield  {author} {\bibinfo {author} {\bibfnamefont {M.}~\bibnamefont
  {Sch{\"u}tz}}, \bibinfo {author} {\bibfnamefont {G.}~\bibnamefont {Hetzer}},
  \ and\ \bibinfo {author} {\bibfnamefont {H.-J.}\ \bibnamefont {Werner}},\
  }\href {\doibase 10.1063/1.479957} {\bibfield  {journal} {\bibinfo  {journal}
  {J. Chem. Phys.}\ }\textbf {\bibinfo {volume} {111}},\ \bibinfo {pages}
  {5691} (\bibinfo {year} {1999})}\BibitemShut {NoStop}%
\bibitem [{\citenamefont {Ma}\ and\ \citenamefont
  {Werner}(2018)}]{ma_explicitly_2018}%
  \BibitemOpen
  \bibfield  {author} {\bibinfo {author} {\bibfnamefont {Q.}~\bibnamefont
  {Ma}}\ and\ \bibinfo {author} {\bibfnamefont {H.-J.}\ \bibnamefont
  {Werner}},\ }\href {\doibase 10.1002/wcms.1371} {\bibfield  {journal}
  {\bibinfo  {journal} {WIREs Comput. Mol. Sci.}\ ,\ \bibinfo {pages} {e1371}}
  (\bibinfo {year} {2018})}\BibitemShut {NoStop}%
\bibitem [{\citenamefont {Riplinger}\ \emph {et~al.}(2016)\citenamefont
  {Riplinger}, \citenamefont {Pinski}, \citenamefont {Becker}, \citenamefont
  {Valeev},\ and\ \citenamefont {Neese}}]{riplinger_sparse_2016}%
  \BibitemOpen
  \bibfield  {author} {\bibinfo {author} {\bibfnamefont {C.}~\bibnamefont
  {Riplinger}}, \bibinfo {author} {\bibfnamefont {P.}~\bibnamefont {Pinski}},
  \bibinfo {author} {\bibfnamefont {U.}~\bibnamefont {Becker}}, \bibinfo
  {author} {\bibfnamefont {E.~F.}\ \bibnamefont {Valeev}}, \ and\ \bibinfo
  {author} {\bibfnamefont {F.}~\bibnamefont {Neese}},\ }\href {\doibase
  10.1063/1.4939030} {\bibfield  {journal} {\bibinfo  {journal} {J. Chem.
  Phys.}\ }\textbf {\bibinfo {volume} {144}},\ \bibinfo {pages} {024109}
  (\bibinfo {year} {2016})}\BibitemShut {NoStop}%
\bibitem [{\citenamefont {Del~Ben}, \citenamefont {Hutter},\ and\ \citenamefont
  {VandeVondele}(2013)}]{del_ben_electron_2013}%
  \BibitemOpen
  \bibfield  {author} {\bibinfo {author} {\bibfnamefont {M.}~\bibnamefont
  {Del~Ben}}, \bibinfo {author} {\bibfnamefont {J.}~\bibnamefont {Hutter}}, \
  and\ \bibinfo {author} {\bibfnamefont {J.}~\bibnamefont {VandeVondele}},\
  }\href {\doibase 10.1021/ct4002202} {\bibfield  {journal} {\bibinfo
  {journal} {J. Chem. Theory Comput.}\ }\textbf {\bibinfo {volume} {9}},\
  \bibinfo {pages} {2654} (\bibinfo {year} {2013})}\BibitemShut {NoStop}%
\bibitem [{\citenamefont {Wilhelm}\ \emph {et~al.}(2016)\citenamefont
  {Wilhelm}, \citenamefont {Seewald}, \citenamefont {Del~Ben},\ and\
  \citenamefont {Hutter}}]{wilhelm_large-scale_2016}%
  \BibitemOpen
  \bibfield  {author} {\bibinfo {author} {\bibfnamefont {J.}~\bibnamefont
  {Wilhelm}}, \bibinfo {author} {\bibfnamefont {P.}~\bibnamefont {Seewald}},
  \bibinfo {author} {\bibfnamefont {M.}~\bibnamefont {Del~Ben}}, \ and\
  \bibinfo {author} {\bibfnamefont {J.}~\bibnamefont {Hutter}},\ }\href
  {\doibase 10.1021/acs.jctc.6b00840} {\bibfield  {journal} {\bibinfo
  {journal} {J. Chem. Theory Comput.}\ }\textbf {\bibinfo {volume} {12}},\
  \bibinfo {pages} {5851} (\bibinfo {year} {2016})}\BibitemShut {NoStop}%
\bibitem [{\citenamefont {Grüneis}\ \emph {et~al.}(2011)\citenamefont
  {Grüneis}, \citenamefont {Booth}, \citenamefont {Marsman}, \citenamefont
  {Spencer}, \citenamefont {Alavi},\ and\ \citenamefont
  {Kresse}}]{gruneis_natural_2011}%
  \BibitemOpen
  \bibfield  {author} {\bibinfo {author} {\bibfnamefont {A.}~\bibnamefont
  {Grüneis}}, \bibinfo {author} {\bibfnamefont {G.~H.}\ \bibnamefont {Booth}},
  \bibinfo {author} {\bibfnamefont {M.}~\bibnamefont {Marsman}}, \bibinfo
  {author} {\bibfnamefont {J.}~\bibnamefont {Spencer}}, \bibinfo {author}
  {\bibfnamefont {A.}~\bibnamefont {Alavi}}, \ and\ \bibinfo {author}
  {\bibfnamefont {G.}~\bibnamefont {Kresse}},\ }\href {\doibase
  10.1021/ct200263g} {\bibfield  {journal} {\bibinfo  {journal} {J. Chem.
  Theory Comput.}\ }\textbf {\bibinfo {volume} {7}},\ \bibinfo {pages} {2780}
  (\bibinfo {year} {2011})}\BibitemShut {NoStop}%
\bibitem [{\citenamefont {Kaltak}, \citenamefont {Klimeš},\ and\ \citenamefont
  {Kresse}(2014)}]{kaltak_low_2014}%
  \BibitemOpen
  \bibfield  {author} {\bibinfo {author} {\bibfnamefont {M.}~\bibnamefont
  {Kaltak}}, \bibinfo {author} {\bibfnamefont {J.}~\bibnamefont {Klimeš}}, \
  and\ \bibinfo {author} {\bibfnamefont {G.}~\bibnamefont {Kresse}},\ }\href
  {\doibase 10.1021/ct5001268} {\bibfield  {journal} {\bibinfo  {journal} {J.
  Chem. Theory Comput.}\ }\textbf {\bibinfo {volume} {10}},\ \bibinfo {pages}
  {2498} (\bibinfo {year} {2014})}\BibitemShut {NoStop}%
\bibitem [{\citenamefont {Marsili}\ \emph {et~al.}(2017)\citenamefont
  {Marsili}, \citenamefont {Mosconi}, \citenamefont {De~Angelis},\ and\
  \citenamefont {Umari}}]{marsili_large-scale_2017}%
  \BibitemOpen
  \bibfield  {author} {\bibinfo {author} {\bibfnamefont {M.}~\bibnamefont
  {Marsili}}, \bibinfo {author} {\bibfnamefont {E.}~\bibnamefont {Mosconi}},
  \bibinfo {author} {\bibfnamefont {F.}~\bibnamefont {De~Angelis}}, \ and\
  \bibinfo {author} {\bibfnamefont {P.}~\bibnamefont {Umari}},\ }\href
  {\doibase 10.1103/PhysRevB.95.075415} {\bibfield  {journal} {\bibinfo
  {journal} {Phys. Rev. B}\ }\textbf {\bibinfo {volume} {95}},\ \bibinfo
  {pages} {075415} (\bibinfo {year} {2017})}\BibitemShut {NoStop}%
\bibitem [{\citenamefont {Umari}\ \emph {et~al.}(2011)\citenamefont {Umari},
  \citenamefont {Qian}, \citenamefont {Marzari}, \citenamefont {Stenuit},
  \citenamefont {Giacomazzi},\ and\ \citenamefont
  {Baroni}}]{umari_accelerating_2011}%
  \BibitemOpen
  \bibfield  {author} {\bibinfo {author} {\bibfnamefont {P.}~\bibnamefont
  {Umari}}, \bibinfo {author} {\bibfnamefont {X.}~\bibnamefont {Qian}},
  \bibinfo {author} {\bibfnamefont {N.}~\bibnamefont {Marzari}}, \bibinfo
  {author} {\bibfnamefont {G.}~\bibnamefont {Stenuit}}, \bibinfo {author}
  {\bibfnamefont {L.}~\bibnamefont {Giacomazzi}}, \ and\ \bibinfo {author}
  {\bibfnamefont {S.}~\bibnamefont {Baroni}},\ }\href {\doibase
  10.1002/pssb.201046264} {\bibfield  {journal} {\bibinfo  {journal} {Phys.
  Status Solidi B}\ }\textbf {\bibinfo {volume} {248}},\ \bibinfo {pages} {527}
  (\bibinfo {year} {2011})}\BibitemShut {NoStop}%
\bibitem [{\citenamefont {Neuhauser}\ \emph {et~al.}(2014)\citenamefont
  {Neuhauser}, \citenamefont {Gao}, \citenamefont {Arntsen}, \citenamefont
  {Karshenas}, \citenamefont {Rabani},\ and\ \citenamefont
  {Baer}}]{neuhauser_breaking_2014}%
  \BibitemOpen
  \bibfield  {author} {\bibinfo {author} {\bibfnamefont {D.}~\bibnamefont
  {Neuhauser}}, \bibinfo {author} {\bibfnamefont {Y.}~\bibnamefont {Gao}},
  \bibinfo {author} {\bibfnamefont {C.}~\bibnamefont {Arntsen}}, \bibinfo
  {author} {\bibfnamefont {C.}~\bibnamefont {Karshenas}}, \bibinfo {author}
  {\bibfnamefont {E.}~\bibnamefont {Rabani}}, \ and\ \bibinfo {author}
  {\bibfnamefont {R.}~\bibnamefont {Baer}},\ }\href {\doibase
  10.1103/PhysRevLett.113.076402} {\bibfield  {journal} {\bibinfo  {journal}
  {Phys. Rev. Lett.}\ }\textbf {\bibinfo {volume} {113}},\ \bibinfo {pages}
  {076402} (\bibinfo {year} {2014})}\BibitemShut {NoStop}%
\bibitem [{\citenamefont {Ljungberg}\ \emph {et~al.}(2015)\citenamefont
  {Ljungberg}, \citenamefont {Koval}, \citenamefont {Ferrari}, \citenamefont
  {Foerster},\ and\ \citenamefont
  {Sánchez-Portal}}]{ljungberg_cubic-scaling_2015}%
  \BibitemOpen
  \bibfield  {author} {\bibinfo {author} {\bibfnamefont {M.~P.}\ \bibnamefont
  {Ljungberg}}, \bibinfo {author} {\bibfnamefont {P.}~\bibnamefont {Koval}},
  \bibinfo {author} {\bibfnamefont {F.}~\bibnamefont {Ferrari}}, \bibinfo
  {author} {\bibfnamefont {D.}~\bibnamefont {Foerster}}, \ and\ \bibinfo
  {author} {\bibfnamefont {D.}~\bibnamefont {Sánchez-Portal}},\ }\href
  {\doibase 10.1103/PhysRevB.92.075422} {\bibfield  {journal} {\bibinfo
  {journal} {Phys. Rev. B}\ }\textbf {\bibinfo {volume} {92}},\ \bibinfo
  {pages} {075422} (\bibinfo {year} {2015})}\BibitemShut {NoStop}%
\bibitem [{\citenamefont {Booth}\ \emph {et~al.}(2013)\citenamefont {Booth},
  \citenamefont {Grüneis}, \citenamefont {Kresse},\ and\ \citenamefont
  {Alavi}}]{booth_towards_2013}%
  \BibitemOpen
  \bibfield  {author} {\bibinfo {author} {\bibfnamefont {G.~H.}\ \bibnamefont
  {Booth}}, \bibinfo {author} {\bibfnamefont {A.}~\bibnamefont {Grüneis}},
  \bibinfo {author} {\bibfnamefont {G.}~\bibnamefont {Kresse}}, \ and\ \bibinfo
  {author} {\bibfnamefont {A.}~\bibnamefont {Alavi}},\ }\href {\doibase
  10.1038/nature11770} {\bibfield  {journal} {\bibinfo  {journal} {Nature}\
  }\textbf {\bibinfo {volume} {493}},\ \bibinfo {pages} {365} (\bibinfo {year}
  {2013})}\BibitemShut {NoStop}%
\bibitem [{\citenamefont {Ihrig}\ \emph {et~al.}(2015)\citenamefont {Ihrig},
  \citenamefont {Wieferink}, \citenamefont {Zhang}, \citenamefont {Ropo},
  \citenamefont {Ren}, \citenamefont {Rinke}, \citenamefont {Scheffler},\ and\
  \citenamefont {Blum}}]{ihrig_accurate_2015}%
  \BibitemOpen
  \bibfield  {author} {\bibinfo {author} {\bibfnamefont {A.~C.}\ \bibnamefont
  {Ihrig}}, \bibinfo {author} {\bibfnamefont {J.}~\bibnamefont {Wieferink}},
  \bibinfo {author} {\bibfnamefont {I.~Y.}\ \bibnamefont {Zhang}}, \bibinfo
  {author} {\bibfnamefont {M.}~\bibnamefont {Ropo}}, \bibinfo {author}
  {\bibfnamefont {X.}~\bibnamefont {Ren}}, \bibinfo {author} {\bibfnamefont
  {P.}~\bibnamefont {Rinke}}, \bibinfo {author} {\bibfnamefont
  {M.}~\bibnamefont {Scheffler}}, \ and\ \bibinfo {author} {\bibfnamefont
  {V.}~\bibnamefont {Blum}},\ }\href {\doibase 10.1088/1367-2630/17/9/093020}
  {\bibfield  {journal} {\bibinfo  {journal} {New J. Phys.}\ }\textbf {\bibinfo
  {volume} {17}},\ \bibinfo {pages} {093020} (\bibinfo {year}
  {2015})}\BibitemShut {NoStop}%
\bibitem [{\citenamefont {Hu}\ \emph {et~al.}(2017)\citenamefont {Hu},
  \citenamefont {Lin}, \citenamefont {Banerjee}, \citenamefont {Vecharynski},\
  and\ \citenamefont {Yang}}]{hu_adaptively_2017}%
  \BibitemOpen
  \bibfield  {author} {\bibinfo {author} {\bibfnamefont {W.}~\bibnamefont
  {Hu}}, \bibinfo {author} {\bibfnamefont {L.}~\bibnamefont {Lin}}, \bibinfo
  {author} {\bibfnamefont {A.~S.}\ \bibnamefont {Banerjee}}, \bibinfo {author}
  {\bibfnamefont {E.}~\bibnamefont {Vecharynski}}, \ and\ \bibinfo {author}
  {\bibfnamefont {C.}~\bibnamefont {Yang}},\ }\href {\doibase
  10.1021/acs.jctc.6b01184} {\bibfield  {journal} {\bibinfo  {journal} {J.
  Chem. Theory Comput.}\ }\textbf {\bibinfo {volume} {13}},\ \bibinfo {pages}
  {1188} (\bibinfo {year} {2017})}\BibitemShut {NoStop}%
\bibitem [{\citenamefont {Ayala}, \citenamefont {Kudin},\ and\ \citenamefont
  {Scuseria}(2001)}]{ayala_atomic_2001}%
  \BibitemOpen
  \bibfield  {author} {\bibinfo {author} {\bibfnamefont {P.~Y.}\ \bibnamefont
  {Ayala}}, \bibinfo {author} {\bibfnamefont {K.~N.}\ \bibnamefont {Kudin}}, \
  and\ \bibinfo {author} {\bibfnamefont {G.~E.}\ \bibnamefont {Scuseria}},\
  }\href {\doibase 10.1063/1.1414369} {\bibfield  {journal} {\bibinfo
  {journal} {J. Chem. Phys.}\ }\textbf {\bibinfo {volume} {115}},\ \bibinfo
  {pages} {9698} (\bibinfo {year} {2001})}\BibitemShut {NoStop}%
\bibitem [{\citenamefont {Izmaylov}\ and\ \citenamefont
  {Scuseria}(2008)}]{f.izmaylov_resolution_2008}%
  \BibitemOpen
  \bibfield  {author} {\bibinfo {author} {\bibfnamefont {A.~F.}\ \bibnamefont
  {Izmaylov}}\ and\ \bibinfo {author} {\bibfnamefont {G.~E.}\ \bibnamefont
  {Scuseria}},\ }\href {\doibase 10.1039/B803274M} {\bibfield  {journal}
  {\bibinfo  {journal} {Phys. Chem. Chem. Phys.}\ }\textbf {\bibinfo {volume}
  {10}},\ \bibinfo {pages} {3421} (\bibinfo {year} {2008})}\BibitemShut
  {NoStop}%
\bibitem [{\citenamefont {Pisani}\ \emph {et~al.}(2005)\citenamefont {Pisani},
  \citenamefont {Busso}, \citenamefont {Capecchi}, \citenamefont {Casassa},
  \citenamefont {Dovesi}, \citenamefont {Maschio}, \citenamefont
  {Zicovich-Wilson},\ and\ \citenamefont {Sch{\"u}tz}}]{pisani_local-mp2_2005}%
  \BibitemOpen
  \bibfield  {author} {\bibinfo {author} {\bibfnamefont {C.}~\bibnamefont
  {Pisani}}, \bibinfo {author} {\bibfnamefont {M.}~\bibnamefont {Busso}},
  \bibinfo {author} {\bibfnamefont {G.}~\bibnamefont {Capecchi}}, \bibinfo
  {author} {\bibfnamefont {S.}~\bibnamefont {Casassa}}, \bibinfo {author}
  {\bibfnamefont {R.}~\bibnamefont {Dovesi}}, \bibinfo {author} {\bibfnamefont
  {L.}~\bibnamefont {Maschio}}, \bibinfo {author} {\bibfnamefont
  {C.}~\bibnamefont {Zicovich-Wilson}}, \ and\ \bibinfo {author} {\bibfnamefont
  {M.}~\bibnamefont {Sch{\"u}tz}},\ }\href {\doibase 10.1063/1.1857479}
  {\bibfield  {journal} {\bibinfo  {journal} {J. Chem. Phys.}\ }\textbf
  {\bibinfo {volume} {122}},\ \bibinfo {pages} {094113} (\bibinfo {year}
  {2005})}\BibitemShut {NoStop}%
\bibitem [{\citenamefont {Lorenz}, \citenamefont {Usvyat},\ and\ \citenamefont
  {Sch{\"u}tz}(2011)}]{lorenz_local_2011}%
  \BibitemOpen
  \bibfield  {author} {\bibinfo {author} {\bibfnamefont {M.}~\bibnamefont
  {Lorenz}}, \bibinfo {author} {\bibfnamefont {D.}~\bibnamefont {Usvyat}}, \
  and\ \bibinfo {author} {\bibfnamefont {M.}~\bibnamefont {Sch{\"u}tz}},\
  }\href {\doibase 10.1063/1.3554209} {\bibfield  {journal} {\bibinfo
  {journal} {J. Chem. Phys.}\ }\textbf {\bibinfo {volume} {134}},\ \bibinfo
  {pages} {094101} (\bibinfo {year} {2011})}\BibitemShut {NoStop}%
\bibitem [{\citenamefont {Pisani}\ \emph {et~al.}(2012)\citenamefont {Pisani},
  \citenamefont {Sch{\"u}tz}, \citenamefont {Casassa}, \citenamefont {Usvyat},
  \citenamefont {Maschio}, \citenamefont {Lorenz},\ and\ \citenamefont
  {Erba}}]{pisani_cryscor:_2012}%
  \BibitemOpen
  \bibfield  {author} {\bibinfo {author} {\bibfnamefont {C.}~\bibnamefont
  {Pisani}}, \bibinfo {author} {\bibfnamefont {M.}~\bibnamefont {Sch{\"u}tz}},
  \bibinfo {author} {\bibfnamefont {S.}~\bibnamefont {Casassa}}, \bibinfo
  {author} {\bibfnamefont {D.}~\bibnamefont {Usvyat}}, \bibinfo {author}
  {\bibfnamefont {L.}~\bibnamefont {Maschio}}, \bibinfo {author} {\bibfnamefont
  {M.}~\bibnamefont {Lorenz}}, \ and\ \bibinfo {author} {\bibfnamefont
  {A.}~\bibnamefont {Erba}},\ }\href {\doibase 10.1039/c2cp23927b} {\bibfield
  {journal} {\bibinfo  {journal} {Phys. Chem. Chem. Phys.}\ }\textbf {\bibinfo
  {volume} {14}},\ \bibinfo {pages} {7615} (\bibinfo {year}
  {2012})}\BibitemShut {NoStop}%
\bibitem [{\citenamefont {Sch{\"a}fer}, \citenamefont {Ramberger},\ and\
  \citenamefont {Kresse}(2017)}]{schafer_quartic_2017}%
  \BibitemOpen
  \bibfield  {author} {\bibinfo {author} {\bibfnamefont {T.}~\bibnamefont
  {Sch{\"a}fer}}, \bibinfo {author} {\bibfnamefont {B.}~\bibnamefont
  {Ramberger}}, \ and\ \bibinfo {author} {\bibfnamefont {G.}~\bibnamefont
  {Kresse}},\ }\href {\doibase 10.1063/1.4976937} {\bibfield  {journal}
  {\bibinfo  {journal} {J. Chem. Phys.}\ }\textbf {\bibinfo {volume} {146}},\
  \bibinfo {pages} {104101} (\bibinfo {year} {2017})}\BibitemShut {NoStop}%
\bibitem [{\citenamefont {Rebolini}\ \emph {et~al.}(2018)\citenamefont
  {Rebolini}, \citenamefont {Baardsen}, \citenamefont {Hansen}, \citenamefont
  {Leikanger},\ and\ \citenamefont
  {Pedersen}}]{rebolini_divideexpandconsolidate_2018}%
  \BibitemOpen
  \bibfield  {author} {\bibinfo {author} {\bibfnamefont {E.}~\bibnamefont
  {Rebolini}}, \bibinfo {author} {\bibfnamefont {G.}~\bibnamefont {Baardsen}},
  \bibinfo {author} {\bibfnamefont {A.~S.}\ \bibnamefont {Hansen}}, \bibinfo
  {author} {\bibfnamefont {K.~R.}\ \bibnamefont {Leikanger}}, \ and\ \bibinfo
  {author} {\bibfnamefont {T.~B.}\ \bibnamefont {Pedersen}},\ }\href {\doibase
  10.1021/acs.jctc.8b00021} {\bibfield  {journal} {\bibinfo  {journal} {J.
  Chem. Theory Comput.}\ }\textbf {\bibinfo {volume} {14}},\ \bibinfo {pages}
  {2427} (\bibinfo {year} {2018})}\BibitemShut {NoStop}%
\bibitem [{\citenamefont {Usvyat}, \citenamefont {Maschio},\ and\ \citenamefont
  {Sch{\"u}tz}(2018)}]{usvyat_periodic_2018}%
  \BibitemOpen
  \bibfield  {author} {\bibinfo {author} {\bibfnamefont {D.}~\bibnamefont
  {Usvyat}}, \bibinfo {author} {\bibfnamefont {L.}~\bibnamefont {Maschio}}, \
  and\ \bibinfo {author} {\bibfnamefont {M.}~\bibnamefont {Sch{\"u}tz}},\
  }\href {\doibase 10.1002/wcms.1357} {\bibfield  {journal} {\bibinfo
  {journal} {WIREs Comput. Mol. Sci.}\ ,\ \bibinfo {pages} {e1357}} (\bibinfo
  {year} {2018})}\BibitemShut {NoStop}%
\bibitem [{\citenamefont {Skriver}(1984)}]{skriver_lmto_1984}%
  \BibitemOpen
  \bibfield  {author} {\bibinfo {author} {\bibfnamefont {H.~L.}\ \bibnamefont
  {Skriver}},\ }\href@noop {} {\emph {\bibinfo {title} {The {LMTO} {Method}:
  {Muffin}-{Tin} {Orbitals} and {Electronic} {Structure}}}}\ (\bibinfo
  {publisher} {Springer-Verlag, Berlin},\ \bibinfo {year} {1984})\BibitemShut
  {NoStop}%
\bibitem [{\citenamefont {Jia}\ \emph {et~al.}(2009)\citenamefont {Jia},
  \citenamefont {Hill}, \citenamefont {Beylkin},\ and\ \citenamefont
  {Harrison}}]{Jia:2009vm}%
  \BibitemOpen
  \bibfield  {author} {\bibinfo {author} {\bibfnamefont {J.}~\bibnamefont
  {Jia}}, \bibinfo {author} {\bibfnamefont {J.~C.}\ \bibnamefont {Hill}},
  \bibinfo {author} {\bibfnamefont {G.}~\bibnamefont {Beylkin}}, \ and\
  \bibinfo {author} {\bibfnamefont {R.~J.}\ \bibnamefont {Harrison}},\ }in\
  \href@noop {} {\emph {\bibinfo {booktitle} {Proceedings of the 8th
  International Symposium on Distributed Computing and Applications to
  Business, Engineering and Science}}}\ (\bibinfo {year} {2009})\ pp.\ \bibinfo
  {pages} {13--16}\BibitemShut {NoStop}%
\bibitem [{\citenamefont {Blum}\ \emph {et~al.}(2009)\citenamefont {Blum},
  \citenamefont {Gehrke}, \citenamefont {Hanke}, \citenamefont {Havu},
  \citenamefont {Havu}, \citenamefont {Ren}, \citenamefont {Reuter},\ and\
  \citenamefont {Scheffler}}]{blum_ab_2009}%
  \BibitemOpen
  \bibfield  {author} {\bibinfo {author} {\bibfnamefont {V.}~\bibnamefont
  {Blum}}, \bibinfo {author} {\bibfnamefont {R.}~\bibnamefont {Gehrke}},
  \bibinfo {author} {\bibfnamefont {F.}~\bibnamefont {Hanke}}, \bibinfo
  {author} {\bibfnamefont {P.}~\bibnamefont {Havu}}, \bibinfo {author}
  {\bibfnamefont {V.}~\bibnamefont {Havu}}, \bibinfo {author} {\bibfnamefont
  {X.}~\bibnamefont {Ren}}, \bibinfo {author} {\bibfnamefont {K.}~\bibnamefont
  {Reuter}}, \ and\ \bibinfo {author} {\bibfnamefont {M.}~\bibnamefont
  {Scheffler}},\ }\href {\doibase 10.1016/j.cpc.2009.06.022} {\bibfield
  {journal} {\bibinfo  {journal} {Comput. Phys. Commun.}\ }\textbf {\bibinfo
  {volume} {180}},\ \bibinfo {pages} {2175} (\bibinfo {year}
  {2009})}\BibitemShut {NoStop}%
\bibitem [{\citenamefont {Levchenko}\ \emph {et~al.}(2015)\citenamefont
  {Levchenko}, \citenamefont {Ren}, \citenamefont {Wieferink}, \citenamefont
  {Johanni}, \citenamefont {Rinke}, \citenamefont {Blum},\ and\ \citenamefont
  {Scheffler}}]{levchenko_hybrid_2015}%
  \BibitemOpen
  \bibfield  {author} {\bibinfo {author} {\bibfnamefont {S.~V.}\ \bibnamefont
  {Levchenko}}, \bibinfo {author} {\bibfnamefont {X.}~\bibnamefont {Ren}},
  \bibinfo {author} {\bibfnamefont {J.}~\bibnamefont {Wieferink}}, \bibinfo
  {author} {\bibfnamefont {R.}~\bibnamefont {Johanni}}, \bibinfo {author}
  {\bibfnamefont {P.}~\bibnamefont {Rinke}}, \bibinfo {author} {\bibfnamefont
  {V.}~\bibnamefont {Blum}}, \ and\ \bibinfo {author} {\bibfnamefont
  {M.}~\bibnamefont {Scheffler}},\ }\href {\doibase 10.1016/j.cpc.2015.02.021}
  {\bibfield  {journal} {\bibinfo  {journal} {Comput. Phys. Commun.}\ }\textbf
  {\bibinfo {volume} {192}},\ \bibinfo {pages} {60} (\bibinfo {year}
  {2015})}\BibitemShut {NoStop}%
\bibitem [{\citenamefont {Ladik}\ and\ \citenamefont
  {Martino}(1973)}]{Ladik:1973il}%
  \BibitemOpen
  \bibfield  {author} {\bibinfo {author} {\bibfnamefont {J.}~\bibnamefont
  {Ladik}}\ and\ \bibinfo {author} {\bibfnamefont {F.}~\bibnamefont
  {Martino}},\ }\href@noop {} {\bibfield  {journal} {\bibinfo  {journal} {Acta
  Physica}\ }\textbf {\bibinfo {volume} {34}},\ \bibinfo {pages} {67} (\bibinfo
  {year} {1973})}\BibitemShut {NoStop}%
\bibitem [{\citenamefont {Harris}(1975)}]{HARRIS1975147}%
  \BibitemOpen
  \bibfield  {author} {\bibinfo {author} {\bibfnamefont {F.~E.}\ \bibnamefont
  {Harris}},\ }in\ \href {\doibase
  https://doi.org/10.1016/B978-0-12-681901-4.50011-8} {\emph {\bibinfo
  {booktitle} {Theoretical Chemistry}}},\ \bibinfo {series} {Theoretical
  Chemistry}, Vol.~\bibinfo {volume} {1},\ \bibinfo {editor} {edited by\
  \bibinfo {editor} {\bibfnamefont {H.}~\bibnamefont {EYRING}}\ and\ \bibinfo
  {editor} {\bibfnamefont {D.}~\bibnamefont {HENDERSON}}}\ (\bibinfo
  {publisher} {Elsevier},\ \bibinfo {year} {1975})\ pp.\ \bibinfo {pages} {147
  -- 218}\BibitemShut {NoStop}%
\bibitem [{\citenamefont {Pisani}\ and\ \citenamefont
  {Dovesi}(1980)}]{pisani_exact-exchange_1980}%
  \BibitemOpen
  \bibfield  {author} {\bibinfo {author} {\bibfnamefont {C.}~\bibnamefont
  {Pisani}}\ and\ \bibinfo {author} {\bibfnamefont {R.}~\bibnamefont
  {Dovesi}},\ }\href {\doibase 10.1002/qua.560170311} {\bibfield  {journal}
  {\bibinfo  {journal} {Int. J. Quantum Chem.}\ }\textbf {\bibinfo {volume}
  {17}},\ \bibinfo {pages} {501} (\bibinfo {year} {1980})}\BibitemShut
  {NoStop}%
\bibitem [{\citenamefont {Saunders}\ \emph {et~al.}(1992)\citenamefont
  {Saunders}, \citenamefont {Freyria-Fava}, \citenamefont {Dovesi},
  \citenamefont {Salasco},\ and\ \citenamefont {Roetti}}]{Saunders:1992iz}%
  \BibitemOpen
  \bibfield  {author} {\bibinfo {author} {\bibfnamefont {V.~R.}\ \bibnamefont
  {Saunders}}, \bibinfo {author} {\bibfnamefont {C.}~\bibnamefont
  {Freyria-Fava}}, \bibinfo {author} {\bibfnamefont {R.}~\bibnamefont
  {Dovesi}}, \bibinfo {author} {\bibfnamefont {L.}~\bibnamefont {Salasco}}, \
  and\ \bibinfo {author} {\bibfnamefont {C.}~\bibnamefont {Roetti}},\
  }\href@noop {} {\bibfield  {journal} {\bibinfo  {journal} {Molecular
  Physics}\ }\textbf {\bibinfo {volume} {77}},\ \bibinfo {pages} {629}
  (\bibinfo {year} {1992})}\BibitemShut {NoStop}%
\bibitem [{\citenamefont {Varga}(2008)}]{varga_long-range_2008}%
  \BibitemOpen
  \bibfield  {author} {\bibinfo {author} {\bibfnamefont {{\u S}.}~\bibnamefont
  {Varga}},\ }\href {\doibase 10.1002/qua.21682} {\bibfield  {journal}
  {\bibinfo  {journal} {Int. J. Quantum Chem.}\ }\textbf {\bibinfo {volume}
  {108}},\ \bibinfo {pages} {1518} (\bibinfo {year} {2008})}\BibitemShut
  {NoStop}%
\bibitem [{\citenamefont {Burow}, \citenamefont {Sierka},\ and\ \citenamefont
  {Mohamed}(2009)}]{burow_resolution_2009}%
  \BibitemOpen
  \bibfield  {author} {\bibinfo {author} {\bibfnamefont {A.~M.}\ \bibnamefont
  {Burow}}, \bibinfo {author} {\bibfnamefont {M.}~\bibnamefont {Sierka}}, \
  and\ \bibinfo {author} {\bibfnamefont {F.}~\bibnamefont {Mohamed}},\ }\href
  {\doibase 10.1063/1.3267858} {\bibfield  {journal} {\bibinfo  {journal} {J.
  Chem. Phys.}\ }\textbf {\bibinfo {volume} {131}},\ \bibinfo {pages} {214101}
  (\bibinfo {year} {2009})}\BibitemShut {NoStop}%
\bibitem [{\citenamefont {Maschio}\ \emph {et~al.}(2007)\citenamefont
  {Maschio}, \citenamefont {Usvyat}, \citenamefont {Manby}, \citenamefont
  {Casassa}, \citenamefont {Pisani},\ and\ \citenamefont
  {Sch{\"u}tz}}]{maschio_fast_2007}%
  \BibitemOpen
  \bibfield  {author} {\bibinfo {author} {\bibfnamefont {L.}~\bibnamefont
  {Maschio}}, \bibinfo {author} {\bibfnamefont {D.}~\bibnamefont {Usvyat}},
  \bibinfo {author} {\bibfnamefont {F.~R.}\ \bibnamefont {Manby}}, \bibinfo
  {author} {\bibfnamefont {S.}~\bibnamefont {Casassa}}, \bibinfo {author}
  {\bibfnamefont {C.}~\bibnamefont {Pisani}}, \ and\ \bibinfo {author}
  {\bibfnamefont {M.}~\bibnamefont {Sch{\"u}tz}},\ }\href {\doibase
  10.1103/PhysRevB.76.075101} {\bibfield  {journal} {\bibinfo  {journal} {Phys.
  Rev. B}\ }\textbf {\bibinfo {volume} {76}},\ \bibinfo {pages} {075101}
  (\bibinfo {year} {2007})}\BibitemShut {NoStop}%
\bibitem [{\citenamefont {Usvyat}\ \emph {et~al.}(2007)\citenamefont {Usvyat},
  \citenamefont {Maschio}, \citenamefont {Manby}, \citenamefont {Casassa},
  \citenamefont {Sch{\"u}tz},\ and\ \citenamefont {Pisani}}]{usvyat_fast_2007}%
  \BibitemOpen
  \bibfield  {author} {\bibinfo {author} {\bibfnamefont {D.}~\bibnamefont
  {Usvyat}}, \bibinfo {author} {\bibfnamefont {L.}~\bibnamefont {Maschio}},
  \bibinfo {author} {\bibfnamefont {F.~R.}\ \bibnamefont {Manby}}, \bibinfo
  {author} {\bibfnamefont {S.}~\bibnamefont {Casassa}}, \bibinfo {author}
  {\bibfnamefont {M.}~\bibnamefont {Sch{\"u}tz}}, \ and\ \bibinfo {author}
  {\bibfnamefont {C.}~\bibnamefont {Pisani}},\ }\href {\doibase
  10.1103/PhysRevB.76.075102} {\bibfield  {journal} {\bibinfo  {journal} {Phys.
  Rev. B}\ }\textbf {\bibinfo {volume} {76}},\ \bibinfo {pages} {075102}
  (\bibinfo {year} {2007})}\BibitemShut {NoStop}%
\bibitem [{\citenamefont {Sun}\ \emph {et~al.}(2017)\citenamefont {Sun},
  \citenamefont {Berkelbach}, \citenamefont {McClain},\ and\ \citenamefont
  {Chan}}]{sun_gaussian_2017}%
  \BibitemOpen
  \bibfield  {author} {\bibinfo {author} {\bibfnamefont {Q.}~\bibnamefont
  {Sun}}, \bibinfo {author} {\bibfnamefont {T.~C.}\ \bibnamefont {Berkelbach}},
  \bibinfo {author} {\bibfnamefont {J.~D.}\ \bibnamefont {McClain}}, \ and\
  \bibinfo {author} {\bibfnamefont {G.~K.-L.}\ \bibnamefont {Chan}},\ }\href
  {\doibase 10.1063/1.4998644} {\bibfield  {journal} {\bibinfo  {journal} {J.
  Chem. Phys.}\ }\textbf {\bibinfo {volume} {147}},\ \bibinfo {pages} {164119}
  (\bibinfo {year} {2017})}\BibitemShut {NoStop}%
\bibitem [{\citenamefont {Werner}, \citenamefont {Manby},\ and\ \citenamefont
  {Knowles}(2003)}]{Werner03}%
  \BibitemOpen
  \bibfield  {author} {\bibinfo {author} {\bibfnamefont {H.-J.}\ \bibnamefont
  {Werner}}, \bibinfo {author} {\bibfnamefont {F.~R.}\ \bibnamefont {Manby}}, \
  and\ \bibinfo {author} {\bibfnamefont {P.~J.}\ \bibnamefont {Knowles}},\
  }\href@noop {} {\bibfield  {journal} {\bibinfo  {journal} {J. Chem. Phys.}\
  }\textbf {\bibinfo {volume} {118}},\ \bibinfo {pages} {8149} (\bibinfo {year}
  {2003})}\BibitemShut {NoStop}%
\bibitem [{\citenamefont {Neese}, \citenamefont {Wennmohs},\ and\ \citenamefont
  {Hansen}(2009)}]{Neese2009}%
  \BibitemOpen
  \bibfield  {author} {\bibinfo {author} {\bibfnamefont {F.}~\bibnamefont
  {Neese}}, \bibinfo {author} {\bibfnamefont {F.}~\bibnamefont {Wennmohs}}, \
  and\ \bibinfo {author} {\bibfnamefont {A.}~\bibnamefont {Hansen}},\ }\href
  {\doibase 10.1063/1.3086717} {\bibfield  {journal} {\bibinfo  {journal} {J.
  Chem. Phys.}\ }\textbf {\bibinfo {volume} {130}},\ \bibinfo {pages} {114108}
  (\bibinfo {year} {2009})}\BibitemShut {NoStop}%
\bibitem [{\citenamefont {Pinski}\ \emph {et~al.}(2015)\citenamefont {Pinski},
  \citenamefont {Riplinger}, \citenamefont {Valeev},\ and\ \citenamefont
  {Neese}}]{Pinski2015}%
  \BibitemOpen
  \bibfield  {author} {\bibinfo {author} {\bibfnamefont {P.}~\bibnamefont
  {Pinski}}, \bibinfo {author} {\bibfnamefont {C.}~\bibnamefont {Riplinger}},
  \bibinfo {author} {\bibfnamefont {E.~F.}\ \bibnamefont {Valeev}}, \ and\
  \bibinfo {author} {\bibfnamefont {F.}~\bibnamefont {Neese}},\ }\href
  {\doibase 10.1063/1.4926879} {\bibfield  {journal} {\bibinfo  {journal} {J.
  Chem. Phys.}\ }\textbf {\bibinfo {volume} {143}},\ \bibinfo {pages} {034108}
  (\bibinfo {year} {2015})}\BibitemShut {NoStop}%
\bibitem [{\citenamefont {Hollman}, \citenamefont {Schaefer},\ and\
  \citenamefont {Valeev}(2014)}]{hollman_semi-exact_2014}%
  \BibitemOpen
  \bibfield  {author} {\bibinfo {author} {\bibfnamefont {D.~S.}\ \bibnamefont
  {Hollman}}, \bibinfo {author} {\bibfnamefont {H.~F.}\ \bibnamefont
  {Schaefer}}, \ and\ \bibinfo {author} {\bibfnamefont {E.~F.}\ \bibnamefont
  {Valeev}},\ }\href {\doibase 10.1063/1.4864755} {\bibfield  {journal}
  {\bibinfo  {journal} {J. Chem. Phys.}\ }\textbf {\bibinfo {volume} {140}},\
  \bibinfo {pages} {064109} (\bibinfo {year} {2014})}\BibitemShut {NoStop}%
\bibitem [{\citenamefont {Hollman}, \citenamefont {Schaefer},\ and\
  \citenamefont {Valeev}(2017)}]{hollman_fast_2017}%
  \BibitemOpen
  \bibfield  {author} {\bibinfo {author} {\bibfnamefont {D.~S.}\ \bibnamefont
  {Hollman}}, \bibinfo {author} {\bibfnamefont {H.~F.}\ \bibnamefont
  {Schaefer}}, \ and\ \bibinfo {author} {\bibfnamefont {E.~F.}\ \bibnamefont
  {Valeev}},\ }\href {\doibase 10.1080/00268976.2017.1346312} {\bibfield
  {journal} {\bibinfo  {journal} {Mol. Phys.}\ }\textbf {\bibinfo {volume}
  {115}},\ \bibinfo {pages} {2065} (\bibinfo {year} {2017})}\BibitemShut
  {NoStop}%
\bibitem [{\citenamefont {Merlot}\ \emph {et~al.}(2013)\citenamefont {Merlot},
  \citenamefont {Kj{\ae}rgaard}, \citenamefont {Helgaker}, \citenamefont
  {Lindh}, \citenamefont {Aquilante}, \citenamefont {Reine},\ and\
  \citenamefont {Pedersen}}]{merlot_attractive_2013}%
  \BibitemOpen
  \bibfield  {author} {\bibinfo {author} {\bibfnamefont {P.}~\bibnamefont
  {Merlot}}, \bibinfo {author} {\bibfnamefont {T.}~\bibnamefont
  {Kj{\ae}rgaard}}, \bibinfo {author} {\bibfnamefont {T.}~\bibnamefont
  {Helgaker}}, \bibinfo {author} {\bibfnamefont {R.}~\bibnamefont {Lindh}},
  \bibinfo {author} {\bibfnamefont {F.}~\bibnamefont {Aquilante}}, \bibinfo
  {author} {\bibfnamefont {S.}~\bibnamefont {Reine}}, \ and\ \bibinfo {author}
  {\bibfnamefont {T.~B.}\ \bibnamefont {Pedersen}},\ }\href {\doibase
  10.1002/jcc.23284} {\bibfield  {journal} {\bibinfo  {journal} {J. Comput.
  Chem.}\ }\textbf {\bibinfo {volume} {34}},\ \bibinfo {pages} {1486} (\bibinfo
  {year} {2013})}\BibitemShut {NoStop}%
\bibitem [{\citenamefont {Manzer}, \citenamefont {Epifanovsky},\ and\
  \citenamefont {Head-Gordon}(2015)}]{manzer_efficient_2015}%
  \BibitemOpen
  \bibfield  {author} {\bibinfo {author} {\bibfnamefont {S.~F.}\ \bibnamefont
  {Manzer}}, \bibinfo {author} {\bibfnamefont {E.}~\bibnamefont {Epifanovsky}},
  \ and\ \bibinfo {author} {\bibfnamefont {M.}~\bibnamefont {Head-Gordon}},\
  }\href {\doibase 10.1021/ct5008586} {\bibfield  {journal} {\bibinfo
  {journal} {J. Chem. Theory Comput.}\ }\textbf {\bibinfo {volume} {11}},\
  \bibinfo {pages} {518} (\bibinfo {year} {2015})}\BibitemShut {NoStop}%
\bibitem [{\citenamefont {Rebolini}\ \emph {et~al.}(2016)\citenamefont
  {Rebolini}, \citenamefont {Izsá{\'a}k}, \citenamefont {Reine}, \citenamefont
  {Helgaker},\ and\ \citenamefont {Pedersen}}]{rebolini_comparison_2016}%
  \BibitemOpen
  \bibfield  {author} {\bibinfo {author} {\bibfnamefont {E.}~\bibnamefont
  {Rebolini}}, \bibinfo {author} {\bibfnamefont {R.}~\bibnamefont
  {Izsá{\'a}k}}, \bibinfo {author} {\bibfnamefont {S.~S.}\ \bibnamefont
  {Reine}}, \bibinfo {author} {\bibfnamefont {T.}~\bibnamefont {Helgaker}}, \
  and\ \bibinfo {author} {\bibfnamefont {T.~B.}\ \bibnamefont {Pedersen}},\
  }\href {\doibase 10.1021/acs.jctc.6b00074} {\bibfield  {journal} {\bibinfo
  {journal} {J. Chem. Theory Comput.}\ }\textbf {\bibinfo {volume} {12}},\
  \bibinfo {pages} {3514} (\bibinfo {year} {2016})}\BibitemShut {NoStop}%
\bibitem [{\citenamefont {Whitten}(1973)}]{whitten_coulombic_1973}%
  \BibitemOpen
  \bibfield  {author} {\bibinfo {author} {\bibfnamefont {J.~L.}\ \bibnamefont
  {Whitten}},\ }\href {\doibase 10.1063/1.1679012} {\bibfield  {journal}
  {\bibinfo  {journal} {J. Chem. Phys.}\ }\textbf {\bibinfo {volume} {58}},\
  \bibinfo {pages} {4496} (\bibinfo {year} {1973})}\BibitemShut {NoStop}%
\bibitem [{\citenamefont {Jafri}\ and\ \citenamefont
  {Whitten}(1974)}]{jafri_electron_1974}%
  \BibitemOpen
  \bibfield  {author} {\bibinfo {author} {\bibfnamefont {J.~A.}\ \bibnamefont
  {Jafri}}\ and\ \bibinfo {author} {\bibfnamefont {J.~L.}\ \bibnamefont
  {Whitten}},\ }\href {\doibase 10.1063/1.1682222} {\bibfield  {journal}
  {\bibinfo  {journal} {J. Chem. Phys.}\ }\textbf {\bibinfo {volume} {61}},\
  \bibinfo {pages} {2116} (\bibinfo {year} {1974})}\BibitemShut {NoStop}%
\bibitem [{\citenamefont {Harris}\ and\ \citenamefont
  {Rein}(1966)}]{Harris:1966ky}%
  \BibitemOpen
  \bibfield  {author} {\bibinfo {author} {\bibfnamefont {F.~E.}\ \bibnamefont
  {Harris}}\ and\ \bibinfo {author} {\bibfnamefont {R.}~\bibnamefont {Rein}},\
  }\href@noop {} {\bibfield  {journal} {\bibinfo  {journal} {Theor. Chim.
  Acta}\ }\textbf {\bibinfo {volume} {6}},\ \bibinfo {pages} {73} (\bibinfo
  {year} {1966})}\BibitemShut {NoStop}%
\bibitem [{\citenamefont {Baerends}, \citenamefont {Ellis},\ and\ \citenamefont
  {Ros}(1973)}]{baerends_self-consistent_1973}%
  \BibitemOpen
  \bibfield  {author} {\bibinfo {author} {\bibfnamefont {E.~J.}\ \bibnamefont
  {Baerends}}, \bibinfo {author} {\bibfnamefont {D.~E.}\ \bibnamefont {Ellis}},
  \ and\ \bibinfo {author} {\bibfnamefont {P.}~\bibnamefont {Ros}},\ }\href
  {\doibase 10.1016/0301-0104(73)80059-X} {\bibfield  {journal} {\bibinfo
  {journal} {Chem. Phys.}\ }\textbf {\bibinfo {volume} {2}},\ \bibinfo {pages}
  {41} (\bibinfo {year} {1973})}\BibitemShut {NoStop}%
\bibitem [{\citenamefont {Sambe}\ and\ \citenamefont
  {Felton}(1975)}]{sambe_new_1975}%
  \BibitemOpen
  \bibfield  {author} {\bibinfo {author} {\bibfnamefont {H.}~\bibnamefont
  {Sambe}}\ and\ \bibinfo {author} {\bibfnamefont {R.~H.}\ \bibnamefont
  {Felton}},\ }\href {\doibase 10.1063/1.430555} {\bibfield  {journal}
  {\bibinfo  {journal} {J. Chem. Phys.}\ }\textbf {\bibinfo {volume} {62}},\
  \bibinfo {pages} {1122} (\bibinfo {year} {1975})}\BibitemShut {NoStop}%
\bibitem [{\citenamefont {Beebe}\ and\ \citenamefont
  {Linderberg}(1977)}]{beebe_simplifications_1977}%
  \BibitemOpen
  \bibfield  {author} {\bibinfo {author} {\bibfnamefont {N.~H.~F.}\
  \bibnamefont {Beebe}}\ and\ \bibinfo {author} {\bibfnamefont
  {J.}~\bibnamefont {Linderberg}},\ }\href {\doibase 10.1002/qua.560120408}
  {\bibfield  {journal} {\bibinfo  {journal} {Int. J. Quantum Chem.}\ }\textbf
  {\bibinfo {volume} {12}},\ \bibinfo {pages} {683} (\bibinfo {year}
  {1977})}\BibitemShut {NoStop}%
\bibitem [{\citenamefont {Dunlap}, \citenamefont {Connolly},\ and\
  \citenamefont {Sabin}(1979{\natexlab{a}})}]{dunlap_approximations_1979}%
  \BibitemOpen
  \bibfield  {author} {\bibinfo {author} {\bibfnamefont {B.~I.}\ \bibnamefont
  {Dunlap}}, \bibinfo {author} {\bibfnamefont {J.~W.~D.}\ \bibnamefont
  {Connolly}}, \ and\ \bibinfo {author} {\bibfnamefont {J.~R.}\ \bibnamefont
  {Sabin}},\ }\href {\doibase 10.1063/1.438728} {\bibfield  {journal} {\bibinfo
   {journal} {J. Chem. Phys.}\ }\textbf {\bibinfo {volume} {71}},\ \bibinfo
  {pages} {3396} (\bibinfo {year} {1979}{\natexlab{a}})}\BibitemShut {NoStop}%
\bibitem [{\citenamefont {Dunlap}, \citenamefont {Connolly},\ and\
  \citenamefont {Sabin}(1979{\natexlab{b}})}]{dunlap_firstrow_1979}%
  \BibitemOpen
  \bibfield  {author} {\bibinfo {author} {\bibfnamefont {B.~I.}\ \bibnamefont
  {Dunlap}}, \bibinfo {author} {\bibfnamefont {J.~W.~D.}\ \bibnamefont
  {Connolly}}, \ and\ \bibinfo {author} {\bibfnamefont {J.~R.}\ \bibnamefont
  {Sabin}},\ }\href {\doibase 10.1063/1.438313} {\bibfield  {journal} {\bibinfo
   {journal} {J. Chem. Phys.}\ }\textbf {\bibinfo {volume} {71}},\ \bibinfo
  {pages} {4993} (\bibinfo {year} {1979}{\natexlab{b}})}\BibitemShut {NoStop}%
\bibitem [{\citenamefont {Feyereisen}, \citenamefont {Fitzgerald},\ and\
  \citenamefont {Komornicki}(1993)}]{feyereisen_use_1993}%
  \BibitemOpen
  \bibfield  {author} {\bibinfo {author} {\bibfnamefont {M.}~\bibnamefont
  {Feyereisen}}, \bibinfo {author} {\bibfnamefont {G.}~\bibnamefont
  {Fitzgerald}}, \ and\ \bibinfo {author} {\bibfnamefont {A.}~\bibnamefont
  {Komornicki}},\ }\href {\doibase 10.1016/0009-2614(93)87156-W} {\bibfield
  {journal} {\bibinfo  {journal} {Chem. Phys. Lett.}\ }\textbf {\bibinfo
  {volume} {208}},\ \bibinfo {pages} {359} (\bibinfo {year}
  {1993})}\BibitemShut {NoStop}%
\bibitem [{\citenamefont {Vahtras}, \citenamefont {Alml{\"o}f},\ and\
  \citenamefont {Feyereisen}(1993)}]{vahtras_integral_1993}%
  \BibitemOpen
  \bibfield  {author} {\bibinfo {author} {\bibfnamefont {O.}~\bibnamefont
  {Vahtras}}, \bibinfo {author} {\bibfnamefont {J.}~\bibnamefont {Alml{\"o}f}},
  \ and\ \bibinfo {author} {\bibfnamefont {M.~W.}\ \bibnamefont {Feyereisen}},\
  }\href {\doibase 10.1016/0009-2614(93)89151-7} {\bibfield  {journal}
  {\bibinfo  {journal} {Chem. Phys. Lett.}\ }\textbf {\bibinfo {volume}
  {213}},\ \bibinfo {pages} {514} (\bibinfo {year} {1993})}\BibitemShut
  {NoStop}%
\bibitem [{\citenamefont {Eichkorn}\ \emph {et~al.}(1995)\citenamefont
  {Eichkorn}, \citenamefont {Treutler}, \citenamefont {{\"O}hm}, \citenamefont
  {H{\"a}ser},\ and\ \citenamefont {Ahlrichs}}]{eichkorn_auxiliary_1995}%
  \BibitemOpen
  \bibfield  {author} {\bibinfo {author} {\bibfnamefont {K.}~\bibnamefont
  {Eichkorn}}, \bibinfo {author} {\bibfnamefont {O.}~\bibnamefont {Treutler}},
  \bibinfo {author} {\bibfnamefont {H.}~\bibnamefont {{\"O}hm}}, \bibinfo
  {author} {\bibfnamefont {M.}~\bibnamefont {H{\"a}ser}}, \ and\ \bibinfo
  {author} {\bibfnamefont {R.}~\bibnamefont {Ahlrichs}},\ }\href {\doibase
  10.1016/0009-2614(95)00621-A} {\bibfield  {journal} {\bibinfo  {journal}
  {Chem. Phys. Lett.}\ }\textbf {\bibinfo {volume} {240}},\ \bibinfo {pages}
  {283} (\bibinfo {year} {1995})}\BibitemShut {NoStop}%
\bibitem [{\citenamefont {Eichkorn}\ \emph {et~al.}(1997)\citenamefont
  {Eichkorn}, \citenamefont {Weigend}, \citenamefont {Treutler},\ and\
  \citenamefont {Ahlrichs}}]{eichkorn_auxiliary_1997}%
  \BibitemOpen
  \bibfield  {author} {\bibinfo {author} {\bibfnamefont {K.}~\bibnamefont
  {Eichkorn}}, \bibinfo {author} {\bibfnamefont {F.}~\bibnamefont {Weigend}},
  \bibinfo {author} {\bibfnamefont {O.}~\bibnamefont {Treutler}}, \ and\
  \bibinfo {author} {\bibfnamefont {R.}~\bibnamefont {Ahlrichs}},\ }\href
  {http://link.springer.com/article/10.1007/s002140050244} {\bibfield
  {journal} {\bibinfo  {journal} {Theor. Chem. Acc.}\ }\textbf {\bibinfo
  {volume} {97}},\ \bibinfo {pages} {119} (\bibinfo {year} {1997})}\BibitemShut
  {NoStop}%
\bibitem [{\citenamefont {Bauernschmitt}\ \emph {et~al.}(1997)\citenamefont
  {Bauernschmitt}, \citenamefont {H{\"a}ser}, \citenamefont {Treutler},\ and\
  \citenamefont {Ahlrichs}}]{bauernschmitt_calculation_1997}%
  \BibitemOpen
  \bibfield  {author} {\bibinfo {author} {\bibfnamefont {R.}~\bibnamefont
  {Bauernschmitt}}, \bibinfo {author} {\bibfnamefont {M.}~\bibnamefont
  {H{\"a}ser}}, \bibinfo {author} {\bibfnamefont {O.}~\bibnamefont {Treutler}},
  \ and\ \bibinfo {author} {\bibfnamefont {R.}~\bibnamefont {Ahlrichs}},\
  }\href {\doibase 10.1016/S0009-2614(96)01343-7} {\bibfield  {journal}
  {\bibinfo  {journal} {Chem. Phys. Lett.}\ }\textbf {\bibinfo {volume}
  {264}},\ \bibinfo {pages} {573} (\bibinfo {year} {1997})}\BibitemShut
  {NoStop}%
\bibitem [{\citenamefont {Weigend}\ \emph {et~al.}(1998)\citenamefont
  {Weigend}, \citenamefont {H{\"a}ser}, \citenamefont {Patzelt},\ and\
  \citenamefont {Ahlrichs}}]{weigend_ri-mp2:_1998}%
  \BibitemOpen
  \bibfield  {author} {\bibinfo {author} {\bibfnamefont {F.}~\bibnamefont
  {Weigend}}, \bibinfo {author} {\bibfnamefont {M.}~\bibnamefont {H{\"a}ser}},
  \bibinfo {author} {\bibfnamefont {H.}~\bibnamefont {Patzelt}}, \ and\
  \bibinfo {author} {\bibfnamefont {R.}~\bibnamefont {Ahlrichs}},\ }\href
  {\doibase 10.1016/S0009-2614(98)00862-8} {\bibfield  {journal} {\bibinfo
  {journal} {Chem. Phys. Lett.}\ }\textbf {\bibinfo {volume} {294}},\ \bibinfo
  {pages} {143} (\bibinfo {year} {1998})}\BibitemShut {NoStop}%
\bibitem [{\citenamefont {Weigend}, \citenamefont {K{\"o}hn},\ and\
  \citenamefont {H{\"a}ttig}(2002)}]{weigend_efficient_2002}%
  \BibitemOpen
  \bibfield  {author} {\bibinfo {author} {\bibfnamefont {F.}~\bibnamefont
  {Weigend}}, \bibinfo {author} {\bibfnamefont {A.}~\bibnamefont {K{\"o}hn}}, \
  and\ \bibinfo {author} {\bibfnamefont {C.}~\bibnamefont {H{\"a}ttig}},\
  }\href
  {http://scitation.aip.org/content/aip/journal/jcp/116/8/10.1063/1.1445115}
  {\bibfield  {journal} {\bibinfo  {journal} {J. Chem. Phys.}\ }\textbf
  {\bibinfo {volume} {116}},\ \bibinfo {pages} {3175} (\bibinfo {year}
  {2002})}\BibitemShut {NoStop}%
\bibitem [{\citenamefont {Weigend}, \citenamefont {Kattannek},\ and\
  \citenamefont {Ahlrichs}(2009)}]{weigend_approximated_2009}%
  \BibitemOpen
  \bibfield  {author} {\bibinfo {author} {\bibfnamefont {F.}~\bibnamefont
  {Weigend}}, \bibinfo {author} {\bibfnamefont {M.}~\bibnamefont {Kattannek}},
  \ and\ \bibinfo {author} {\bibfnamefont {R.}~\bibnamefont {Ahlrichs}},\
  }\href {\doibase 10.1063/1.3116103} {\bibfield  {journal} {\bibinfo
  {journal} {J. Chem. Phys.}\ }\textbf {\bibinfo {volume} {130}},\ \bibinfo
  {pages} {164106} (\bibinfo {year} {2009})}\BibitemShut {NoStop}%
\bibitem [{\citenamefont {Bostr{\"o}m}\ \emph {et~al.}(2009)\citenamefont
  {Bostr{\"o}m}, \citenamefont {Aquilante}, \citenamefont {Pedersen},\ and\
  \citenamefont {Lindh}}]{bostrom_ab_2009}%
  \BibitemOpen
  \bibfield  {author} {\bibinfo {author} {\bibfnamefont {J.}~\bibnamefont
  {Bostr{\"o}m}}, \bibinfo {author} {\bibfnamefont {F.}~\bibnamefont
  {Aquilante}}, \bibinfo {author} {\bibfnamefont {T.~B.}\ \bibnamefont
  {Pedersen}}, \ and\ \bibinfo {author} {\bibfnamefont {R.}~\bibnamefont
  {Lindh}},\ }\href {\doibase 10.1021/ct9000284} {\bibfield  {journal}
  {\bibinfo  {journal} {J. Chem. Theory Comput.}\ }\textbf {\bibinfo {volume}
  {5}},\ \bibinfo {pages} {1545} (\bibinfo {year} {2009})}\BibitemShut
  {NoStop}%
\bibitem [{\citenamefont {Dunlap}(2000{\natexlab{a}})}]{dunlap_robust_2000}%
  \BibitemOpen
  \bibfield  {author} {\bibinfo {author} {\bibfnamefont {B.~I.}\ \bibnamefont
  {Dunlap}},\ }\href {\doibase 10.1016/S0166-1280(00)00528-5} {\bibfield
  {journal} {\bibinfo  {journal} {J. Mol. Struct. (THEOCHEM)}\ }\textbf
  {\bibinfo {volume} {529}},\ \bibinfo {pages} {37} (\bibinfo {year}
  {2000}{\natexlab{a}})}\BibitemShut {NoStop}%
\bibitem [{\citenamefont {Dunlap}(2000{\natexlab{b}})}]{Dunlap:2000iw}%
  \BibitemOpen
  \bibfield  {author} {\bibinfo {author} {\bibfnamefont {B.~I.}\ \bibnamefont
  {Dunlap}},\ }\href@noop {} {\bibfield  {journal} {\bibinfo  {journal} {Phys.
  Chem. Chem. Phys.}\ }\textbf {\bibinfo {volume} {2}},\ \bibinfo {pages}
  {2113} (\bibinfo {year} {2000}{\natexlab{b}})}\BibitemShut {NoStop}%
\bibitem [{\citenamefont {Sch{\"u}tz}\ and\ \citenamefont
  {Manby}(2003)}]{schutz_linear_2003}%
  \BibitemOpen
  \bibfield  {author} {\bibinfo {author} {\bibfnamefont {M.}~\bibnamefont
  {Sch{\"u}tz}}\ and\ \bibinfo {author} {\bibfnamefont {F.~R.}\ \bibnamefont
  {Manby}},\ }\href {\doibase 10.1039/b304550a} {\bibfield  {journal} {\bibinfo
   {journal} {Phys. Chem. Chem. Phys.}\ }\textbf {\bibinfo {volume} {5}},\
  \bibinfo {pages} {3349} (\bibinfo {year} {2003})}\BibitemShut {NoStop}%
\bibitem [{\citenamefont {Pisani}\ \emph {et~al.}(2008)\citenamefont {Pisani},
  \citenamefont {Maschio}, \citenamefont {Casassa}, \citenamefont {Halo},
  \citenamefont {Sch{\"u}tz},\ and\ \citenamefont
  {Usvyat}}]{pisani_periodic_2008}%
  \BibitemOpen
  \bibfield  {author} {\bibinfo {author} {\bibfnamefont {C.}~\bibnamefont
  {Pisani}}, \bibinfo {author} {\bibfnamefont {L.}~\bibnamefont {Maschio}},
  \bibinfo {author} {\bibfnamefont {S.}~\bibnamefont {Casassa}}, \bibinfo
  {author} {\bibfnamefont {M.}~\bibnamefont {Halo}}, \bibinfo {author}
  {\bibfnamefont {M.}~\bibnamefont {Sch{\"u}tz}}, \ and\ \bibinfo {author}
  {\bibfnamefont {D.}~\bibnamefont {Usvyat}},\ }\href {\doibase
  10.1002/jcc.20975} {\bibfield  {journal} {\bibinfo  {journal} {J. Comput.
  Chem.}\ }\textbf {\bibinfo {volume} {29}},\ \bibinfo {pages} {2113} (\bibinfo
  {year} {2008})}\BibitemShut {NoStop}%
\bibitem [{\citenamefont {Maschio}\ and\ \citenamefont
  {Usvyat}(2008)}]{maschio_fitting_2008}%
  \BibitemOpen
  \bibfield  {author} {\bibinfo {author} {\bibfnamefont {L.}~\bibnamefont
  {Maschio}}\ and\ \bibinfo {author} {\bibfnamefont {D.}~\bibnamefont
  {Usvyat}},\ }\href {\doibase 10.1103/PhysRevB.78.073102} {\bibfield
  {journal} {\bibinfo  {journal} {Phys. Rev. B}\ }\textbf {\bibinfo {volume}
  {78}},\ \bibinfo {pages} {073102} (\bibinfo {year} {2008})}\BibitemShut
  {NoStop}%
\bibitem [{\citenamefont {Maschio}(2011)}]{maschio_local_2011}%
  \BibitemOpen
  \bibfield  {author} {\bibinfo {author} {\bibfnamefont {L.}~\bibnamefont
  {Maschio}},\ }\href {\doibase 10.1021/ct200352g} {\bibfield  {journal}
  {\bibinfo  {journal} {J. Chem. Theory Comput.}\ }\textbf {\bibinfo {volume}
  {7}},\ \bibinfo {pages} {2818} (\bibinfo {year} {2011})}\BibitemShut
  {NoStop}%
\bibitem [{\citenamefont {Gill}\ \emph {et~al.}(2005)\citenamefont {Gill},
  \citenamefont {Gilbert}, \citenamefont {Taylor}, \citenamefont {Friesecke},\
  and\ \citenamefont {Head-Gordon}}]{Gill:2005jw}%
  \BibitemOpen
  \bibfield  {author} {\bibinfo {author} {\bibfnamefont {P.~M.~W.}\
  \bibnamefont {Gill}}, \bibinfo {author} {\bibfnamefont {A.~T.~B.}\
  \bibnamefont {Gilbert}}, \bibinfo {author} {\bibfnamefont {S.~W.}\
  \bibnamefont {Taylor}}, \bibinfo {author} {\bibfnamefont {G.}~\bibnamefont
  {Friesecke}}, \ and\ \bibinfo {author} {\bibfnamefont {M.}~\bibnamefont
  {Head-Gordon}},\ }\href@noop {} {\bibfield  {journal} {\bibinfo  {journal}
  {J. Chem. Phys.}\ }\textbf {\bibinfo {volume} {123}},\ \bibinfo {pages}
  {061101} (\bibinfo {year} {2005})}\BibitemShut {NoStop}%
\bibitem [{\citenamefont {Tew}(2018)}]{Tew:2018cf}%
  \BibitemOpen
  \bibfield  {author} {\bibinfo {author} {\bibfnamefont {D.~P.}\ \bibnamefont
  {Tew}},\ }\href@noop {} {\bibfield  {journal} {\bibinfo  {journal} {J. Chem.
  Phys.}\ }\textbf {\bibinfo {volume} {148}},\ \bibinfo {pages} {011102}
  (\bibinfo {year} {2018})}\BibitemShut {NoStop}%
\bibitem [{\citenamefont {Jaffe}\ and\ \citenamefont
  {Hess}(1996)}]{jaffe_gaussian_1996}%
  \BibitemOpen
  \bibfield  {author} {\bibinfo {author} {\bibfnamefont {J.~E.}\ \bibnamefont
  {Jaffe}}\ and\ \bibinfo {author} {\bibfnamefont {A.~C.}\ \bibnamefont
  {Hess}},\ }\href {\doibase 10.1063/1.472866} {\bibfield  {journal} {\bibinfo
  {journal} {J. Chem. Phys.}\ }\textbf {\bibinfo {volume} {105}},\ \bibinfo
  {pages} {10983} (\bibinfo {year} {1996})}\BibitemShut {NoStop}%
\bibitem [{\citenamefont {Varga}(2005)}]{varga_density_2005}%
  \BibitemOpen
  \bibfield  {author} {\bibinfo {author} {\bibfnamefont {{\u S}.}~\bibnamefont
  {Varga}},\ }\href {\doibase 10.1103/PhysRevB.71.073103} {\bibfield  {journal}
  {\bibinfo  {journal} {Phys. Rev. B}\ }\textbf {\bibinfo {volume} {71}},\
  \bibinfo {pages} {073103} (\bibinfo {year} {2005})}\BibitemShut {NoStop}%
\bibitem [{\citenamefont {{\L}azarski}, \citenamefont {Burow},\ and\
  \citenamefont {Sierka}(2015)}]{lazarski_density_2015}%
  \BibitemOpen
  \bibfield  {author} {\bibinfo {author} {\bibfnamefont {R.}~\bibnamefont
  {{\L}azarski}}, \bibinfo {author} {\bibfnamefont {A.~M.}\ \bibnamefont
  {Burow}}, \ and\ \bibinfo {author} {\bibfnamefont {M.}~\bibnamefont
  {Sierka}},\ }\href {\doibase 10.1021/acs.jctc.5b00252} {\bibfield  {journal}
  {\bibinfo  {journal} {J. Chem. Theory Comput.}\ }\textbf {\bibinfo {volume}
  {11}},\ \bibinfo {pages} {3029} (\bibinfo {year} {2015})}\BibitemShut
  {NoStop}%
\bibitem [{\citenamefont {Franchini}\ \emph {et~al.}(2014)\citenamefont
  {Franchini}, \citenamefont {Philipsen}, \citenamefont {van Lenthe},\ and\
  \citenamefont {Visscher}}]{franchini_accurate_2014}%
  \BibitemOpen
  \bibfield  {author} {\bibinfo {author} {\bibfnamefont {M.}~\bibnamefont
  {Franchini}}, \bibinfo {author} {\bibfnamefont {P.~H.~T.}\ \bibnamefont
  {Philipsen}}, \bibinfo {author} {\bibfnamefont {E.}~\bibnamefont {van
  Lenthe}}, \ and\ \bibinfo {author} {\bibfnamefont {L.}~\bibnamefont
  {Visscher}},\ }\href {\doibase 10.1021/ct500172n} {\bibfield  {journal}
  {\bibinfo  {journal} {J. Chem. Theory Comput.}\ }\textbf {\bibinfo {volume}
  {10}},\ \bibinfo {pages} {1994} (\bibinfo {year} {2014})}\BibitemShut
  {NoStop}%
\bibitem [{\citenamefont {Pisani}, \citenamefont {Dovesi},\ and\ \citenamefont
  {Roetti}(1988)}]{pisani_hartree-fock_1988}%
  \BibitemOpen
  \bibfield  {author} {\bibinfo {author} {\bibfnamefont {C.}~\bibnamefont
  {Pisani}}, \bibinfo {author} {\bibfnamefont {R.}~\bibnamefont {Dovesi}}, \
  and\ \bibinfo {author} {\bibfnamefont {C.}~\bibnamefont {Roetti}},\ }\href
  {http://link.springer.com/10.1007/978-3-642-93385-1} {\emph {\bibinfo {title}
  {Hartree-{Fock} {Ab} {Initio} {Treatment} of {Crystalline} {Systems}}}},\
  \bibinfo {series} {Lecture {Notes} in {Chemistry}}, Vol.~\bibinfo {volume}
  {48}\ (\bibinfo  {publisher} {Springer-Verlag, Berlin},\ \bibinfo {year}
  {1988})\BibitemShut {NoStop}%
\bibitem [{\citenamefont {Dovesi}\ \emph {et~al.}(2000)\citenamefont {Dovesi},
  \citenamefont {Orlando}, \citenamefont {Roetti}, \citenamefont {Pisani},\
  and\ \citenamefont {Saunders}}]{dovesi_periodic_2000}%
  \BibitemOpen
  \bibfield  {author} {\bibinfo {author} {\bibfnamefont {R.}~\bibnamefont
  {Dovesi}}, \bibinfo {author} {\bibfnamefont {R.}~\bibnamefont {Orlando}},
  \bibinfo {author} {\bibfnamefont {C.}~\bibnamefont {Roetti}}, \bibinfo
  {author} {\bibfnamefont {C.}~\bibnamefont {Pisani}}, \ and\ \bibinfo {author}
  {\bibfnamefont {V.}~\bibnamefont {Saunders}},\ }\href {\doibase
  10.1002/(SICI)1521-3951(200001)217:1<63::AID-PSSB63>3.0.CO;2-F} {\bibfield
  {journal} {\bibinfo  {journal} {Phys. Stat. Sol. (b)}\ }\textbf {\bibinfo
  {volume} {217}},\ \bibinfo {pages} {63} (\bibinfo {year} {2000})}\BibitemShut
  {NoStop}%
\bibitem [{\citenamefont {Kohn}(1959)}]{kohn_analytic_1959}%
  \BibitemOpen
  \bibfield  {author} {\bibinfo {author} {\bibfnamefont {W.}~\bibnamefont
  {Kohn}},\ }\href {\doibase 10.1103/PhysRev.115.809} {\bibfield  {journal}
  {\bibinfo  {journal} {Phys. Rev.}\ }\textbf {\bibinfo {volume} {115}},\
  \bibinfo {pages} {809} (\bibinfo {year} {1959})}\BibitemShut {NoStop}%
\bibitem [{\citenamefont {Cloizeaux}(1964)}]{cloizeaux_analytical_1964}%
  \BibitemOpen
  \bibfield  {author} {\bibinfo {author} {\bibfnamefont {J.~D.}\ \bibnamefont
  {Cloizeaux}},\ }\href {\doibase 10.1103/PhysRev.135.A698} {\bibfield
  {journal} {\bibinfo  {journal} {Phys. Rev.}\ }\textbf {\bibinfo {volume}
  {135}},\ \bibinfo {pages} {A698} (\bibinfo {year} {1964})}\BibitemShut
  {NoStop}%
\bibitem [{\citenamefont {He}\ and\ \citenamefont
  {Vanderbilt}(2001)}]{he_exponential_2001}%
  \BibitemOpen
  \bibfield  {author} {\bibinfo {author} {\bibfnamefont {L.}~\bibnamefont
  {He}}\ and\ \bibinfo {author} {\bibfnamefont {D.}~\bibnamefont
  {Vanderbilt}},\ }\href {\doibase 10.1103/PhysRevLett.86.5341} {\bibfield
  {journal} {\bibinfo  {journal} {Phys. Rev. Lett.}\ }\textbf {\bibinfo
  {volume} {86}},\ \bibinfo {pages} {5341} (\bibinfo {year}
  {2001})}\BibitemShut {NoStop}%
\bibitem [{\citenamefont {Taraskin}, \citenamefont {Drabold},\ and\
  \citenamefont {Elliott}(2002)}]{taraskin_spatial_2002}%
  \BibitemOpen
  \bibfield  {author} {\bibinfo {author} {\bibfnamefont {S.~N.}\ \bibnamefont
  {Taraskin}}, \bibinfo {author} {\bibfnamefont {D.~A.}\ \bibnamefont
  {Drabold}}, \ and\ \bibinfo {author} {\bibfnamefont {S.~R.}\ \bibnamefont
  {Elliott}},\ }\href {\doibase 10.1103/PhysRevLett.88.196405} {\bibfield
  {journal} {\bibinfo  {journal} {Phys. Rev. Lett.}\ }\textbf {\bibinfo
  {volume} {88}},\ \bibinfo {pages} {196405} (\bibinfo {year}
  {2002})}\BibitemShut {NoStop}%
\bibitem [{\citenamefont {Goedecker}(1998)}]{goedecker_decay_1998}%
  \BibitemOpen
  \bibfield  {author} {\bibinfo {author} {\bibfnamefont {S.}~\bibnamefont
  {Goedecker}},\ }\href {\doibase 10.1103/PhysRevB.58.3501} {\bibfield
  {journal} {\bibinfo  {journal} {Phys. Rev. B}\ }\textbf {\bibinfo {volume}
  {58}},\ \bibinfo {pages} {3501} (\bibinfo {year} {1998})}\BibitemShut
  {NoStop}%
\bibitem [{\citenamefont {Ismail-Beigi}\ and\ \citenamefont
  {Arias}(1999)}]{ismail-beigi_locality_1999}%
  \BibitemOpen
  \bibfield  {author} {\bibinfo {author} {\bibfnamefont {S.}~\bibnamefont
  {Ismail-Beigi}}\ and\ \bibinfo {author} {\bibfnamefont {T.~A.}\ \bibnamefont
  {Arias}},\ }\href {\doibase 10.1103/PhysRevLett.82.2127} {\bibfield
  {journal} {\bibinfo  {journal} {Phys. Rev. Lett.}\ }\textbf {\bibinfo
  {volume} {82}},\ \bibinfo {pages} {2127} (\bibinfo {year}
  {1999})}\BibitemShut {NoStop}%
\bibitem [{\citenamefont {Taraskin}\ \emph {et~al.}(2002)\citenamefont
  {Taraskin}, \citenamefont {Fry}, \citenamefont {Zhang}, \citenamefont
  {Drabold},\ and\ \citenamefont {Elliott}}]{taraskin_spatial_2002-1}%
  \BibitemOpen
  \bibfield  {author} {\bibinfo {author} {\bibfnamefont {S.~N.}\ \bibnamefont
  {Taraskin}}, \bibinfo {author} {\bibfnamefont {P.~A.}\ \bibnamefont {Fry}},
  \bibinfo {author} {\bibfnamefont {X.}~\bibnamefont {Zhang}}, \bibinfo
  {author} {\bibfnamefont {D.~A.}\ \bibnamefont {Drabold}}, \ and\ \bibinfo
  {author} {\bibfnamefont {S.~R.}\ \bibnamefont {Elliott}},\ }\href {\doibase
  10.1103/PhysRevB.66.233101} {\bibfield  {journal} {\bibinfo  {journal} {Phys.
  Rev. B}\ }\textbf {\bibinfo {volume} {66}},\ \bibinfo {pages} {233101}
  (\bibinfo {year} {2002})}\BibitemShut {NoStop}%
\bibitem [{\citenamefont {H\"{a}ser}\ and\ \citenamefont
  {Ahlrichs}(1989)}]{haser_improvements_1989}%
  \BibitemOpen
  \bibfield  {author} {\bibinfo {author} {\bibfnamefont {M.}~\bibnamefont
  {H\"{a}ser}}\ and\ \bibinfo {author} {\bibfnamefont {R.}~\bibnamefont
  {Ahlrichs}},\ }\href {\doibase 10.1002/jcc.540100111} {\bibfield  {journal}
  {\bibinfo  {journal} {J. Comput. Chem.}\ }\textbf {\bibinfo {volume} {10}},\
  \bibinfo {pages} {104} (\bibinfo {year} {1989})}\BibitemShut {NoStop}%
\bibitem [{\citenamefont {Hollman}, \citenamefont {Schaefer},\ and\
  \citenamefont {Valeev}(2015)}]{Hollman:2015ca}%
  \BibitemOpen
  \bibfield  {author} {\bibinfo {author} {\bibfnamefont {D.~S.}\ \bibnamefont
  {Hollman}}, \bibinfo {author} {\bibfnamefont {H.~F.}\ \bibnamefont
  {Schaefer}}, \ and\ \bibinfo {author} {\bibfnamefont {E.~F.}\ \bibnamefont
  {Valeev}},\ }\href@noop {} {\bibfield  {journal} {\bibinfo  {journal} {J Chem
  Phys}\ }\textbf {\bibinfo {volume} {142}} (\bibinfo {year}
  {2015})}\BibitemShut {NoStop}%
\bibitem [{\citenamefont {Calvin}, \citenamefont {Lewis},\ and\ \citenamefont
  {Valeev}(2015)}]{Calvin:2015ka}%
  \BibitemOpen
  \bibfield  {author} {\bibinfo {author} {\bibfnamefont {J.~A.}\ \bibnamefont
  {Calvin}}, \bibinfo {author} {\bibfnamefont {C.~A.}\ \bibnamefont {Lewis}}, \
  and\ \bibinfo {author} {\bibfnamefont {E.~F.}\ \bibnamefont {Valeev}},\ }in\
  \href@noop {} {\emph {\bibinfo {booktitle} {IA3 '15, The 5th Workshop on
  Irregular Applications: Architectures and Algorithms}}}\ (\bibinfo
  {publisher} {ACM Press},\ \bibinfo {address} {New York, New York, USA},\
  \bibinfo {year} {2015})\ pp.\ \bibinfo {pages} {1--8}\BibitemShut {NoStop}%
\bibitem [{\citenamefont {Peng}\ \emph {et~al.}()\citenamefont {Peng},
  \citenamefont {Lewis}, \citenamefont {Wang}, \citenamefont {Clement},
  \citenamefont {Pierce}, \citenamefont {Rishi}, \citenamefont
  {Pavo\v{s}evi\'c}, \citenamefont {Slattery}, \citenamefont {Zhang},
  \citenamefont {Teke}, \citenamefont {Kumar}, \citenamefont {Masteran},
  \citenamefont {Asadchev}, \citenamefont {Calvin},\ and\ \citenamefont
  {Valeev}}]{mpqc4paper}%
  \BibitemOpen
  \bibfield  {author} {\bibinfo {author} {\bibfnamefont {C.}~\bibnamefont
  {Peng}}, \bibinfo {author} {\bibfnamefont {C.~A.}\ \bibnamefont {Lewis}},
  \bibinfo {author} {\bibfnamefont {X.}~\bibnamefont {Wang}}, \bibinfo {author}
  {\bibfnamefont {M.~C.}\ \bibnamefont {Clement}}, \bibinfo {author}
  {\bibfnamefont {K.}~\bibnamefont {Pierce}}, \bibinfo {author} {\bibfnamefont
  {V.}~\bibnamefont {Rishi}}, \bibinfo {author} {\bibfnamefont
  {F.}~\bibnamefont {Pavo\v{s}evi\'c}}, \bibinfo {author} {\bibfnamefont
  {S.}~\bibnamefont {Slattery}}, \bibinfo {author} {\bibfnamefont
  {J.}~\bibnamefont {Zhang}}, \bibinfo {author} {\bibfnamefont
  {N.}~\bibnamefont {Teke}}, \bibinfo {author} {\bibfnamefont {A.}~\bibnamefont
  {Kumar}}, \bibinfo {author} {\bibfnamefont {C.}~\bibnamefont {Masteran}},
  \bibinfo {author} {\bibfnamefont {A.}~\bibnamefont {Asadchev}}, \bibinfo
  {author} {\bibfnamefont {J.~A.}\ \bibnamefont {Calvin}}, \ and\ \bibinfo
  {author} {\bibfnamefont {E.~F.}\ \bibnamefont {Valeev}},\ }\href@noop {}
  {\bibfield  {journal} {\bibinfo  {journal} {J. Chem. Phys.}\ }}\bibinfo
  {note} {Submitted}\BibitemShut {NoStop}%
\bibitem [{cal()}]{calvin_tiledarray:_2016}%
  \BibitemOpen
  \href@noop {} {}\bibinfo {howpublished} {Calvin, J. A.; Peng, C.; Lewis, C.
  A.; Zhang, J.; Valeev, E. F. \textit{``TiledArray: A Framework for
  Distributed-Memory Block-Sparse Tensor Computation''},
  https://github.com/valeevgroup/tiledarray.}\BibitemShut {Stop}%
\bibitem [{\citenamefont {Harrison}\ \emph {et~al.}(2016)\citenamefont
  {Harrison}, \citenamefont {Beylkin}, \citenamefont {Bischoff}, \citenamefont
  {Calvin}, \citenamefont {Fann}, \citenamefont {Fosso-Tande}, \citenamefont
  {Galindo}, \citenamefont {Hammond}, \citenamefont {Hartman-Baker},
  \citenamefont {Hill}, \citenamefont {Jia}, \citenamefont {Kottmann},
  \citenamefont {Yvonne~Ou}, \citenamefont {Pei}, \citenamefont {Ratcliff},
  \citenamefont {Reuter}, \citenamefont {Richie-Halford}, \citenamefont
  {Romero}, \citenamefont {Sekino}, \citenamefont {Shelton}, \citenamefont
  {Sundahl}, \citenamefont {Thornton}, \citenamefont {Valeev}, \citenamefont
  {V{\'a}zquez-Mayagoitia}, \citenamefont {Vence}, \citenamefont {Yanai},\ and\
  \citenamefont {Yokoi}}]{harrison_madness:_2016}%
  \BibitemOpen
  \bibfield  {author} {\bibinfo {author} {\bibfnamefont {R.}~\bibnamefont
  {Harrison}}, \bibinfo {author} {\bibfnamefont {G.}~\bibnamefont {Beylkin}},
  \bibinfo {author} {\bibfnamefont {F.}~\bibnamefont {Bischoff}}, \bibinfo
  {author} {\bibfnamefont {J.}~\bibnamefont {Calvin}}, \bibinfo {author}
  {\bibfnamefont {G.}~\bibnamefont {Fann}}, \bibinfo {author} {\bibfnamefont
  {J.}~\bibnamefont {Fosso-Tande}}, \bibinfo {author} {\bibfnamefont
  {D.}~\bibnamefont {Galindo}}, \bibinfo {author} {\bibfnamefont
  {J.}~\bibnamefont {Hammond}}, \bibinfo {author} {\bibfnamefont
  {R.}~\bibnamefont {Hartman-Baker}}, \bibinfo {author} {\bibfnamefont
  {J.}~\bibnamefont {Hill}}, \bibinfo {author} {\bibfnamefont {J.}~\bibnamefont
  {Jia}}, \bibinfo {author} {\bibfnamefont {J.}~\bibnamefont {Kottmann}},
  \bibinfo {author} {\bibfnamefont {M.}~\bibnamefont {Yvonne~Ou}}, \bibinfo
  {author} {\bibfnamefont {J.}~\bibnamefont {Pei}}, \bibinfo {author}
  {\bibfnamefont {L.}~\bibnamefont {Ratcliff}}, \bibinfo {author}
  {\bibfnamefont {M.}~\bibnamefont {Reuter}}, \bibinfo {author} {\bibfnamefont
  {A.}~\bibnamefont {Richie-Halford}}, \bibinfo {author} {\bibfnamefont
  {N.}~\bibnamefont {Romero}}, \bibinfo {author} {\bibfnamefont
  {H.}~\bibnamefont {Sekino}}, \bibinfo {author} {\bibfnamefont
  {W.}~\bibnamefont {Shelton}}, \bibinfo {author} {\bibfnamefont
  {B.}~\bibnamefont {Sundahl}}, \bibinfo {author} {\bibfnamefont
  {W.}~\bibnamefont {Thornton}}, \bibinfo {author} {\bibfnamefont
  {E.}~\bibnamefont {Valeev}}, \bibinfo {author} {\bibfnamefont
  {{\'A}.}~\bibnamefont {V{\'a}zquez-Mayagoitia}}, \bibinfo {author}
  {\bibfnamefont {N.}~\bibnamefont {Vence}}, \bibinfo {author} {\bibfnamefont
  {T.}~\bibnamefont {Yanai}}, \ and\ \bibinfo {author} {\bibfnamefont
  {Y.}~\bibnamefont {Yokoi}},\ }\href {\doibase 10.1137/15M1026171} {\bibfield
  {journal} {\bibinfo  {journal} {SIAM J. Sci. Comput.}\ }\textbf {\bibinfo
  {volume} {38}},\ \bibinfo {pages} {S123} (\bibinfo {year}
  {2016})}\BibitemShut {NoStop}%
\bibitem [{\citenamefont {Avitabile}\ \emph {et~al.}(1975)\citenamefont
  {Avitabile}, \citenamefont {Napolitano}, \citenamefont {Pirozzi},
  \citenamefont {Rouse}, \citenamefont {Thomas},\ and\ \citenamefont
  {Willis}}]{avitabile_low_1975}%
  \BibitemOpen
  \bibfield  {author} {\bibinfo {author} {\bibfnamefont {G.}~\bibnamefont
  {Avitabile}}, \bibinfo {author} {\bibfnamefont {R.}~\bibnamefont
  {Napolitano}}, \bibinfo {author} {\bibfnamefont {B.}~\bibnamefont {Pirozzi}},
  \bibinfo {author} {\bibfnamefont {K.~D.}\ \bibnamefont {Rouse}}, \bibinfo
  {author} {\bibfnamefont {M.~W.}\ \bibnamefont {Thomas}}, \ and\ \bibinfo
  {author} {\bibfnamefont {B.~T.~M.}\ \bibnamefont {Willis}},\ }\href {\doibase
  10.1002/pol.1975.130130607} {\bibfield  {journal} {\bibinfo  {journal} {J.
  Polym. Sci.: Polym. Lett. Ed.}\ }\textbf {\bibinfo {volume} {13}},\ \bibinfo
  {pages} {351} (\bibinfo {year} {1975})}\BibitemShut {NoStop}%
\bibitem [{\citenamefont {F{\"o}rner}\ \emph {et~al.}(1997)\citenamefont
  {F{\"o}rner}, \citenamefont {Knab}, \citenamefont {{\v C}{\'i}{\v z}ek},\
  and\ \citenamefont {Ladik}}]{forner_numerical_1997}%
  \BibitemOpen
  \bibfield  {author} {\bibinfo {author} {\bibfnamefont {W.}~\bibnamefont
  {F{\"o}rner}}, \bibinfo {author} {\bibfnamefont {R.}~\bibnamefont {Knab}},
  \bibinfo {author} {\bibfnamefont {J.}~\bibnamefont {{\v C}{\'i}{\v z}ek}}, \
  and\ \bibinfo {author} {\bibfnamefont {J.}~\bibnamefont {Ladik}},\ }\href
  {\doibase 10.1063/1.474051} {\bibfield  {journal} {\bibinfo  {journal} {J.
  Chem. Phys.}\ }\textbf {\bibinfo {volume} {106}},\ \bibinfo {pages} {10248}
  (\bibinfo {year} {1997})}\BibitemShut {NoStop}%
\bibitem [{\citenamefont {Dresselhaus}, \citenamefont {Dresselhaus},\ and\
  \citenamefont {Saito}(1995)}]{dresselhaus_physics_1995}%
  \BibitemOpen
  \bibfield  {author} {\bibinfo {author} {\bibfnamefont {M.~S.}\ \bibnamefont
  {Dresselhaus}}, \bibinfo {author} {\bibfnamefont {G.}~\bibnamefont
  {Dresselhaus}}, \ and\ \bibinfo {author} {\bibfnamefont {R.}~\bibnamefont
  {Saito}},\ }\href {\doibase 10.1016/0008-6223(95)00017-8} {\bibfield
  {journal} {\bibinfo  {journal} {Carbon}\ }\bibinfo {series} {Nanotubes},\
  \textbf {\bibinfo {volume} {33}},\ \bibinfo {pages} {883} (\bibinfo {year}
  {1995})}\BibitemShut {NoStop}%
\bibitem [{\citenamefont {Kim}\ \emph {et~al.}(2017)\citenamefont {Kim},
  \citenamefont {Kim}, \citenamefont {Song}, \citenamefont {Lee},\ and\
  \citenamefont {Lee}}]{kim_geometric_2017}%
  \BibitemOpen
  \bibfield  {author} {\bibinfo {author} {\bibfnamefont {D.-H.}\ \bibnamefont
  {Kim}}, \bibinfo {author} {\bibfnamefont {H.-S.}\ \bibnamefont {Kim}},
  \bibinfo {author} {\bibfnamefont {M.~W.}\ \bibnamefont {Song}}, \bibinfo
  {author} {\bibfnamefont {S.}~\bibnamefont {Lee}}, \ and\ \bibinfo {author}
  {\bibfnamefont {S.~Y.}\ \bibnamefont {Lee}},\ }\href {\doibase
  10.1186/s40580-017-0107-0} {\bibfield  {journal} {\bibinfo  {journal} {Nano
  Convergence}\ }\textbf {\bibinfo {volume} {4}},\ \bibinfo {pages} {13}
  (\bibinfo {year} {2017})}\BibitemShut {NoStop}%
\bibitem [{\citenamefont {Swaminathan}, \citenamefont {Craven},\ and\
  \citenamefont {McMullan}(1984)}]{swaminathan_crystal_1984}%
  \BibitemOpen
  \bibfield  {author} {\bibinfo {author} {\bibfnamefont {S.}~\bibnamefont
  {Swaminathan}}, \bibinfo {author} {\bibfnamefont {B.~M.}\ \bibnamefont
  {Craven}}, \ and\ \bibinfo {author} {\bibfnamefont {R.~K.}\ \bibnamefont
  {McMullan}},\ }\href {\doibase 10.1107/S0108768184002135} {\bibfield
  {journal} {\bibinfo  {journal} {Acta Crystallogr. Sec. B: Struct. Sci.}\
  }\textbf {\bibinfo {volume} {40}},\ \bibinfo {pages} {300} (\bibinfo {year}
  {1984})}\BibitemShut {NoStop}%
\bibitem [{\citenamefont {Persson}(2014{\natexlab{a}})}]{osti_1281384}%
  \BibitemOpen
  \bibfield  {author} {\bibinfo {author} {\bibfnamefont {K.}~\bibnamefont
  {Persson}},\ }\href {\doibase 10.17188/1281384} {\enquote {\bibinfo {title}
  {Materials data on c (sg:227) by materials project},}\ } (\bibinfo {year}
  {2014}{\natexlab{a}})\BibitemShut {NoStop}%
\bibitem [{\citenamefont {Persson}(2014{\natexlab{b}})}]{osti_1199672}%
  \BibitemOpen
  \bibfield  {author} {\bibinfo {author} {\bibfnamefont {K.}~\bibnamefont
  {Persson}},\ }\href {\doibase 10.17188/1199672} {\enquote {\bibinfo {title}
  {Materials data on lih (sg:225) by materials project},}\ } (\bibinfo {year}
  {2014}{\natexlab{b}})\BibitemShut {NoStop}%
\bibitem [{\citenamefont {Weigend}\ and\ \citenamefont
  {Ahlrichs}(2005)}]{weigend_balanced_2005}%
  \BibitemOpen
  \bibfield  {author} {\bibinfo {author} {\bibfnamefont {F.}~\bibnamefont
  {Weigend}}\ and\ \bibinfo {author} {\bibfnamefont {R.}~\bibnamefont
  {Ahlrichs}},\ }\href {\doibase 10.1039/b508541a} {\bibfield  {journal}
  {\bibinfo  {journal} {Phys. Chem. Chem. Phys.}\ }\textbf {\bibinfo {volume}
  {7}},\ \bibinfo {pages} {3297} (\bibinfo {year} {2005})}\BibitemShut
  {NoStop}%
\bibitem [{\citenamefont {Weigend}(2006)}]{weigend_accurate_2006}%
  \BibitemOpen
  \bibfield  {author} {\bibinfo {author} {\bibfnamefont {F.}~\bibnamefont
  {Weigend}},\ }\href {\doibase 10.1039/b515623h} {\bibfield  {journal}
  {\bibinfo  {journal} {Phys. Chem. Chem. Phys.}\ }\textbf {\bibinfo {volume}
  {8}},\ \bibinfo {pages} {1057} (\bibinfo {year} {2006})}\BibitemShut
  {NoStop}%
\bibitem [{\citenamefont {Dunning}(1989)}]{dunning_gaussian_1989}%
  \BibitemOpen
  \bibfield  {author} {\bibinfo {author} {\bibfnamefont {T.~H.}\ \bibnamefont
  {Dunning}},\ }\href {\doibase 10.1063/1.456153} {\bibfield  {journal}
  {\bibinfo  {journal} {J. Chem. Phys.}\ }\textbf {\bibinfo {volume} {90}},\
  \bibinfo {pages} {1007} (\bibinfo {year} {1989})}\BibitemShut {NoStop}%
\bibitem [{\citenamefont {Lorenz}\ \emph {et~al.}(2012)\citenamefont {Lorenz},
  \citenamefont {Maschio}, \citenamefont {Sch\"{u}tz},\ and\ \citenamefont
  {Usvyat}}]{lorenz_local_2012}%
  \BibitemOpen
  \bibfield  {author} {\bibinfo {author} {\bibfnamefont {M.}~\bibnamefont
  {Lorenz}}, \bibinfo {author} {\bibfnamefont {L.}~\bibnamefont {Maschio}},
  \bibinfo {author} {\bibfnamefont {M.}~\bibnamefont {Sch\"{u}tz}}, \ and\
  \bibinfo {author} {\bibfnamefont {D.}~\bibnamefont {Usvyat}},\ }\href
  {\doibase 10.1063/1.4767775} {\bibfield  {journal} {\bibinfo  {journal} {J.
  Chem. Phys.}\ }\textbf {\bibinfo {volume} {137}},\ \bibinfo {pages} {204119}
  (\bibinfo {year} {2012})}\BibitemShut {NoStop}%
\bibitem [{\citenamefont {Binkley}, \citenamefont {Pople},\ and\ \citenamefont
  {Hehre}(1980)}]{binkley_self-consistent_1980}%
  \BibitemOpen
  \bibfield  {author} {\bibinfo {author} {\bibfnamefont {J.~S.}\ \bibnamefont
  {Binkley}}, \bibinfo {author} {\bibfnamefont {J.~A.}\ \bibnamefont {Pople}},
  \ and\ \bibinfo {author} {\bibfnamefont {W.~J.}\ \bibnamefont {Hehre}},\
  }\href {\doibase 10.1021/ja00523a008} {\bibfield  {journal} {\bibinfo
  {journal} {J. Am. Chem. Soc.}\ }\textbf {\bibinfo {volume} {102}},\ \bibinfo
  {pages} {939} (\bibinfo {year} {1980})}\BibitemShut {NoStop}%
\bibitem [{\citenamefont {H\"{a}ttig}(2005)}]{hattig_optimization_2005}%
  \BibitemOpen
  \bibfield  {author} {\bibinfo {author} {\bibfnamefont {C.}~\bibnamefont
  {H\"{a}ttig}},\ }\href
  {http://pubs.rsc.org/en/content/articlehtml/2005/cp/b415208e} {\bibfield
  {journal} {\bibinfo  {journal} {Phys. Chem. Chem. Phys.}\ }\textbf {\bibinfo
  {volume} {7}},\ \bibinfo {pages} {59} (\bibinfo {year} {2005})}\BibitemShut
  {NoStop}%
\bibitem [{Note1()}]{Note1}%
  \BibitemOpen
  \bibinfo {note} {Since we only focus on the most time-consuming step, namely
  the construction of direct-space exchange matrix (\protect \cref {eq:K}), for
  a chosen unit cell, we do not analyze the cost with respect to the number of
  sampled \protect \ensuremath {\protect \bm {k}} points in the first Brillouin
  zone ({\protect \it N}\protect \ensuremath {_{\protect \text {k}}}, see, for
  example, \protect \cref {eq:D}); in any case, the number of \protect
  \ensuremath {\protect \bm {k}} points can always be reduced to 1 by
  increasing unit cell size appropriately.}\BibitemShut {Stop}%
\bibitem [{\citenamefont {Ouyang}\ \emph {et~al.}(2001)\citenamefont {Ouyang},
  \citenamefont {Huang}, \citenamefont {Cheung},\ and\ \citenamefont
  {Lieber}}]{ouyang_energy_2001}%
  \BibitemOpen
  \bibfield  {author} {\bibinfo {author} {\bibfnamefont {M.}~\bibnamefont
  {Ouyang}}, \bibinfo {author} {\bibfnamefont {J.-L.}\ \bibnamefont {Huang}},
  \bibinfo {author} {\bibfnamefont {C.~L.}\ \bibnamefont {Cheung}}, \ and\
  \bibinfo {author} {\bibfnamefont {C.~M.}\ \bibnamefont {Lieber}},\ }\href
  {\doibase 10.1126/science.1058853} {\bibfield  {journal} {\bibinfo  {journal}
  {Science}\ }\textbf {\bibinfo {volume} {292}},\ \bibinfo {pages} {702}
  (\bibinfo {year} {2001})}\BibitemShut {NoStop}%
\bibitem [{\citenamefont {Matsuda}, \citenamefont {Tahir-Kheli},\ and\
  \citenamefont {Goddard}(2010)}]{matsuda_definitive_2010}%
  \BibitemOpen
  \bibfield  {author} {\bibinfo {author} {\bibfnamefont {Y.}~\bibnamefont
  {Matsuda}}, \bibinfo {author} {\bibfnamefont {J.}~\bibnamefont
  {Tahir-Kheli}}, \ and\ \bibinfo {author} {\bibfnamefont {W.~A.}\ \bibnamefont
  {Goddard}},\ }\href {\doibase 10.1021/jz100889u} {\bibfield  {journal}
  {\bibinfo  {journal} {J. Phys. Chem. Lett.}\ }\textbf {\bibinfo {volume}
  {1}},\ \bibinfo {pages} {2946} (\bibinfo {year} {2010})}\BibitemShut
  {NoStop}%
\bibitem [{\citenamefont {Park}, \citenamefont {Terakura},\ and\ \citenamefont
  {Hamada}(1987)}]{park_band-structure_1987}%
  \BibitemOpen
  \bibfield  {author} {\bibinfo {author} {\bibfnamefont {K.~T.}\ \bibnamefont
  {Park}}, \bibinfo {author} {\bibfnamefont {K.}~\bibnamefont {Terakura}}, \
  and\ \bibinfo {author} {\bibfnamefont {N.}~\bibnamefont {Hamada}},\ }\href
  {\doibase 10.1088/0022-3719/20/9/014} {\bibfield  {journal} {\bibinfo
  {journal} {J. Phys. C: Solid State Phys.}\ }\textbf {\bibinfo {volume}
  {20}},\ \bibinfo {pages} {1241} (\bibinfo {year} {1987})}\BibitemShut
  {NoStop}%
\bibitem [{\citenamefont {Si}\ \emph {et~al.}(2009)\citenamefont {Si},
  \citenamefont {Li}, \citenamefont {Shi}, \citenamefont {Niu},\ and\
  \citenamefont {Xue}}]{si_divacancies_2009}%
  \BibitemOpen
  \bibfield  {author} {\bibinfo {author} {\bibfnamefont {M.~S.}\ \bibnamefont
  {Si}}, \bibinfo {author} {\bibfnamefont {J.~Y.}\ \bibnamefont {Li}}, \bibinfo
  {author} {\bibfnamefont {H.~G.}\ \bibnamefont {Shi}}, \bibinfo {author}
  {\bibfnamefont {X.~N.}\ \bibnamefont {Niu}}, \ and\ \bibinfo {author}
  {\bibfnamefont {D.~S.}\ \bibnamefont {Xue}},\ }\href {\doibase
  10.1209/0295-5075/86/46002} {\bibfield  {journal} {\bibinfo  {journal} {EPL}\
  }\textbf {\bibinfo {volume} {86}},\ \bibinfo {pages} {46002} (\bibinfo {year}
  {2009})}\BibitemShut {NoStop}%
\bibitem [{\citenamefont {Huang}\ and\ \citenamefont
  {Lee}(2012)}]{huang_defect_2012}%
  \BibitemOpen
  \bibfield  {author} {\bibinfo {author} {\bibfnamefont {B.}~\bibnamefont
  {Huang}}\ and\ \bibinfo {author} {\bibfnamefont {H.}~\bibnamefont {Lee}},\
  }\href {\doibase 10.1103/PhysRevB.86.245406} {\bibfield  {journal} {\bibinfo
  {journal} {Phys. Rev. B}\ }\textbf {\bibinfo {volume} {86}},\ \bibinfo
  {pages} {245406} (\bibinfo {year} {2012})}\BibitemShut {NoStop}%
\bibitem [{\citenamefont {Dovesi}\ \emph {et~al.}(2018)\citenamefont {Dovesi},
  \citenamefont {Erba}, \citenamefont {Orlando}, \citenamefont
  {Zicovich-Wilson}, \citenamefont {Civalleri}, \citenamefont {Maschio},
  \citenamefont {R{\'e}rat}, \citenamefont {Casassa}, \citenamefont {Baima},
  \citenamefont {Salustro},\ and\ \citenamefont
  {Kirtman}}]{dovesi_quantum-mechanical_2018}%
  \BibitemOpen
  \bibfield  {author} {\bibinfo {author} {\bibfnamefont {R.}~\bibnamefont
  {Dovesi}}, \bibinfo {author} {\bibfnamefont {A.}~\bibnamefont {Erba}},
  \bibinfo {author} {\bibfnamefont {R.}~\bibnamefont {Orlando}}, \bibinfo
  {author} {\bibfnamefont {C.~M.}\ \bibnamefont {Zicovich-Wilson}}, \bibinfo
  {author} {\bibfnamefont {B.}~\bibnamefont {Civalleri}}, \bibinfo {author}
  {\bibfnamefont {L.}~\bibnamefont {Maschio}}, \bibinfo {author} {\bibfnamefont
  {M.}~\bibnamefont {R{\'e}rat}}, \bibinfo {author} {\bibfnamefont
  {S.}~\bibnamefont {Casassa}}, \bibinfo {author} {\bibfnamefont
  {J.}~\bibnamefont {Baima}}, \bibinfo {author} {\bibfnamefont
  {S.}~\bibnamefont {Salustro}}, \ and\ \bibinfo {author} {\bibfnamefont
  {B.}~\bibnamefont {Kirtman}},\ }\href {\doibase 10.1002/wcms.1360} {\bibfield
   {journal} {\bibinfo  {journal} {WIREs Comput. Mol. Sci.}\ ,\ \bibinfo
  {pages} {e1360}} (\bibinfo {year} {2018})}\BibitemShut {NoStop}%
\bibitem [{fri()}]{frisch_gaussian_manual}%
  \BibitemOpen
  \href@noop {} {}\bibinfo {howpublished} {Gaussian 09, Revision A.02, M. J.
  Frisch, G. W. Trucks, H. B. Schlegel, G. E. Scuseria, M. A. Robb, J. R.
  Cheeseman, G. Scalmani, V. Barone, G. A. Petersson, H. Nakatsuji, X. Li, M.
  Caricato, A. Marenich, J. Bloino, B. G. Janesko, R. Gomperts, B. Mennucci, H.
  P. Hratchian, J. V. Ortiz, A. F. Izmaylov, J. L. Sonnenberg, D.
  Williams-Young, F. Ding, F. Lipparini, F. Egidi, J. Goings, B. Peng, A.
  Petrone, T. Henderson, D. Ranasinghe, V. G. Zakrzewski, J. Gao, N. Rega, G.
  Zheng, W. Liang, M. Hada, M. Ehara, K. Toyota, R. Fukuda, J. Hasegawa, M.
  Ishida, T. Nakajima, Y. Honda, O. Kitao, H. Nakai, T. Vreven, K. Throssell,
  J. A. Montgomery, Jr., J. E. Peralta, F. Ogliaro, M. Bearpark, J. J. Heyd, E.
  Brothers, K. N. Kudin, V. N. Staroverov, T. Keith, R. Kobayashi, J. Normand,
  K. Raghavachari, A. Rendell, J. C. Burant, S. S. Iyengar, J. Tomasi, M.
  Cossi, J. M. Millam, M. Klene, C. Adamo, R. Cammi, J. W. Ochterski, R. L.
  Martin, K. Morokuma, O. Farkas, J. B. Foresman, and D. J. Fox, Gaussian,
  Inc., Wallingford CT, 2016.}\BibitemShut {Stop}%
\bibitem [{\citenamefont {Izmaylov}, \citenamefont {Scuseria},\ and\
  \citenamefont {Frisch}(2006)}]{izmaylov_efficient_2006}%
  \BibitemOpen
  \bibfield  {author} {\bibinfo {author} {\bibfnamefont {A.~F.}\ \bibnamefont
  {Izmaylov}}, \bibinfo {author} {\bibfnamefont {G.~E.}\ \bibnamefont
  {Scuseria}}, \ and\ \bibinfo {author} {\bibfnamefont {M.~J.}\ \bibnamefont
  {Frisch}},\ }\href {\doibase 10.1063/1.2347713} {\bibfield  {journal}
  {\bibinfo  {journal} {J. Chem. Phys.}\ }\textbf {\bibinfo {volume} {125}},\
  \bibinfo {pages} {104103} (\bibinfo {year} {2006})}\BibitemShut {NoStop}%
\bibitem [{\citenamefont {Hutter}\ \emph {et~al.}(2014)\citenamefont {Hutter},
  \citenamefont {Iannuzzi}, \citenamefont {Schiffmann},\ and\ \citenamefont
  {VandeVondele}}]{hutter_cp2k:_2014}%
  \BibitemOpen
  \bibfield  {author} {\bibinfo {author} {\bibfnamefont {J.}~\bibnamefont
  {Hutter}}, \bibinfo {author} {\bibfnamefont {M.}~\bibnamefont {Iannuzzi}},
  \bibinfo {author} {\bibfnamefont {F.}~\bibnamefont {Schiffmann}}, \ and\
  \bibinfo {author} {\bibfnamefont {J.}~\bibnamefont {VandeVondele}},\ }\href
  {\doibase 10.1002/wcms.1159} {\bibfield  {journal} {\bibinfo  {journal}
  {WIREs Comput. Mol. Sci.}\ }\textbf {\bibinfo {volume} {4}},\ \bibinfo
  {pages} {15} (\bibinfo {year} {2014})}\BibitemShut {NoStop}%
\bibitem [{\citenamefont {Sun}\ \emph {et~al.}(2018)\citenamefont {Sun},
  \citenamefont {Berkelbach}, \citenamefont {Blunt}, \citenamefont {Booth},
  \citenamefont {Guo}, \citenamefont {Li}, \citenamefont {Liu}, \citenamefont
  {McClain}, \citenamefont {Sayfutyarova}, \citenamefont {Sharma},
  \citenamefont {Wouters},\ and\ \citenamefont {Chan}}]{sun_pyscf:_2018}%
  \BibitemOpen
  \bibfield  {author} {\bibinfo {author} {\bibfnamefont {Q.}~\bibnamefont
  {Sun}}, \bibinfo {author} {\bibfnamefont {T.~C.}\ \bibnamefont {Berkelbach}},
  \bibinfo {author} {\bibfnamefont {N.~S.}\ \bibnamefont {Blunt}}, \bibinfo
  {author} {\bibfnamefont {G.~H.}\ \bibnamefont {Booth}}, \bibinfo {author}
  {\bibfnamefont {S.}~\bibnamefont {Guo}}, \bibinfo {author} {\bibfnamefont
  {Z.}~\bibnamefont {Li}}, \bibinfo {author} {\bibfnamefont {J.}~\bibnamefont
  {Liu}}, \bibinfo {author} {\bibfnamefont {J.~D.}\ \bibnamefont {McClain}},
  \bibinfo {author} {\bibfnamefont {E.~R.}\ \bibnamefont {Sayfutyarova}},
  \bibinfo {author} {\bibfnamefont {S.}~\bibnamefont {Sharma}}, \bibinfo
  {author} {\bibfnamefont {S.}~\bibnamefont {Wouters}}, \ and\ \bibinfo
  {author} {\bibfnamefont {G.~K.-L.}\ \bibnamefont {Chan}},\ }\href {\doibase
  10.1002/wcms.1340} {\bibfield  {journal} {\bibinfo  {journal} {WIREs Comput.
  Mol. Sci.}\ }\textbf {\bibinfo {volume} {8}},\ \bibinfo {pages} {e1340}
  (\bibinfo {year} {2018})}\BibitemShut {NoStop}%
\bibitem [{\citenamefont {Zicovich-Wilson}, \citenamefont {Dovesi},\ and\
  \citenamefont {Saunders}(2001)}]{zicovich-wilson_general_2001}%
  \BibitemOpen
  \bibfield  {author} {\bibinfo {author} {\bibfnamefont {C.~M.}\ \bibnamefont
  {Zicovich-Wilson}}, \bibinfo {author} {\bibfnamefont {R.}~\bibnamefont
  {Dovesi}}, \ and\ \bibinfo {author} {\bibfnamefont {V.~R.}\ \bibnamefont
  {Saunders}},\ }\href {\doibase 10.1063/1.1415745} {\bibfield  {journal}
  {\bibinfo  {journal} {J. Chem. Phys.}\ }\textbf {\bibinfo {volume} {115}},\
  \bibinfo {pages} {9708} (\bibinfo {year} {2001})}\BibitemShut {NoStop}%
\bibitem [{\citenamefont {Casassa}, \citenamefont {Zicovich-Wilson},\ and\
  \citenamefont {Pisani}(2006)}]{casassa_symmetry-adapted_2006}%
  \BibitemOpen
  \bibfield  {author} {\bibinfo {author} {\bibfnamefont {S.}~\bibnamefont
  {Casassa}}, \bibinfo {author} {\bibfnamefont {C.~M.}\ \bibnamefont
  {Zicovich-Wilson}}, \ and\ \bibinfo {author} {\bibfnamefont {C.}~\bibnamefont
  {Pisani}},\ }\href {\doibase 10.1007/s00214-006-0119-z} {\bibfield  {journal}
  {\bibinfo  {journal} {Theor. Chem. Acc.}\ }\textbf {\bibinfo {volume}
  {116}},\ \bibinfo {pages} {726} (\bibinfo {year} {2006})}\BibitemShut
  {NoStop}%
\bibitem [{\citenamefont {Marzari}\ and\ \citenamefont
  {Vanderbilt}(1997)}]{marzari_maximally_1997}%
  \BibitemOpen
  \bibfield  {author} {\bibinfo {author} {\bibfnamefont {N.}~\bibnamefont
  {Marzari}}\ and\ \bibinfo {author} {\bibfnamefont {D.}~\bibnamefont
  {Vanderbilt}},\ }\href {\doibase 10.1103/PhysRevB.56.12847} {\bibfield
  {journal} {\bibinfo  {journal} {Phys. Rev. B}\ }\textbf {\bibinfo {volume}
  {56}},\ \bibinfo {pages} {12847} (\bibinfo {year} {1997})}\BibitemShut
  {NoStop}%
\bibitem [{\citenamefont {Zeiner}, \citenamefont {Dirl},\ and\ \citenamefont
  {Davies}(1998)}]{zeiner_bloch_1998}%
  \BibitemOpen
  \bibfield  {author} {\bibinfo {author} {\bibfnamefont {P.}~\bibnamefont
  {Zeiner}}, \bibinfo {author} {\bibfnamefont {R.}~\bibnamefont {Dirl}}, \ and\
  \bibinfo {author} {\bibfnamefont {B.~L.}\ \bibnamefont {Davies}},\ }\href
  {\doibase 10.1063/1.532297} {\bibfield  {journal} {\bibinfo  {journal} {J.
  Math. Phys.}\ }\textbf {\bibinfo {volume} {39}},\ \bibinfo {pages} {2437}
  (\bibinfo {year} {1998})}\BibitemShut {NoStop}%
\bibitem [{\citenamefont {Ewald}(1921)}]{ewald_evaluation_1921}%
  \BibitemOpen
  \bibfield  {author} {\bibinfo {author} {\bibfnamefont {P.}~\bibnamefont
  {Ewald}},\ }\href@noop {} {\bibfield  {journal} {\bibinfo  {journal} {Ann.
  Phys.}\ }\textbf {\bibinfo {volume} {64}},\ \bibinfo {pages} {253} (\bibinfo
  {year} {1921})}\BibitemShut {NoStop}%
\bibitem [{\citenamefont {Toukmaji}\ and\ \citenamefont
  {Board}(1996)}]{toukmaji_ewald_1996}%
  \BibitemOpen
  \bibfield  {author} {\bibinfo {author} {\bibfnamefont {A.~Y.}\ \bibnamefont
  {Toukmaji}}\ and\ \bibinfo {author} {\bibfnamefont {J.~A.}\ \bibnamefont
  {Board}},\ }\href {\doibase 10.1016/0010-4655(96)00016-1} {\bibfield
  {journal} {\bibinfo  {journal} {Comput. Phys. Commun.}\ }\textbf {\bibinfo
  {volume} {95}},\ \bibinfo {pages} {73} (\bibinfo {year} {1996})}\BibitemShut
  {NoStop}%
\bibitem [{\citenamefont {Kittel}(1996)}]{kittel_introduction_1996}%
  \BibitemOpen
  \bibfield  {author} {\bibinfo {author} {\bibfnamefont {C.}~\bibnamefont
  {Kittel}},\ }\href@noop {} {\emph {\bibinfo {title} {Introduction to Solid
  State Physics}}}\ (\bibinfo  {publisher} {John Wiley \& Sons, New York},\
  \bibinfo {year} {1996})\BibitemShut {NoStop}%
\bibitem [{\citenamefont {Greengard}\ and\ \citenamefont
  {Rokhlin}(1987)}]{greengard_fast_1987}%
  \BibitemOpen
  \bibfield  {author} {\bibinfo {author} {\bibfnamefont {L.}~\bibnamefont
  {Greengard}}\ and\ \bibinfo {author} {\bibfnamefont {V.}~\bibnamefont
  {Rokhlin}},\ }\href {\doibase 10.1016/0021-9991(87)90140-9} {\bibfield
  {journal} {\bibinfo  {journal} {J. Comput. Phys.}\ }\textbf {\bibinfo
  {volume} {73}},\ \bibinfo {pages} {325} (\bibinfo {year} {1987})}\BibitemShut
  {NoStop}%
\bibitem [{\citenamefont {White}\ and\ \citenamefont
  {Head-Gordon}(1994)}]{white_derivation_1994}%
  \BibitemOpen
  \bibfield  {author} {\bibinfo {author} {\bibfnamefont {C.~A.}\ \bibnamefont
  {White}}\ and\ \bibinfo {author} {\bibfnamefont {M.}~\bibnamefont
  {Head-Gordon}},\ }\href {\doibase 10.1063/1.468354} {\bibfield  {journal}
  {\bibinfo  {journal} {J. Chem. Phys.}\ }\textbf {\bibinfo {volume} {101}},\
  \bibinfo {pages} {6593} (\bibinfo {year} {1994})}\BibitemShut {NoStop}%
\bibitem [{\citenamefont {Strain}, \citenamefont {Scuseria},\ and\
  \citenamefont {Frisch}(1996)}]{strain_achieving_1996}%
  \BibitemOpen
  \bibfield  {author} {\bibinfo {author} {\bibfnamefont {M.~C.}\ \bibnamefont
  {Strain}}, \bibinfo {author} {\bibfnamefont {G.~E.}\ \bibnamefont
  {Scuseria}}, \ and\ \bibinfo {author} {\bibfnamefont {M.~J.}\ \bibnamefont
  {Frisch}},\ }\href
  {http://search.proquest.com/docview/213567345/abstract/92454C57DAC3445BPQ/1}
  {\bibfield  {journal} {\bibinfo  {journal} {Science}\ }\textbf {\bibinfo
  {volume} {271}},\ \bibinfo {pages} {51} (\bibinfo {year} {1996})}\BibitemShut
  {NoStop}%
\bibitem [{\citenamefont {Challacombe}, \citenamefont {White},\ and\
  \citenamefont {Head-Gordon}(1997)}]{challacombe_periodic_1997}%
  \BibitemOpen
  \bibfield  {author} {\bibinfo {author} {\bibfnamefont {M.}~\bibnamefont
  {Challacombe}}, \bibinfo {author} {\bibfnamefont {C.}~\bibnamefont {White}},
  \ and\ \bibinfo {author} {\bibfnamefont {M.}~\bibnamefont {Head-Gordon}},\
  }\href {\doibase 10.1063/1.474150} {\bibfield  {journal} {\bibinfo  {journal}
  {J. Chem. Phys.}\ }\textbf {\bibinfo {volume} {107}},\ \bibinfo {pages}
  {10131} (\bibinfo {year} {1997})}\BibitemShut {NoStop}%
\bibitem [{\citenamefont {Kudin}\ and\ \citenamefont
  {Scuseria}(2000)}]{kudin_linear-scaling_2000}%
  \BibitemOpen
  \bibfield  {author} {\bibinfo {author} {\bibfnamefont {K.~N.}\ \bibnamefont
  {Kudin}}\ and\ \bibinfo {author} {\bibfnamefont {G.~E.}\ \bibnamefont
  {Scuseria}},\ }\href {\doibase 10.1103/PhysRevB.61.16440} {\bibfield
  {journal} {\bibinfo  {journal} {Phys. Rev. B}\ }\textbf {\bibinfo {volume}
  {61}},\ \bibinfo {pages} {16440} (\bibinfo {year} {2000})}\BibitemShut
  {NoStop}%
\bibitem [{\citenamefont {Kudin}\ and\ \citenamefont
  {Scuseria}(2004)}]{kudin_revisiting_2004}%
  \BibitemOpen
  \bibfield  {author} {\bibinfo {author} {\bibfnamefont {K.~N.}\ \bibnamefont
  {Kudin}}\ and\ \bibinfo {author} {\bibfnamefont {G.~E.}\ \bibnamefont
  {Scuseria}},\ }\href {\doibase 10.1063/1.1771634} {\bibfield  {journal}
  {\bibinfo  {journal} {J. Chem. Phys.}\ }\textbf {\bibinfo {volume} {121}},\
  \bibinfo {pages} {2886} (\bibinfo {year} {2004})}\BibitemShut {NoStop}%
\bibitem [{\citenamefont {Genovese}\ \emph {et~al.}(2006)\citenamefont
  {Genovese}, \citenamefont {Deutsch}, \citenamefont {Neelov}, \citenamefont
  {Goedecker},\ and\ \citenamefont {Beylkin}}]{Genovese:2006eb}%
  \BibitemOpen
  \bibfield  {author} {\bibinfo {author} {\bibfnamefont {L.}~\bibnamefont
  {Genovese}}, \bibinfo {author} {\bibfnamefont {T.}~\bibnamefont {Deutsch}},
  \bibinfo {author} {\bibfnamefont {A.}~\bibnamefont {Neelov}}, \bibinfo
  {author} {\bibfnamefont {S.}~\bibnamefont {Goedecker}}, \ and\ \bibinfo
  {author} {\bibfnamefont {G.}~\bibnamefont {Beylkin}},\ }\href@noop {}
  {\bibfield  {journal} {\bibinfo  {journal} {J Chem Phys}\ }\textbf {\bibinfo
  {volume} {125}},\ \bibinfo {pages} {074105} (\bibinfo {year}
  {2006})}\BibitemShut {NoStop}%
\bibitem [{\citenamefont {Beylkin}, \citenamefont {Cheruvu},\ and\
  \citenamefont {Perez}(2008)}]{Beylkin:2008im}%
  \BibitemOpen
  \bibfield  {author} {\bibinfo {author} {\bibfnamefont {G.}~\bibnamefont
  {Beylkin}}, \bibinfo {author} {\bibfnamefont {V.}~\bibnamefont {Cheruvu}}, \
  and\ \bibinfo {author} {\bibfnamefont {F.}~\bibnamefont {Perez}},\
  }\href@noop {} {\bibfield  {journal} {\bibinfo  {journal} {Applied and
  Computational Harmonic Analysis}\ }\textbf {\bibinfo {volume} {24}},\
  \bibinfo {pages} {354} (\bibinfo {year} {2008})}\BibitemShut {NoStop}%
\bibitem [{\citenamefont {Sierka}, \citenamefont {Hogekamp},\ and\
  \citenamefont {Ahlrichs}(2003)}]{Sierka:2003bs}%
  \BibitemOpen
  \bibfield  {author} {\bibinfo {author} {\bibfnamefont {M.}~\bibnamefont
  {Sierka}}, \bibinfo {author} {\bibfnamefont {A.}~\bibnamefont {Hogekamp}}, \
  and\ \bibinfo {author} {\bibfnamefont {R.}~\bibnamefont {Ahlrichs}},\
  }\href@noop {} {\bibfield  {journal} {\bibinfo  {journal} {J. Chem. Phys.}\
  }\textbf {\bibinfo {volume} {118}},\ \bibinfo {pages} {9136} (\bibinfo {year}
  {2003})}\BibitemShut {NoStop}%
\bibitem [{\citenamefont {P{\'e}rez‐Jord{\'a}}\ and\ \citenamefont
  {Yang}(1996)}]{perezjorda_concise_1996}%
  \BibitemOpen
  \bibfield  {author} {\bibinfo {author} {\bibfnamefont {J.~M.}\ \bibnamefont
  {P{\'e}rez‐Jord{\'a}}}\ and\ \bibinfo {author} {\bibfnamefont
  {W.}~\bibnamefont {Yang}},\ }\href {\doibase 10.1063/1.471517} {\bibfield
  {journal} {\bibinfo  {journal} {J. Chem. Phys.}\ }\textbf {\bibinfo {volume}
  {104}},\ \bibinfo {pages} {8003} (\bibinfo {year} {1996})}\BibitemShut
  {NoStop}%
\bibitem [{\citenamefont {Carlson}\ and\ \citenamefont
  {Rushbrooke}(1950)}]{Carlson:1950ca}%
  \BibitemOpen
  \bibfield  {author} {\bibinfo {author} {\bibfnamefont {B.~C.}\ \bibnamefont
  {Carlson}}\ and\ \bibinfo {author} {\bibfnamefont {G.~S.}\ \bibnamefont
  {Rushbrooke}},\ }\href@noop {} {\bibfield  {journal} {\bibinfo  {journal}
  {Math. Proc. Camb. Phil. Soc.}\ }\textbf {\bibinfo {volume} {46}},\ \bibinfo
  {pages} {626} (\bibinfo {year} {1950})}\BibitemShut {NoStop}%
\bibitem [{\citenamefont {Buehler}\ and\ \citenamefont
  {Hirschfelder}(1951)}]{Buehler:1951ca}%
  \BibitemOpen
  \bibfield  {author} {\bibinfo {author} {\bibfnamefont {R.~J.}\ \bibnamefont
  {Buehler}}\ and\ \bibinfo {author} {\bibfnamefont {J.~O.}\ \bibnamefont
  {Hirschfelder}},\ }\href@noop {} {\bibfield  {journal} {\bibinfo  {journal}
  {Phys. Rev.}\ }\textbf {\bibinfo {volume} {83}},\ \bibinfo {pages} {628}
  (\bibinfo {year} {1951})}\BibitemShut {NoStop}%
\bibitem [{\citenamefont {Solov'yov}\ \emph {et~al.}(2007)\citenamefont
  {Solov'yov}, \citenamefont {Yakubovich}, \citenamefont {Solov'yov},\ and\
  \citenamefont {Greiner}}]{Solovyov:2007bk}%
  \BibitemOpen
  \bibfield  {author} {\bibinfo {author} {\bibfnamefont {I.~A.}\ \bibnamefont
  {Solov'yov}}, \bibinfo {author} {\bibfnamefont {A.~V.}\ \bibnamefont
  {Yakubovich}}, \bibinfo {author} {\bibfnamefont {A.~V.}\ \bibnamefont
  {Solov'yov}}, \ and\ \bibinfo {author} {\bibfnamefont {W.}~\bibnamefont
  {Greiner}},\ }\href@noop {} {\bibfield  {journal} {\bibinfo  {journal} {Phys.
  Rev. E}\ }\textbf {\bibinfo {volume} {75}},\ \bibinfo {pages} {437} (\bibinfo
  {year} {2007})}\BibitemShut {NoStop}%
\bibitem [{\citenamefont {Kudin}\ and\ \citenamefont
  {Scuseria}(1998{\natexlab{a}})}]{kudin_fast_1998}%
  \BibitemOpen
  \bibfield  {author} {\bibinfo {author} {\bibfnamefont {K.~N.}\ \bibnamefont
  {Kudin}}\ and\ \bibinfo {author} {\bibfnamefont {G.~E.}\ \bibnamefont
  {Scuseria}},\ }\href {\doibase 10.1016/S0009-2614(97)01329-8} {\bibfield
  {journal} {\bibinfo  {journal} {Chem. Phys. Lett.}\ }\textbf {\bibinfo
  {volume} {283}},\ \bibinfo {pages} {61} (\bibinfo {year}
  {1998}{\natexlab{a}})}\BibitemShut {NoStop}%
\bibitem [{\citenamefont {Kudin}\ and\ \citenamefont
  {Scuseria}(1998{\natexlab{b}})}]{kudin_fast_1998-1}%
  \BibitemOpen
  \bibfield  {author} {\bibinfo {author} {\bibfnamefont {K.~N.}\ \bibnamefont
  {Kudin}}\ and\ \bibinfo {author} {\bibfnamefont {G.~E.}\ \bibnamefont
  {Scuseria}},\ }\href {\doibase 10.1016/S0009-2614(98)00468-0} {\bibfield
  {journal} {\bibinfo  {journal} {Chem. Phys. Lett.}\ }\textbf {\bibinfo
  {volume} {289}},\ \bibinfo {pages} {611} (\bibinfo {year}
  {1998}{\natexlab{b}})}\BibitemShut {NoStop}%
\bibitem [{\citenamefont {Valeev}(2018)}]{Libint2}%
  \BibitemOpen
  \bibfield  {author} {\bibinfo {author} {\bibfnamefont {E.~F.}\ \bibnamefont
  {Valeev}},\ }\href {http://libint.valeyev.net/} {\enquote {\bibinfo {title}
  {Libint: A library for the evaluation of molecular integrals of many-body
  operators over gaussian functions},}\ }\bibinfo {howpublished}
  {http://libint.valeyev.net/} (\bibinfo {year} {2018}),\ \bibinfo {note}
  {version 2.5.0-beta.1}\BibitemShut {NoStop}%
\bibitem [{\citenamefont {de~Leeuw}, \citenamefont {Perram},\ and\
  \citenamefont {Smith}(1980)}]{deLeeuw:1980kw}%
  \BibitemOpen
  \bibfield  {author} {\bibinfo {author} {\bibfnamefont {S.~W.}\ \bibnamefont
  {de~Leeuw}}, \bibinfo {author} {\bibfnamefont {J.~W.}\ \bibnamefont
  {Perram}}, \ and\ \bibinfo {author} {\bibfnamefont {E.~R.}\ \bibnamefont
  {Smith}},\ }\href@noop {} {\bibfield  {journal} {\bibinfo  {journal} {Proc.
  R. Soc. Lond. A}\ }\textbf {\bibinfo {volume} {373}},\ \bibinfo {pages} {27}
  (\bibinfo {year} {1980})}\BibitemShut {NoStop}%
\bibitem [{\citenamefont {Euwema}\ and\ \citenamefont
  {Surratt}(1975)}]{Euwema:1975ee}%
  \BibitemOpen
  \bibfield  {author} {\bibinfo {author} {\bibfnamefont {R.~N.}\ \bibnamefont
  {Euwema}}\ and\ \bibinfo {author} {\bibfnamefont {G.~T.}\ \bibnamefont
  {Surratt}},\ }\href@noop {} {\bibfield  {journal} {\bibinfo  {journal}
  {Journal of Physics and Chemistry of Solids}\ }\textbf {\bibinfo {volume}
  {36}},\ \bibinfo {pages} {67} (\bibinfo {year} {1975})}\BibitemShut {NoStop}%
\end{thebibliography}%

\end{document}